\pdfoutput=1
\RequirePackage{ifpdf}
\ifpdf
\documentclass[pdftex]{sigma}
\else
\documentclass{sigma}
\fi

\usepackage[shortlabels]{enumitem}
\usepackage{mathtools,amsfonts,bm,mathrsfs, bbm}
\usepackage{array,tabularx,booktabs}
\usepackage{multicol}
\usepackage{tikz,tikz-cd,cancel}
\usetikzlibrary{decorations.pathmorphing,decorations}
\usepackage{circuitikz}
\usepackage{esint}
\allowdisplaybreaks
\DeclareGraphicsExtensions{.pdf,.png,.jpg}

\numberwithin{equation}{section}

\newtheorem{Result}{Result}[section]

\theoremstyle{definition}
\newtheorem{Remark}[Result]{Remark}

\begin{document}

\allowdisplaybreaks

\newcommand{\arXivNumber}{2602.03663}

\renewcommand{\PaperNumber}{097}

\FirstPageHeading

\ShortArticleName{Dirac Observables for Gowdy Cosmologies Regular at the Big Bang}
\ArticleName{Dirac Observables for Gowdy Cosmologies\\ Regular at the Big Bang}

\Author{Max NIEDERMAIER and Mahdi SEDIGHI JAFARI}
\AuthorNameForHeading{M.~Niedermaier and M.~Sedighi Jafari}

\Address{Department of Physics and Astronomy, University of Pittsburgh, \\
100 Allen Hall, Pittsburgh, PA 15260, USA}
\Email{\mail{mnie@pitt.edu}, \mail{mas1371@pitt.edu}}

\ArticleDates{Received February 09, 2026, in final form September 21, 2026; Published online October 14, 2026}

\Abstract{Gowdy cosmologies are exact, spatially inhomogeneous solutions of the vacuum Einstein equations which describe nonlinear gravitational waves coalescing at the big bang singularity. With toroidal spatial sections they provenly have the asymptotic velocity domination property, in that close to the big bang dynamical spatial gradients fade out and the dynamics is governed by a~Carroll-type gravity theory. Here we construct an infinite set of exact Dirac observables for Gowdy cosmologies, valid off-shell, strongly, and without gauge fixing. These observables stay regular at the big bang and can be matched to much simpler Dirac observables of the Carroll-type gravity theory. Conversely, in an adapted foliation there is a~systematic anti-Newtonian expansion (in inverse powers of the reduced Newton constant) of the full Dirac observables whose leading terms are the Carroll ones. In particular, this provides an off-shell generalization of the asymptotic velocity domination property.}


\Classification{35Q75; 83C20; 83F05}

\section{Introduction}

In the absence of a~preferred background geometry the notion of
observables in gravitational theories is subtle. In a~Hamiltonian
formulation, Dirac observables $\mathcal{O}$ are defined as quantities
weakly Poisson commuting with the constraints
\begin{equation*}
\{ \mathcal{H}_0, \mathcal{O}\}|_{\mathcal{H}_0 = 0 = \mathcal{H}_a} = 0 =
\{ \mathcal{H}_a, \mathcal{O}\}|_{\mathcal{H}_0 = 0 = \mathcal{H}_a},
\end{equation*}
where $\mathcal{H}_0$ is the Hamiltonian constraint and $\mathcal{H}_a$, $a =1,2,3$,
are the diffeomorphism constraints. A~sufficiently large set of Dirac
observables $\mathcal{O}$ is thought to adequately capture the intrinsic,
gauge invariant content of a~gravity theory. We refer to a~few
review-like articles~\cite{Giddings,Hoehn,Pons,Tambornino} for
pointers to the vast literature. For vacuum gravity, Dirac observables
are exceedingly difficult to construct, so normally one has to
be content with lowest order perturbative constructions, toy models
with a~finite number of degrees of freedom, or schematic sketches.

The subsectors of Einstein gravity with two commuting Killing vectors
offer an interesting middle ground. These are effectively two-dimensional
field theories with a~diffeomorphism-type local gauge invariance whose
classical solutions are also exact solutions of the (four-dimensional)
Einstein field equations~\cite{Alekseev,Gravisolitons,Griffithsbook,Kleinbook}.
In addition, they have a~bonus mathematical structure~\cite{Gravisolitons,BM,Maison},
which we shall employ here
to construct one-parameter families of \textit{exact} Dirac observables
for the so-called Gowdy cosmologies. These describe inhomogeneous
cosmological vacuum spacetimes where nonlinear gravitational waves
coalesce at a~big bang singularity~\cite{Matzner,Gowdy}. The spatial
sections can have different topologies of which three-tori $\bigl(T^3\bigr)$
or a~variant with one non-compact direction $\bigl(\mathbb{R} \times T^2\bigr)$ are considered.

For orientation, we first summarize our main results in schematic form
and then comment on the various points. Let $\theta$ be a~complex
(`spectral') parameter with $\operatorname{Im} \theta \neq 0$. We construct
several one-parameter families of exact Dirac observables $\mathcal{O}(\theta)$
for the (`unpolarized') $\mathbb{R} \times T^2$ and $T^3$ Gowdy cosmologies
with the following properties:
\begin{itemize}\itemsep=0pt
\item[(a)] $\mathcal{O}(\theta)$ is a~functional on the full
phase space which Poisson commutes strongly with the constraints
\begin{gather}
\{ \mathcal{H}_0, \mathcal{O}(\theta)\}= 0 = \{ \mathcal{H}_1, \mathcal{O}(\theta)\}.
\label{i2}
\end{gather}
\item[(b)] The property~\eqref{i2} holds without using any equations
 of motion and without invoking a~gauge fixing procedure or related conduits.
\item[(c)] $\mathcal{O}(\theta)$ is spatially nonlocal but refers to a~fixed time;
 it remains regular at the big bang.
\item[(d)]$\mathcal{O}(\theta)$
admits a~convergent `anti-Newtonian' expansion in \textit{inverse} powers
of the reduced Newton constant $\lambda_{\text{\sc n}}>0$, whose leading
term corresponds to a~one-parameter family of Dirac observables
$\mathcal{O}^{\text{\sc vd}}(\theta)$ (with properties (a), (b), (c)) of the
velocity dominated $\mathbb{R} \times T^2$ or~$T^3$ Gowdy system
\begin{gather}
\mathcal{O}(\theta) = \mathcal{O}_0(\theta) + O\bigl(\rho^2/\lambda_{\text{\sc n}}^2\bigr),
 \qquad \mathcal{O}_0(\theta)\big|_{\text{\sc vd} \mathrm{ phase space}} =
\mathcal{O}^{\text{\sc vd}}(\theta),
 \qquad \operatorname{Im} \theta \neq 0.
\label{i3}
\end{gather}
Here $\rho>0$ is a~temporal function whose $\rho=0$ level set can be
identified with the big bang.\looseness=-1
\end{itemize}
These results of course have precursors
\cite{AHusain,Husain,KSGowdyNC,KSYangian,ManoS,Gerochreview}
and we shall comment on the precise relation individually later on.

Concerning points (a), (b), most of the literature on 2-Killing vector
reductions aims at the construction and characterization of classical
solutions~\cite{Gravisolitons,Griffithsbook,IsenMonc,KRend,Kleinbook,Li,%
RingstroemInitial,Ringstroemproof}. To this end, typically a~complete or partial gauge fixing
condition is imposed and the gravitational constraints~${\mathcal{H}_0 = 0 = \mathcal{H}_1}$
are solved for one of the degrees of freedom (leading to a~``reduced phase
space''). In the present context, we
neither want to solve the constraints nor do we want to impose any gauge
condition. The phase space consists of four conjugate field pairs and
$\mathcal{H}_0$, $\mathcal{H}_1$ are local functions of these pairs which generate
(in the absence of gauge fixing!) infinitesimal diffeomorphisms in
the two non-Killing directions. The Dirac observables $\mathcal{O}(\theta)$
sought are functionals on the same phase space and~\eqref{i2}
should hold without use of equations of motion.\looseness=-1

Concerning point (c), there are no-go results~\cite{TorreNoGo} that
preclude the existence of local Dirac observables in vacuum Einstein
gravity with compact spatial sections. This result is expected to carry
over to the Gowdy cosmologies, so the spatial nonlocality of the
$\mathcal{O}(\theta)$ we find should not be surprising. Nevertheless, our
$\mathcal{O}(\theta)$ do not involve temporal integrals and refer to a~fixed coordinate time (at which also the equal-time Poisson
brackets in~\eqref{i2} are taken). This coordinate time can
(but need not) be related to the level surfaces of a~scalar function with timelike gradient, i.e.,~a temporal function.
In 2-Killing vector reductions, such a~function exists and is traditionally denoted by $\rho$ (though $T$ would
be clearer in this context), is strictly positive, and for Gowdy cosmologies is such that
the $\rho =0$ level set can be identified with the big bang.
Further, $\rho$ and its conjugate momentum form one of the four
phase space pairs, and our Dirac observables are also functionals
of $\rho$. They have the remarkable property of staying regular
as~${\rho \rightarrow 0}$.\looseness=-1

The limiting quantity $\mathcal{O}_0(\theta)$ has an intriguing interpretation
alluded to in (d): it is (upon identification of the phase space
variables) a~Dirac observable of the 2-Killing vector reduction of
another gravity theory, which we shall refer to as velocity dominated
(VD) gravity. This gravity theory has a~long history dating back to
the 1970s~\cite{Henn1,Isham}, and is now often
called ``electric Carroll gravity'' \cite{Carrollgrav1}. It can be
thought of as an
ultrarelativistic limit of Einstein gravity, where dynamical
spatial gradients are absent. The limit can be interpreted
in different ways: as a~``speed of light to zero limit'', as a~``Newton constant to infinity limit'', or as a~``scale
of the spatial metric to infinity limit''. We prefer the latter, and
after multiplying the spatial metric by~$\ell^2$ send $\ell \rightarrow \infty$
at the end.
In Figure~\ref{2Kdiagram} below, this is the upper horizontal arrow,
starting from the~${1 + 3}$ form of the Einstein--Hilbert action.
Next, we perform a~2-Killing vector reduction of this VD gravity action
to find a~reduced gravity system with action $S_{\text{\sc vd2k}}$.
This begs the question of what happens if the two operations
are performed in the opposite order, giving~$S_{\text{\sc 2kvd}}$.
Both actions are found to coincide, $S_{\text{\sc vd2k}} =
S_{\text{\sc 2kvd}}$, turning Figure~\ref{2Kdiagram} into a~commutative diagram.
The details are relegated to Appendix~\ref{appA}.

These comments explain the $\mathcal{O}_0(\theta)
|_{\text{\sc vd} \  \mathrm{phase} \  \mathrm{space}} =\mathcal{O}^{\text{\sc vd}}(\theta)$
part of the assertion in (d). What about the subleading terms?
In gravity theories, one almost always uses ``post-Newtonian''
expansions (in inverse powers of the speed of light) or the
related ``post-Minkowskian'' expansions (in positive powers
of Newton's constant). The same could be done in the 2-Killing subsector,
where an expansion in positive powers of the reduced Newton constant $\lambda_{\text{\sc n}}$
would arise. In view of the above interpretation of $\mathcal{O}_0(\theta)$,
the natural expansion here is, however, one
away from the big bang, i.e.,~in powers of the temporal function $\rho$.
Our Dirac observables turn out to admit a~\textit{convergent} expansion
of this form. Moreover, the powers of $\rho$
turn out to correlate with \textit{inverse} powers of Newton's constant
and~\eqref{i3} can be viewed as an anti-Newtonian expansion (opposite
to the post-Newtonian one). In view of the interpretation of VD gravity
as a~``Newton constant to infinity limit'' this pattern is not
surprising. Although our considerations are all off-shell and
related to Dirac observables, we should also mention that these results
tie in nicely with a~remarkable dynamical (on-shell) property of
$T^3$ Gowdy cosmologies, known as asymptotic velocity domination (AVD)
\cite{IsenMonc,KRend,Li,Ringstroemproof,RingstroemLR}.

\begin{figure}[!th]
 \centering
\resizebox{0.77\textwidth}{!}{%
\begin{circuitikz}
\tikzstyle{every node}=[font=\LARGE]


\draw[->, >=Stealth] (4.25,11.15) -- (10.25,11.15)
 node[midway, below=2pt, font=\normalsize] {$\ell \rightarrow \infty$};

\node [font=\normalsize] at (11.75,11.25) {$S^L_{\text{\sc vd}}$};

\node [font=\normalsize, color={rgb,400:red,30; green,250; blue,100}]
at (2.75,12.5) {Full Einstein gravity};
\node [font=\normalsize, color={rgb,400:red,30; green,250; blue,100}]
at (11.75,12.5) {VD gravity};

\node [font=\normalsize, color={rgb,400:red,30; green,250; blue,100}]
at (2.75,4.5) {Full 2K reduction};
\node [font=\normalsize, color={rgb,400:red,30; green,250; blue,100}]
at (11.75,4.5) {VD 2K reduction};

\node [font=\normalsize] at (2.75,11.25) {$S^L_{1+3}$};

\draw[->, >=Stealth] (2.75,10.5) -- (2.75,6.5)
 node[midway, right=3pt, font=\normalsize] {Two Killing};
\node [font=\normalsize] at (2.75,6) {$S^L_{\text{\sc 2k}}$};

\draw[->, >=Stealth] (4.25,5.90) -- (10.25,5.90)
 node[midway, below=2pt, font=\normalsize] {$\ell \rightarrow \infty$};

\draw[->, >=Stealth] (11.75,10.5) -- (11.75,6.5)
 node[midway, right=3pt, font=\normalsize] {Two VD Killing};
 \node [font=\normalsize] at (11.75,6) {$S^L_{\text{\sc 2kvd}}
 = S^L_{\text{\sc vd2k}}$};

\end{circuitikz}
}%
\caption{Illustration of the commutativity of the operations
 ``taking the ultrarelativistic limit'' and
 ``taking the two-Killing vector reduction''.}
\label{2Kdiagram}
\end{figure}

The article is organized as follows. In Section~\ref{sec2}, we introduce the
subsectors of Einstein gravity with two commuting spacelike Killing
vectors. Their Lax pair and an ensued one-parameter family
of conserved currents are presented without fixing a~gauge,
generalizing earlier treatments~\cite{Gravisolitons,Maison}.
This sets the arena for the
construction of the Dirac observables~\eqref{i2} for Gowdy cosmologies:
strongly, off-shell, and without gauge fixing. The constructions for
topologies $\mathbb{R} \times T^2$ and $T^3$ are substantially different and
are discussed in Sections~\ref{sec3.1} and~\ref{sec3.2}, respectively.
After introducing the velocity dominated Gowdy
system as a~gravity theory in its own right in Section~\ref{sec4.1}, several
one-parameter families of Dirac observables are identified in Section~\ref{sec4.2}.
The link~\eqref{i3} is established by means of a~kinematical scaling
decomposition in Section~\ref{sec4.3}, while the relation to the dynamics of
AVD is outlined in Section~\ref{sec4.4}. Appendix~\ref{appA} contains a~summary of
the relevant action principles, their interplay, and their local
gauge symmetries. A~rather technical construction of a~limiting
transition matrix needed for $T^3$ Gowdy cosmologies is relegated to
Appendix~\ref{appB}. Appendix~\ref{appC} discusses the so-called monodromy matrix. In lieu of a~theorem-proof exposition we
summarize our main results in a~\textbf{Result s.n} format,
where $s$ is the section number and $n$ is a~counter.


\section{Two spacelike Killing vector reduction of Einstein gravity}\label{sec2}

The 2-Killing vector reductions of Einstein gravity comprise
physically very different situations, depending on the nature of
the Killing vectors and the topology or boundary conditions imposed:
(i) stationary axisymmetric solutions, (ii) colliding plane waves,
(iii) cylindrical gravitational waves, or (iv) spatially inhomogeneous
cosmologies. In situation (i) one of the Killing vectors is timelike
and the other spacelike. In situations (ii), (iii) and (iv), both Killing
vectors are spacelike and the reduced system evolves in time.
The colliding plane wave spacetimes (ii) have dynamical singularities
and are defined in quadrants, making them less suited for our present
quest. The sectors (iii), (iv) can to some extent be treated in parallel
and the exposition in this section applies to both.

We thus consider the infinite-dimensional subsector of Einstein gravity
corresponding to metrics $g^\mathrm{2K}$ with
two commuting spacelike Killing vectors $K_1$, $K_2$. In adapted coordinates
$X = \bigl(x^0,x^1,y^1,y^2\bigr)$ any such metric can be written as
\begin{gather}
g^\mathrm{2K}_{IJ}(X) {\rm d}X^I {\rm d}X^J= \gamma_{\mu\nu}(x) {\rm d}x^{\mu} {\rm d}x^{\nu} +
\rho(x) M_{ab}(x) {\rm d}y^a {\rm d}y^b.
\label{2Kmetric}
\end{gather}
Here, all fields only depend on $\bigl(x^0,x^1\bigr)$ while $\bigl(y^1,y^2\bigr)$ are
Killing coordinates, i.e.,~are such that locally $K_a = K_a^I \partial_I =
\partial/\partial y^a$, $a =1,2$. A~derivation of~\eqref{2Kmetric} (for the case of
one timelike and one spacelike Killing vector) can be found in
\cite[Section 7.2 and Appendix C]{Straumannbook}. For cylindrical
gravitational waves (iii), the gradient $\partial_{\mu} \rho$ is globally
spacelike, while Gowdy cosmologies (iv) have $\partial_{\mu}\rho$ globally
timelike, both with respect to the $\gamma^{\mu\nu}$ below.

\textbf{Parameterization of the block matrices.}
The symmetric matrix $\rho M_{ab}$ can be defined in terms of the
Killing vectors
\begin{gather}
K_a^I K_b^J g^\mathrm{2K}_{IJ} = \rho M_{ab}, \qquad \det M =1.
\label{Mdef}
\end{gather}
The factorization into a~unimodular piece $M_{ab}$ and $\rho >0$
is convenient later on. Lowering indices~$\smash{K_{a,I} = g^\mathrm{2K}_{IJ} K_a^J}$
one can introduce
\smash{$
\gamma_{IJ} := g^\mathrm{2K}_{IJ} - \rho^{-1} \bigl(M^{-1}\bigr)^{ab} K_{a,I} K_{b,J}$}.
It satisfies $K_a^I \gamma_{IJ} =0$ and $\mathcal{L}_{K_a} \gamma_{IJ} =0$,
and therefore is the $(1+1)$-dimensional Lorentzian metric induced on
the orbits of the Killing vectors, see, e.g.,~\cite{TorreRom}. In adapted coordi\-nates $\gamma_{IJ} {\rm d}X^I {\rm d}X^J = \gamma_{\mu\nu}(x) {\rm d}x^{\mu} {\rm d}x^{\nu}$, for
which we adopt the following lapse-shift type parameterization
\begin{gather}
\gamma_{\mu\nu}(x) {\rm d}x^{\mu} {\rm d}x^{\nu} =
{\rm e}^{\tilde{\sigma}}\bigl[ - n^2 \bigl({\rm d}x^0\bigr)^2 + \bigl({\rm d}x^1 + s {\rm d}x^0\bigr)^2\bigr],
\label{2Dmetric}
\end{gather}
i.e.,~$\gamma_{00} = {\rm e}^{\tilde{\sigma}}\bigl(-n^2 + s^2\bigr)$,
$\gamma_{01} = {\rm e}^{\tilde{\sigma}}s = \gamma_{10}$,
$\gamma_{11} = {\rm e}^{\tilde{\sigma}}$. By a~slight abuse of terminology,
we shall refer to $n$, $s$ as lapse, shift, respectively.%
\footnote{ For a~two-dimensional Lorentzian metric the
 lapse $N$, shift $N^1$, spatial metric $\gamma_{11}$ proper would be
 $N = {\rm e}^{\tilde{\sigma}/2} n$, $N^1 = s$, $\gamma_{11} = {\rm e}^{ \tilde{\sigma}}$.
 Note that ${\rm e}^{\tilde{\sigma}}$, $n$ are spatial densities of weight $2$, $-1$,
 respectively, and that $s$, being a~one-dimensional spatial vector,
 can also be viewed as a~spatial density of weight $-1$.
}
For the volume element, this gives $\sqrt{-\gamma} = {\rm e}^{\tilde{\sigma}} n$.
As indicated, $\gamma$ has eigenvalues $(-,+)$ and for
\eqref{2Kmetric} to have eigenvalues $(-,+,+,+)$ one needs
$\rho M$ to be positive definite. Since $\det M =1$, the latter
amounts to $\rho, M_{22}>0$ everywhere.
For the matrix $M_{ab}$, often explicit parameterizations are used.
One reads
\begin{gather}
M = \frac{1}{\Delta} \begin{pmatrix} \Delta^2 + \psi^2 & \psi
 \\ \psi & 1 \end{pmatrix} =
\begin{pmatrix} 1 & \psi \\ 0 & 1 \end{pmatrix}
\begin{pmatrix} \Delta & 0 \\ 0 & \Delta^{-1}\end{pmatrix}
\begin{pmatrix} 1 & 0 \\ \psi & 1 \end{pmatrix},
 \qquad \Delta >0,
\label{Mmat1}
\end{gather}
so that $\psi\equiv 0$ corresponds to orthogonal Killing vectors
and the upper half plane variable $\psi +{\rm i} \Delta$ is (in the
axisymmetric case) the traditional ``Ernst potential''
\cite{Griffithsbook,Kleinbook,Straumannbook}.
The form~\eqref{Mmat1} displays the underlying group theory in that
$M = V V^{\mathsf T}$, where $V$ parameterizes an element of the
coset~${\mathrm{SL}(2,\mathbb{R})/\mathrm{SO}(2)}$ and $C \in \mathrm{SL}(2,\mathbb{R})$
acts via $ M \mapsto C M C^{\mathsf T}$. For later use, we note
\begin{gather}
\bigl(\partial_t M M^{-1}\bigr)^2 = \frac{(\partial_t \Delta)^2 + (\partial_t \psi)^2}{\Delta^2}
\begin{pmatrix} 1 & 0 \\ 0 & 1\end{pmatrix}.\label{Mmat2}
\end{gather}

\textbf{The gradient of $\boldsymbol{\rho}$ and topologies.} Within this setting, a~major distinction concerns the sign of $\gamma^{\mu\nu} \partial_{\mu} \rho
\partial_{\nu} \rho$, which is a~scalar under diffeomorphisms in $\bigl(x^0,x^1\bigr)$,
and thus has a~unique value at each point.
 For later reference, we note its
explicit form
\begin{equation*}
\gamma^{\mu\nu}(x) \partial_{\mu} \rho \partial_{\nu} \rho =
\frac{{\rm e}^{-\tilde{\sigma}}}{n^2} \bigl[ - e_0(\rho)^2 + n^2 (\partial_1 \rho)^2\bigr],
\end{equation*}
where $e_0(\rho) = (\partial_t - s \partial_1)\rho$.
As mentioned, the case when $\gamma^{\mu\nu} \partial_{\mu} \rho \partial_{\nu} \rho <0$
everywhere (i.e.,~$\partial_{\mu}\rho$ timelike) gives rise to Gowdy cosmologies.
The case when $\gamma^{\mu\nu} \partial_{\mu} \rho \partial_{\nu} \rho >0$ everywhere
(i.e.,~$\partial_{\mu}\rho$ spacelike) gives rise to cylindrical gravitational waves.
In both situations there is a~special case where the Killing vectors
are everywhere orthogonal to each other, i.e.,~when $M$ in
\eqref{Mdef} is diagonal. The Gowdy
cosmologies with diagonal $M$ are referred to as polarized Gowdy
cosmologies, the cylindrical gravitational waves with diagonal $M$
are known as Beck--Einstein--Rosen waves.

Concerning topology and boundary conditions, the situation for
cylindrical gravitational waves is simple: the spatial
sections in~\eqref{2Dmetric} are diffeomorphic to $\mathbb{R}$, and the
field $\rho$ with its globally spacelike gradient can be used to
set up a~foliation whose level sets are nested timelike surfaces that
are topologically cylinders. Often adapted coordinates $\rho = r >0$
are used and the fields and their time derivatives are assumed to be
regular as $r \rightarrow 0$ \cite{KSYangian,PennaSmat,TorreRom}, while for~${r \rightarrow \infty}$ a~sufficient fall-off is stipulated.

For Gowdy cosmologies, the situation is more complex.
The possible topologies with compact spatial
sections have been classified by Gowdy~\cite{Gowdy}.
The most interesting one among these is the~$T^3$ topology where
the spatial sections in~\eqref{2Dmetric} are diffeomorphic to a~circle.
In addition, there is the
$\mathbb{R} \times T^2$ case~\cite{Matzner} where the spatial sections
are noncompact in the direction transverse to the Killing orbits.
This case is technically similar to that of the cylindrical
gravitational waves in that suitable fall-off conditions in
the noncompact spatial direction $x^1$ can be imposed.\looseness=1

\textbf{AVD for $\boldsymbol{T^3}$ Gowdy.} Another complication for Gowdy cosmologies
concerns the behavior near the big bang. Since a~big bang singularity
is known to be generic some of the fields and their derivatives must
blow up. (In particular, the `regular subsector'
considered for $\mathbb{R} \times T^2$ in~\cite{KSGowdyNC} cannot be generic.)
It turns out that the
matrix $M$ in~\eqref{2Kmetric}, \eqref{Mmat1} is in general singular
at the big bang, while the time component of the Noether current
in~\eqref{JHdef} below stays regular, see Section~\ref{sec4.4}. The spatial
component of the Noether current vanishes at the big bang. The latter is part
of a~general pattern known as asymptotic velocity domination~(AVD).
AVD posits that, when generic cosmological solutions are evolved
backward towards the big bang, they enter a~velocity dominated regime
in which temporal derivatives overwhelm spatial ones, as in the
Belinski--Khalatnikov--Lifshitz scenario~\cite{BelHenbook}.
Whenever AVD applies the solutions are asymptotically governed by a~much simpler system of equations, dubbed ``velocity dominated''~(VD), without dynamical spatial gradients.
Conversely, the full solution can then in principle
be reconstructed from data specified on the big bang ``boundary'';
see~\cite{RingstroemCR} for an overview.
Importantly, for $T^3$ Gowdy cosmologies AVD has been rigorously
proven in~\cite{IsenMonc,Ringstroemproof}, as reviewed in
\cite{RingstroemLR}. Moreover, the full solution can indeed
be reconstructed from the limiting one~\cite{KRend,RingstroemInitial}.
This forms the backdrop for Figure~\ref{2Kdiagram} and we return
to these on-shell aspects in Section~\ref{sec4.4}.

\subsection{The Hamiltonian action and its gauge symmetries}

For the reasons highlighted in the introduction, we shall mostly
be concerned with the off-shell aspects of our 2-Killing vector
reductions. Moreover, the Hamiltonian formulation is directly
relevant for Dirac observables, so we begin with presenting
the Hamiltonian action and its diffeomorphism-induced local
gauge symmetries. Since the timelike/spacelike nature
of $\rho$'s gradient is secondary, the following discussion applies to both Gowdy cosmologies and cylindrical gravitational waves.

On general grounds, ``variation of an action'' and ``reduction with
respect to Killing vectors'' are commuting operations~\cite{Torre}.
It is therefore sufficient to consider the 2-Killing vector reduction
of the Einstein--Hilbert or Gibbons--Hawking action. The
resulting Lagrangian action is~\eqref{2KLaction}. Its Hamiltonian
counterpart is readily computed and is taken here as our point of departure. An alternative exposition of the Hamiltonian
formalism can be found in~\cite{TorreRom}.

\textbf{Hamiltonian action.} The phase space is parameterized
by four canonical pairs, $\bigl(\Delta,\pi^{\Delta}\bigr)$,
$\bigl(\psi,\pi^{\psi}\bigr)$, $\bigl(\rho,\pi^{\rho}\bigr)$, $\bigl(\sigma,\pi^{\sigma}\bigr)$,
where $\sigma = \tilde{\sigma} + \frac{1}{2} \ln \rho$.
The action principle governing their dynamics is
\begin{gather}
S^{H} = \int {\rm d}^2 x \bigl\{ \pi^{\rho} e_{0}(\rho)
+ \pi^{\sigma} e_{0}(\sigma) + \pi^{\psi} e_{0}(\psi)
+ \pi^{\Delta} e_{0}(\Delta) - n\mathcal{H}_{0} \bigr\},\nonumber
\\
\mathcal{H}_{0} := - \lambda_{\text{\sc n}} \pi^{\sigma} \pi^{\rho} -
\frac{1}{\lambda_{\text{\sc n}}} \bigl(\partial_{1} \rho \partial_{1} \sigma
 - 2 \partial_{1}^{2} \rho\bigr) + \frac{\lambda_{\text{\sc n}}}{2\rho} \Delta^2 \bigl(
\bigl(\pi^{\psi}\bigr)^2 + \bigl(\pi^{\Delta}\bigr)^2 \bigr)\nonumber\\
\phantom{\mathcal{H}_{0} :=}{} +
\frac{\rho}{2\lambda_{\text{\sc n}}} \frac{(\partial_{1}\Delta)^2 + (\partial_{1} \psi)^2}{\Delta^2}.\label{Haction}
\end{gather}
Here, $\lambda_{\text{\sc n}}>0$ is the reduced Newton constant.
The dependence on the shift $s$ enters solely through the
$e_0 = \partial_t - \mathcal{L}_{s}$ derivatives, with the Lie derivative
$\mathcal{L}_s$ acting according to the one-dimensional spatial tensor
type. The $s$ dependence can be rendered
explicit by spatial integrations-by-parts. The result is~\eqref{Haction}
with all $e_0$ derivatives replaced by $\partial_t$ derivatives and an extra
term $-s \mathcal{H}_1$ added, where
\begin{equation*}
\mathcal{H}_{1} := \pi^{\rho} \partial_{1} \rho +
\pi^{\sigma} \partial_{1} \sigma
- 2 \partial_{1} \pi^{\sigma} + \pi^{\psi} \partial_{1} \psi
+ \pi^{\Delta} \partial_{1} \Delta,
\end{equation*}
is the diffeomorphism constraint.
In this form, it is clear that $\delta S^H/\delta n = - \mathcal{H}_0$, and
$\delta S^H/\delta s = - \mathcal{H}_1$.
The basic Poisson structure is
\begin{gather}
\bigl\{ \varphi\bigl(x^1\bigr), \pi^{\varphi}\bigl(y^1\bigr) \bigr\} = \delta\bigl(x^1-y^1\bigr),
 \qquad \varphi = \rho,\sigma,\Delta,\psi,
\label{PB}
\end{gather}
where $\delta$ is viewed as a~spatial $+1$ density.

In the present conventions, both $\mathcal{H}_0$ and $\mathcal{H}_1$
are spatial densities
of weight $+2$, while both~$n$ and $s$ are spatial densities of
weight $-1$.%
\footnote{Note that $\mathcal{H}_1$ is a~weight $+1$ spatial covector density,
equivalent to a~spatial $+2$ density in one dimension.
 Its Lie derivative thus is $\mathcal{L}_s \mathcal{H}_1 = (s \partial_1 \mathcal{H}_1 +
 \partial_1 s \mathcal{H}_1) + \partial_1 s \mathcal{H}_1$. Similarly, $\mathcal{H}_0$ is a~scalar
 $+2$ density, resulting in~${\mathcal{L}_s \mathcal{H}_0 = s \partial_1 \mathcal{H}_0 +
 2 \partial_1 s \mathcal{H}_0}$.}
Hence, the averages
$\mathcal{H}_0(n) = \int {\rm d} x^1 n \mathcal{H}_0$ and $\mathcal{H}_1(s) = \int {\rm d} x^1 s \mathcal{H}_1$
are independent of the choice of spatial coordinates and are
the Poisson generators of infinitesimal time translations
and spatial diffeomorphisms, respectively. Let $F$ be a~(not explicitly time-dependent) functional on phase space
that is also a~spatial tensor. Then $\{ F, \mathcal{H}_0(n)\} = e_0(F)$
encodes the evolution equations and $\{ F, \mathcal{H}_1(s)\} = \mathcal{L}_s F$
the spatial transformation law. The constraints also form off-shell a~closed Poisson algebra~\cite{TorreRom}
\begin{gather}
\{ \mathcal{H}_0\,(x), \mathcal{H}_1(y)\} = \partial_1 \mathcal{H}_0 \delta(x-y) + 2 \mathcal{H}_0\,(x)
\delta'(x-y),
\nonumber\\
\{ \mathcal{H}_1\,(x), \mathcal{H}_1(y)\} = \partial_1 \mathcal{H}_1 \delta(x-y) + 2 \mathcal{H}_1\,(x)
\delta'(x-y),
\nonumber\\
\{ \mathcal{H}_0\,(x), \mathcal{H}_0(y)\} = \delta'(x-y)[\mathcal{H}_1\,(x) + \mathcal{H}_1(y)].
\label{constr0}
\end{gather}
The first two relations just express that the density weights
of $\mathcal{H}_0$ and $\mathcal{H}_1$ are $+2$. By specialization
of the general (model independent) gravitational constraint algebra
to $1 + 1$ dimensions one would expect the last relation to contain a~factor explicitly dependent on the spatial metric. Specifically,
with the present density conventions one would expect an additional
$\gamma \gamma^{11}$
term on the right-hand side of the last relation in~\eqref{constr0}.
Using~\eqref{2Dmetric} this evaluates to $\gamma \gamma^{11} =
n^2 {\rm e}^{\tilde{\sigma}}$, which is a~spatial scalar. A~short computation
shows that the redefinition $\mathcal{H}_0^{\text{\sc ADM}} :=
n {\rm e}^{\tilde{\sigma}/2}\mathcal{H}_0$, $\mathcal{H}_1^{\text{\sc ADM}} := \mathcal{H}_1$,
maps~\eqref{constr0} into the expected ADM type constraint algebra.
In other words, the simplification of the constraint algebra to one
of Lie-algebra type (with structure constants rather than structure
functions) is a~specific $(1 + 1)$-dimensional feature.

\textbf{Hamiltonian gauge variations.} Next, we consider the
local gauge symmetries of the action~\eqref{Haction}.
On general grounds
$\mathcal{H}_0(\epsilon) + \mathcal{H}_1\bigl(\epsilon^1\bigr)$ is the generator of Hamiltonian gauge
variations of any (not explicitly time dependent) functional
$F$ on phase space built from the canonical variables, here
$\Delta$, $\psi$, $\rho$, $\sigma$ and their canonical momenta.
The descriptors $\bigl(\epsilon, \epsilon^1\bigr)$ play the role of $(n,s)$ and
can both be viewed as spatial $-1$ densities, see footnote 1.
Further, $\epsilon$, $\epsilon^1$ are in one-to-one correspondence
with a~linearized diffeomorphism, $t' = t - \xi^0(t,x) + O\bigl(\xi^2\bigr)$,
${x'}^1 = x^1 - \xi^1(t,x) + O\bigl(\xi^2\bigr)$, via the following relations:
\begin{gather}
\xi^{0} = \frac{\epsilon}{n},\qquad
\xi^{1} = \epsilon^{1} - s \frac{\epsilon}{n}.
\label{gtrans2}
\end{gather}
For all but $\sigma$ the spatial variations are all of the form
$\mathcal{L}_{\epsilon^1} d = \epsilon^1 \partial_1 d + pd \partial_1 \epsilon^1$, for a~spatial density
$d$ of weight $p$. For ${\rm e}^{\sigma}$ this holds as well with $p=2$,
which gives $\mathcal{L}_{\epsilon^1} \sigma = \epsilon^1 \partial_1 \sigma + 2 \partial_1 \epsilon^1$.
Result~\ref{resultSLgaugeinv} presents the local gauge variations of the
reduced Lagrangian action. Its Hamiltonian counterpart is the following.

\begin{Result} \label{resultSHgaugeinv}
Let $S^H$ be the action~\eqref{Haction} with
the temporal integration restricted to $x^0 = t \in [t_i,t_f]$
and fall-off or boundary conditions in $x^1$ that ensure
the absence of spatial boundary terms. Then
\begin{gather}
\delta^H_{\epsilon} S^H = 0, \qquad \epsilon|_{t_i} = 0 = \epsilon|_{t_f}.
\label{Hginv}
\end{gather}
Here, all but the gauge variations of lapse and shift are canonically
generated via $\delta_{\epsilon}^H F = \bigl\{ F, \mathcal{H}_0(\epsilon) + \mathcal{H}_1\bigl(\epsilon^1\bigr)\bigr\}$.
The gauge variations of lapse and shift are designed such that
\eqref{Hginv} holds and coincide with those of the Lagrangian formalism.
\end{Result}

For completeness' sake, we record the full set of Hamiltonian gauge
variations here
\begin{gather}
\delta^H_{\epsilon} n = \partial_{0} \epsilon
- (s \partial_{1} \epsilon - \epsilon \partial_{1} s)
+ \epsilon^{1} \partial_{1} n - n\partial_{1} \epsilon^{1},
\nonumber\\
\delta^H_{\epsilon} s = \partial_{0} \epsilon^{1}
+\bigl(\epsilon^{1} \partial_{1} s - s\partial_{1} \epsilon^{1}\bigr)
+ \epsilon \partial_{1} n - n \partial_{1} \epsilon,
\nonumber\\
\delta^H_{\epsilon} \sigma = - \lambda_{\text{\sc n}} \epsilon \pi^{\rho}
+ \epsilon^{1} \partial_{1} \sigma + 2\partial_{1} \epsilon^{1},\qquad
\delta^H_{\epsilon} \rho = -\lambda_{\text{\sc n}} \epsilon \pi^{\sigma} + \epsilon^1 \partial_1 \rho,
\nonumber\\
\delta^H_{\epsilon} \Delta = \lambda_{\text{\sc n}} \frac{\epsilon}{\rho} \Delta^2 \pi^{\Delta}
+ \epsilon^{1} \partial_{1} \Delta,
\qquad
\delta^H_{\epsilon} \psi = \lambda_{\text{\sc n}} \frac{\epsilon}{\rho} \Delta^2 \pi^{\psi}
+ \epsilon^{1} \partial_{1} \psi,
\nonumber\\
\delta^H_{\epsilon} \pi^{\sigma} = -\frac{1}{\lambda_{\text{\sc n}}} \partial_1 (\epsilon \partial_1\rho) +\partial_1\bigl(\epsilon^1\pi^\sigma\bigr),
\nonumber\\
\delta^H_{\epsilon} \pi^{\rho} = - \frac{1}{\lambda_{\text{\sc n}}}\partial_{1}(\epsilon \partial_{1} \sigma) - \frac{2}{\lambda_{\text{\sc n}}}\partial_{1}^2 \epsilon
+ \frac{\epsilon \lambda_{\text{\sc n}} }{2\rho^2} \Delta^2 \bigl(\bigl(\pi^\psi\bigr)^2 + \bigl(\pi^\Delta\bigr)^2\bigr)- \frac{\epsilon}{2\lambda_{\text{\sc n}}} \frac{(\partial_{1}\Delta)^2+(\partial_{1}\psi)^2}{\Delta^2}
\nonumber\\
 \phantom{\delta^H_{\epsilon} \pi^{\rho} =}{} + \partial_1\bigl(\epsilon^1 \pi^\rho\bigr),
\nonumber\\
\delta^H_{\epsilon} \pi^\Delta = \frac{\rho}{\lambda_{\text{\sc n}} \Delta^3}
\epsilon \Big( (\partial_{1} \Delta)^2 + (\partial_{1} \psi)^2 \Big) - \frac{\epsilon \lambda_{\text{\sc n}}}{\rho} \Delta \bigl( \bigl(\pi^\psi\bigr)^2 + \bigl(\pi^\Delta\bigr)^2\bigr) +\frac{1}{\lambda_{\text{\sc n}}} \partial_{1} \left( \frac{\rho}{\Delta^2} \epsilon \partial_{1} \Delta \right)
\nonumber\\
 \phantom{\delta^H_{\epsilon} \pi^\Delta =}{} + \partial_1\bigl(\epsilon^1 \pi^\Delta\bigr),
\nonumber\\
\delta^H_{\epsilon} \pi^{\psi} = \frac{1}{\lambda_{\text{\sc n}}} \partial_{1} \left( \frac{\rho}{\Delta^2} \epsilon \partial_{1} \psi \right) + \partial_1\bigl(\epsilon^1 \pi^\psi\bigr).
\label{gtransH}
\end{gather}
These gauge variations hold fully off-shell and express the $(1 + 1)$-dimensional diffeomorphism invariance of the Hamiltonian system.
Recall that the velocity-momentum relations (as recorded below in
\eqref{Momenta}) comprise one-half of the Hamiltonian evolution
equations and their use in~\eqref{Haction} would return one to the
Lagrangian action~\eqref{2KLaction}. Correspondingly, when inserting
the velocity-momentum relations~\eqref{Momenta} on the right-hand side
of the $\delta_{\epsilon}^H$ variations the Lagrangian gauge variations
\eqref{gtrans1} are recovered.

The gauge descriptors $\bigl(\epsilon,\epsilon^1\bigr)$ will always be understood to
reflect the type of linearized diffeomorphisms underlying them
via~\eqref{gtrans2}. Usually, the (smooth) diffeomorphisms that are considered
`gauge transformations' in a~generally covariant system are `small'
in the sense that they reduce to the identity outside some compact
region, see, e.g.,~\cite{Hoehn}. In the $\mathbb{R} \times T^2$ Gowdy model,
where the spatial sections are diffeomorphic to $\mathbb{R}$, this means the
descriptors are smooth and obey
\begin{gather}
\epsilon,\ \epsilon^1 \  \mbox{have compact support in} \ \mathbb{R}.
\label{epsnoncompact}
\end{gather}
This will automatically remove boundary terms at spatial infinity arising
from the integration of~\eqref{Diracobs4}, \eqref{ahdirac}, seemingly
without the need to specify fall-off conditions on the dynamical variables.
The latter, however, are needed to allow for the improper
$\epsilon =n$, $\epsilon^1 =0$ specialization, where the Hamiltonian time
evolution ought to arise. For the temporal descriptors we maintain our
standing assumption $\epsilon|_{t_i} = 0 = \epsilon|_{t_f}$, see~\eqref{ginv},
\eqref{Hginv}.

In the $T^3$ Gowdy system, the spatial sections are diffeomorphic to a~circle.
We then take as a~minimal requirement that the descriptors $\epsilon$, $\epsilon^1$
are smooth and spatially periodic
\begin{gather}
\epsilon\bigl(x^0,x^1 + 2\pi\bigr) = \epsilon\bigl(x^0,x^1\bigr), \qquad
\epsilon^1\bigl(x^0,x^1 + 2\pi\bigr) = \epsilon^1\bigl(x^0,x^1\bigr).
\label{epscompact}
\end{gather}
This is consistent with their interpretation as linearized lifts
of circle diffeomorphisms. Indeed, the lift of some $\chi\colon S^1 \rightarrow S^1$,
\smash{$\chi\bigl({\rm e}^{{\rm i} x^1}\bigr) = {\rm e}^{{\rm i}f(x^1)}$} can be chosen as quasi-periodic, $f\bigl(x^1 + 2\pi\bigr) =
f\bigl(x^1\bigr) + 2\pi$. Upon linearization $f\bigl(x^1\bigr) = x^1 + (\delta f)\bigl(x^1\bigr)
+ O\bigl(\delta f^2\bigr)$, so the quasi-periodicity of $f$ is consistent with
the periodicity of $\delta f$. The periodicity requirement also
allows for the $\epsilon =n$, $\epsilon^1=0$ choice, as $n$ is spatially
periodic. However, it also allows for $\epsilon = \epsilon\bigl(x^0\bigr)$,
$\epsilon^1 = \epsilon^1\bigl(x^0\bigr)$ constant in $x^1$ that do not vanish outside
some coordinate chart of $\mathbb{R} \mod 2\pi$, as they should in the
interpretation as linearized lifts. This `smallness' condition
will only enter the derivation of~\eqref{rhotvar} later on.
Finally, note that since we have to allow $\epsilon$, $\epsilon^1$ to be nonzero
in the vicinity of any particular point in $\mathbb{R} \mod 2\pi$, it
cannot be used to remove boundary terms in the spatial integration
of~\eqref{Diracobs4}, \eqref{ahdirac}, unlike in the noncompact case.

In addition to these local symmetries, the Hamiltonian action also has a~set of global symmetries related to $\mathrm{SL}(2,\mathbb{R})$ transformations of
the two Killing vectors in~\eqref{2Kmetric}. The matrix $M(x)$ in the
line element then transforms according to $M(x) \rightarrow CM(x)C^{\mathsf T}$, $
 C \in \mathrm{SL}(2,\mathbb{R})$. The associated Noether current turns out to
be given by
\begin{gather}
J_0^H := \begin{pmatrix} \Delta \pi^{\Delta} + \psi \pi^{\psi} &
 \bigl(\Delta^2 - \psi^2\bigr) \pi^{\psi} - 2 \psi \Delta \pi^{\Delta}\\
\pi^{\psi} &- \Delta \pi^{\Delta} - \psi \pi^{\psi}
\end{pmatrix},
\nonumber\\
 J_1^H := \frac{n \rho }{\lambda_{\text{\sc n}} \Delta^2} \begin{pmatrix}
 \psi \partial_{1} \psi + \Delta \partial_{1} \Delta
 & \bigl(\Delta^2 - \psi^2\bigr) \partial_{1} \psi
 - 2 \psi \Delta \partial_{1} \Delta \\
\partial_{1} \psi & - \psi \partial_{1} \psi - \Delta \partial_{1} \Delta
\end{pmatrix}.
\label{JHdef}
\end{gather}
It satisfies
\begin{gather}
\delta_{\epsilon}^H J_0^H = \partial_1\left( \frac{\epsilon}{n} J_1^H\right)
+ \partial_1\bigl( \epsilon^1 J_0^H\bigr),
\label{HJvar}
\end{gather}
without using equations of motion. Its on-shell specialization gives
the conservation equation $e_0\bigl(J_0^H\bigr) = \partial_1 J_1^H$ via~\eqref{Hevol} below.
Of course, the conservation also is a~consequence of the Noether
construction, and it can be verified directly using the evolution
equations~\eqref{Momenta} and~\eqref{EOMH} below.

\textbf{Hamiltonian field equations.} As stressed repeatedly, our considerations
are mostly off-shell. Nevertheless, the specialization to on-shell
configurations occasionally provides a~welcome consistency check.
To this end, we note here the Hamiltonian field equations.
Varying the Hamiltonian action~\eqref{Haction} with respect to the momenta
gives the velocity-momentum relations. Explicitly,
\begin{gather}
\pi^{\Delta} = \frac{\rho}{\lambda_{\text{\sc n}} n}
\frac{ e_0(\Delta)}{\Delta^2},
\qquad
\pi^{\psi} = \frac{\rho}{\lambda_{\text{\sc n}} n}
\frac{ e_0(\psi)}{\Delta^2},
\qquad
\pi^{\rho} = - \frac{1}{\lambda_{\text{\sc n}} n} e_0(\sigma),
\nonumber\\
\pi^{\sigma} = - \frac{1}{\lambda_{\text{\sc n}} n} e_0(\rho).
\label{Momenta}
\end{gather}
These are spatial densities of weight $+1$. Varying the action with
respect to the fields produces the Hamiltonian evolution equations
\begin{gather}
\frac{\delta S^H}{ \delta \Delta} = \frac{\rho}{\lambda_{\text{\sc n}} \Delta^3}
n \bigl( (\partial_{1} \Delta)^2 + (\partial_{1} \psi)^2 \bigr)
- \frac{n \lambda_{\text{\sc n}}}{\rho} \Delta \bigl( \bigl(\pi^\psi\bigr)^2 + \bigl(\pi^\Delta\bigr)^2\bigr)
\nonumber\\
\phantom{\frac{\delta S^H}{ \delta \Delta} =}{}- e_{0} \bigl(\pi^\Delta\bigr) +\frac{1}{\lambda_{\text{\sc n}}} \partial_{1}
\left( \frac{\rho}{\Delta^2} n \partial_{1} \Delta \right),
\nonumber\\
\frac{\delta S^H}{ \delta \psi} = - e_{0} \bigl( \pi^\psi\bigr)
+\frac{1}{\lambda_{\text{\sc n}}} \partial_{1}
\left( \frac{\rho}{\Delta^2} n \partial_{1} \psi \right),
\nonumber\\
\frac{\delta S^H}{ \delta \rho} = -e_{0}\bigl(\pi^\rho\bigr)
- \frac{1}{\lambda_{\text{\sc n}}}\partial_{1}(n \partial_{1} \sigma)
- \frac{2}{\lambda_{\text{\sc n}}}\partial_{1}^2 n
+ \frac{n \lambda_{\text{\sc n}} }{2\rho^2} \Delta^2 \bigl(\bigl(\pi^\psi\bigr)^2 + \bigl(\pi^\Delta\bigr)^2\bigr)
\nonumber\\
\phantom{\frac{\delta S^H}{ \delta \rho} =}{}
- \frac{n}{2\lambda_{\text{\sc n}}} \frac{(\partial_{1}\Delta)^2
 +(\partial_{1}\psi)^2}{\Delta^2},
\nonumber\\
\frac{\delta S^H}{ \delta \sigma} =
-e_{0} \bigl( \pi^\sigma\bigr) - \frac{1}{\lambda_{\text{\sc n}}}\partial_{1}(n\partial_{1}\rho).
\label{EOMH}
\end{gather}
The relations~\eqref{Momenta}, \eqref{EOMH} also have an interpretation
in terms of the Poisson structure:
for any not explicitly time-dependent functional $F$ on phase space, one has
\begin{gather}
e_0(F) = \{ F, \mathcal{H}_0(n) \},
\label{Hevol}
\end{gather}
of which~\eqref{Momenta} and~\eqref{EOMH} are special cases. Among the
consequences of~\eqref{constr0} and~\eqref{Hevol} is that the constraint
surface $\mathcal{H}_0 = 0 = \mathcal{H}_1$ is preserved under time evolution
$
\smash{e_0(\mathcal{H}_0) \big|_{\mathcal{H}_0 = 0 = \mathcal{H}_1} }= 0 =
\smash{e_0(\mathcal{H}_1) \big|_{\mathcal{H}_0 = 0 = \mathcal{H}_1}}$.


\subsection{Linear system without gauge fixing}

The evolution equations of two Killing vector reductions are known to be
integrable in the sense of being coded by the compatibility condition of a~Lax pair.
The version due to Belinski and Zakharov (as reviewed in
\cite{Gravisolitons}) uses
differentiation with respect to a~constant spectral parameter, while
the version due to Maison~\cite{Maison} avoids the differentiation by
use of a~spacetime-dependent quantity replacing the original spectral
parameter. In both formulations, the gauge symmetries of the system are
at least partially gauge fixed through the conformal gauge (here:
lapse~${n=1}$, shift $s=0$). This reduces the symmetries of the system
to reparameterizations in the lightcone coordinates. A~linear system
in this gauge and in lightcone coordinates implying also~${\partial_+ \partial_- \rho=0}$ was formulated in~\cite{2Kvertex}. Usually, the
$\partial_+ \partial_- \rho=0$ wave equation is imposed from the outset and
solved in terms of lightcone coordinates~\cite{Gravisolitons,Maison}.
The so-specialized constraint combinations~${(\mathcal{H}_0 \pm \mathcal{H}_1)_\mathrm{specialized}}$
then generate two commuting copies of a~Virasoro--Witt algebra.
Although the classical invariants under their Poisson action can
presumably be related to Dirac observables proper, we prefer the
conceptually cleaner route here, where \textit{no} gauge fixing is used
and one stays fully off-shell.

In the following, we therefore present a~generalization of Maison's
construction valid without gauge fixing. In addition, we
shall carry along the dimensionless Newton constant $\lambda_{\text{\sc n}}$, as it will enter
the anti-Newtonian expansion in Section~\ref{sec4}.
In terms of $J_0^H$, $J_1^H$, we define for $\operatorname{Im}(\theta) \neq 0$
\begin{align}
L_0^{H,\pm}(\theta) &:=
\frac{\pm \lambda_{\text{\sc n}}}{ 2\bigl[ \lambda_{\text{\sc n}}^2 (\theta + \tilde{\rho})^2 - \rho^2\bigr]^{1/2}}
\left\{ \frac{\lambda_{\text{\sc n}} ( \theta + \tilde{\rho})}{\rho} n J_0^H
- J_1^H \right\} - \frac{\lambda_{\text{\sc n}} n}{2\rho} J_0^H,
\nonumber\\
L_1^{H,\pm}(\theta) &:=
\frac{\pm \lambda_{\text{\sc n}}}{ 2\bigl[ \lambda_{\text{\sc n}}^2 (\theta + \tilde{\rho})^2 - \rho^2\bigr]^{1/2}}
\left\{ \frac{\lambda_{\text{\sc n}}( \theta + \tilde{\rho})}{\rho} \frac{J_1^H}{n}
- J_0^H \right\} - \frac{\lambda_{\text{\sc n}}}{2\rho} \frac{J_1^H}{n}.
\label{LHdef}
\end{align}
Here we introduce a~field $\tilde{\rho}$ partially defined
by the requirements
\begin{gather}
\partial_1 \tilde{\rho} = - \pi^{\sigma}, \qquad
e_0(\tilde{\rho})\big|_{ \pi^{\sigma} \mathrm{evol}} =
\frac{n}{\lambda_{\text{\sc n}}} \partial_1 \rho.
\label{rhotpdef}
\end{gather}
The first condition of course determines $\tilde{\rho}$ up
to an $x^0$-dependent constant. Since $e_0(\partial_1 \tilde{\rho}) =
\partial_1 e_0(\tilde{\rho})$, use of the $\pi^{\sigma}$ evolution equation,
i.e.,~$e_0\bigl(\pi^{\sigma}\bigr) = - \partial_1(n \partial_1 \rho)/\lambda_{\text{\sc n}}$, gives the
second relation, again up to an $x^0$-dependent additive constant. We
shall give a~more detailed specification of $\tilde{\rho}$
later on, for now only the properties~\eqref{rhotpdef} are needed.

By definition, for any fixed $\theta$ with $\operatorname{Im}(\theta) \neq 0$,
\begin{gather}
L_0^{H,+}(\theta) + L_0^{H,-}(\theta) = - \frac{\lambda_{\text{\sc n}} n}{\rho} J_0^H,
\qquad
L_1^{H,+}(\theta) + L_1^{H,-}(\theta) = - \frac{\lambda_{\text{\sc n}}}{n\rho} J_1^H,
\label{LHpm1}
\end{gather}
so that for most purposes
one can use either the $+$ version
or the $-$ version. Whenever the distinction is inessential we shall
simply write $L_{0}^H(\theta) = L_0^{H,\pm}(\theta)$,
$L_1^H(\theta) = L_1^{H,\pm}(\theta)$ to unclutter the notation
and carry the two sign options in~\eqref{LHdef} along. A~situation where the signs matter is for the $\rho \rightarrow 0$ limit,
which will become relevant later on. Taking into account the
branch cuts of the square root one finds
\begin{align}
& L_0^{H,\pm}(\theta) = - \frac{1}{2(\theta + \tilde{\rho})}
\left[ J_1^H + O\left( \frac{\rho}{\lambda_{\text{\sc n}} (\tilde{\rho} + \theta)} \right) \right],
 \qquad \pm \operatorname{Re}(\theta + \tilde{\rho}) > 0,
\nonumber\\
& L_1^{H,\pm}(\theta) = - \frac{1}{2(\theta + \tilde{\rho})}
\left[ J_0^H + O\left( \frac{\rho}{\lambda_{\text{\sc n}} (\tilde{\rho} + \theta)} \right) \right],
 \qquad \pm \operatorname{Re}(\theta + \tilde{\rho}) > 0,
\label{LHpm2}
\end{align}
where the subleading terms are displayed in~\eqref{Hscaling8}. Clearly,
for fixed $\theta \in \mathbb{C}$ with $\operatorname{Re}(\theta + \tilde{\rho}) \neq 0$ not both
of $L_0^{H,\pm}(\theta)$, $L_1^{H,\pm}(\theta)$ can be regular at $\rho =0$.

The two sign options in~\eqref{LHdef} can be understood in terms
of the two branches of the function $z \mapsto z^{-p/2}$, for odd
$p \in \mathbb{N}$. The branch cut runs along the negative real
$z$-axis, so the branches $f_{\pm}(z) := \pm |z|^{-p/2} \exp\bigl\{
- {\rm i} \frac{p}{2} \operatorname{Arg}(z) \bigr\}$ are both analytic for
$\operatorname{Arg}(z) \in (-\pi,\pi)$. In particular,
$f_{\pm}(z)^* = f_{\pm}\bigl(z^*\bigr)$ holds for $\operatorname{Arg}(z) \in (-\pi,\pi)$.
The limits from within the $\operatorname{Arg}(z)$
strip, i.e.,~\smash{$\lim_{\operatorname{Arg}(z) \rightarrow \pm \pi} f_+(z) = \mp{\rm i}^p |z|^{-p/2}$}
and \smash{$\lim_{\operatorname{Arg}(z) \rightarrow \pm \pi} f_-(z) = \pm {\rm i}^p |z|^{-p/2}$}
show each function to be discontinuous on the negative real axis.
As usual, this can be partially remedied by extension to a~Riemann surface. Restricting $f_+(z)$ to
$0 < \operatorname{Arg}{(z)} < \pi$ and $f_-(z)$ to
$-\pi < \operatorname{Arg}{(z)} < 0$, one is the continuous extension of the
other, $\lim_{\operatorname{Arg}(z) \rightarrow \pi} f_+(z) =
\lim_{\operatorname{Arg}(z) \rightarrow -\pi} f_-(z)$.

Applied to \smash{$\theta \mapsto \bigl[\lambda_{\text{\sc n}}^2 (\theta + \tilde{\rho})^2 - \rho^2\bigr]^{-p/2}$}
the branch cut occurs when $\lambda_{\text{\sc n}}^2 (\theta + \tilde{\rho})^2 - \rho^2$
is negative, i.e.,~for $\theta \in [ - \tilde{\rho} - \rho/\lambda_{\text{\sc n}},
 - \tilde{\rho} + \rho/\lambda_{\text{\sc n}}]$, see Figure~\ref{fig1}. Note that the branch
cut is $\bigl(x^0,x^1\bigr)$ dependent and lies symmetric to the point singled out
by $\theta + \tilde{\rho} =0$.

\begin{figure}[!ht]
\centering
\resizebox{11cm}{!}{
\begin{tikzpicture}[>=Latex, font=\small]

\coordinate (L) at (0,0);
\coordinate (R) at (12,0);
\coordinate (C) at (6,0);
\coordinate (A) at (3,0);
\coordinate (B) at (9,0);

\draw[line width=0.9pt] (L) -- (A);
\draw[decorate, decoration={snake, amplitude=1.1mm, segment length=4mm}, line width=0.9pt]
 (A) -- (B);
\draw[->, line width=0.9pt] (B) -- (R);

\node[below=4mm] at (3,0) {$-\tilde{\rho}-\rho/\lambda_{\text{\sc n}}$};
\node[below=4mm] at (C) {$-\tilde{\rho}$};
\node[below=4mm] at (9,0) {$-\tilde{\rho}+\rho/\lambda_{\text{\sc n}}$};
\node[right=2mm] at (R) {$\operatorname{Re}(\theta)$};

\end{tikzpicture}
}
 \caption{Branch cut in the complex $\theta$-plane as seen by $L_\mu^H(\theta)$ and $\mathcal{J}_\mu^H(\theta)$, $\mu=0,1$.}
 \label{fig1}
 \end{figure}
The discontinuity across the cut is
\begin{gather}
\lim_{\epsilon \rightarrow 0^+} \operatorname{Arg}\bigl[ \lambda_{\text{\sc n}}^2 (\theta \pm{\rm i} \epsilon + \tilde{\rho})^2
 - \rho^2\bigr] = \pm \operatorname{sign}(\theta + \tilde{\rho}) \pi,
\label{cut1}
\end{gather}
valid for $\theta \in \mathbb{R}$ and $\lambda_{\text{\sc n}}^2 (\theta + \tilde{\rho})^2 - \rho^2<0$.
Here we used that $\lim_{\delta \rightarrow 0^+} \operatorname{Arg}(-1 \pm {\rm i}\delta) =
\pm \pi$. In the special case where $\theta + \tilde{\rho} =0$
the limit evaluates to $+\pi$. Excluding this point one has
for real $\theta$
\begin{gather}
\lim_{\epsilon \rightarrow 0^+} \bigl[\lambda_{\text{\sc n}}^2 (\theta \pm {\rm i}\epsilon + \tilde{\rho})^2 - \rho^2\bigr]^{-p/2}
\big|_{\rho > \lambda_{\text{\sc n}} |\theta + \tilde{\rho}|}
= \pm {\rm i}^{-p} \operatorname{sign}(\theta + \tilde{\rho})
\bigl|\lambda_{\text{\sc n}}^2 (\theta + \tilde{\rho})^2 - \rho^2\bigr|^{-p/2}.
\label{cut2}
\end{gather}
For $\rho < \lambda_{\text{\sc n}} |\theta + \tilde{\rho}|$, the limit is of course continuous
and just gives \smash{$\bigl(\lambda_{\text{\sc n}}^2 (\theta + \tilde{\rho})^2 - \rho^2\bigr)^{-p/2}$}.
We shall return to~\eqref{cut1}, \eqref{cut2} later on. To avoid
complications with the $\bigl(x^0,x^1\bigr)$ dependent branch cut, we
take $\operatorname{Im}\theta \neq 0$ from now on by default. Further, we
adopt the following

\textbf{Notational convention.} Whenever the distinction between
both sign options in~\eqref{LHdef} is inessential we drop
the $\pm$ superscripts and just write $L_0^H(\theta)$, $L_1^H(\theta)$
for the linear system and $U$ for a~solution. Both sign
options are then carried along without notational
distinction.

We now claim that the following is a~Lax pair (linear system) valid
without any gauge fixing
\begin{align}
& e_0(U) = U L_0^H, \qquad \partial_1 U = U L_1^H,
\qquad e_0\bigl(L_1^H\bigr) - \partial_1 L_0^H + \bigl[L_0^H, L_1^H\bigr] =0.
\label{linsystemH}
\end{align}
Here we treat $U$ as a~spatial scalar so that $\partial_1 U$ (as a~one-dimensional covector) can be viewed as a~spatial $+1$ density.
This gives $e_0(\partial_1 U) = \partial_1 e_0(U)$, and the second relation
in~\eqref{linsystemH} is just the compatibility condition needed.
In turn, this compatibility condition expands to
\begin{gather*}
 e_0\bigl(L_1^H\bigr) - \partial_1 L_0^H + \bigl[L_0^H, L_1^H\bigr]\\
 \qquad =
\frac{\pm \lambda_{\text{\sc n}}}{2\bigl[ \lambda_{\text{\sc n}}^2 (\theta + \tilde{\rho})^2 - \rho^2\bigr]^{1/2}}
\bigl\{ \partial_1 J_1^H - e_0\bigl(J_0^H\bigr) \bigr\}\\
\phantom{\qquad =}{} + \lambda_{\text{\sc n}}\left( - \frac{1}{2} \pm
\frac{\lambda_{\text{\sc n}}(\theta + \tilde{\rho})}%
 {2\bigl[ \lambda_{\text{\sc n}}^2 (\theta + \tilde{\rho})^2 - \rho^2\bigr]^{1/2}}
\right)
 \left\{ e_0\left( \frac{J_1^H}{ \rho n} \right)
- \partial_1 \left( \frac{n}{\rho} J_0^H \right) - \lambda_{\text{\sc n}}
\left[ \frac{n}{\rho} J_0^H, \frac{J_1^H}{\rho n} \right] \right\}
\\
\phantom{\qquad =}{}- \frac{\pm \lambda_{\text{\sc n}}}{2\bigl[ \lambda_{\text{\sc n}}^2 (\theta + \tilde{\rho})^2 - \rho^2\bigr]^{3/2}}
\left\{ \bigl( \lambda_{\text{\sc n}} (\theta + \tilde{\rho}) n J_0^H - \rho J_1^H \bigr)
\left( \partial_1\rho - \frac{\lambda_{\text{\sc n}}}{n} e_0(\tilde{\rho}) \right)\right.\\
\left.\phantom{\qquad =}{}
{}+ \bigl( \lambda_{\text{\sc n}} (\theta + \tilde{\rho}) J_1^H - \rho n J_0^H \bigr)
\left( \lambda_{\text{\sc n}} \partial_1\tilde{\rho} - \frac{1}{ n} e_0(\rho) \right)
\right\}.
\end{gather*}
For example, by expanding in powers of $1/(\theta + \tilde{\rho})$
one sees that for~\eqref{linsystemH} to hold each of the terms in
curly brackets has to vanish separately. Since $n J_0^H$ and $J_1^H$
are in general nonzero and functionally independent also the coefficients
in the last curly bracket have to vanish separately. In summary,
$e_0\bigl(L_1^H\bigr) - \partial_1 L_0^H + \bigl[L_0^H, L_1^H\bigr] =0$ is
equivalent to
\begin{subequations}
\label{JHthetaonshell3}
\begin{align}
&\partial_1\rho - \frac{\lambda_{\text{\sc n}}}{n} e_0(\tilde{\rho}) = 0, \qquad
\partial_1\tilde{\rho} - \frac{1}{\lambda_{\text{\sc n}} n} e_0(\rho) =0,
\label{JHthetaonshell3a}\\
& e_0\bigl(J_0^H\bigr) - \partial_1 J_1^H =0,
\label{JHthetaonshell3b}\\
& e_0\left(\frac{J_1^H}{\rho n}\right) - \partial_1 \left(\frac{n}{\rho} J_0^H \right)
- \lambda_{\text{\sc n}} \left[\frac{n}{\rho} J_0^H, \frac{J_1^H}{\rho n}\right] = 0.
\label{JHthetaonshell3c}
\end{align}
\end{subequations}
 Since by definition
$\partial_1 \tilde{\rho} = -\pi^{\sigma}$ the second equation
\eqref{JHthetaonshell3a} is just the velocity momentum
 relation~${0= \delta S^H/\delta \pi^{\rho} = e_0(\rho) + \lambda_{\text{\sc n}}
 n \pi^{\sigma}}$. The first relation in~\eqref{JHthetaonshell3a} is
 equivalent to the $\pi^{\sigma}$ evolution equation, as noted in
\eqref{rhotpdef}.
Cross-differentiating both equations in~\eqref{JHthetaonshell3a} gives~$e_0\bigl( n^{-1} e_0(\rho)\bigr) = \partial_1( n \partial_1 \rho)$ and
$e_0\bigl( n^{-1} e_0(\tilde{\rho})\bigr) = \partial_1( n \partial_1 \tilde{\rho})$.
The first of these is the $0 = \delta S^H/\delta \sigma =
- e_0\bigl(\pi^{\sigma}\bigr) - \partial_1 (n \partial_1 \rho)/\lambda_{\text{\sc n}}$ equation of motion.
The zero curvature condition~\eqref{JHthetaonshell3c} implies
that there exists a~unimodular matrix function $M$ such that
$\lambda_{\text{\sc n}} n J_0^H/\rho = e_0(M) M^{-1}$ and $\lambda_{\text{\sc n}} J_1^H/(\rho n) = \partial_1 M M^{-1}$.
With $J_0^H$ defined by the matrix in~\eqref{JHdef} and $M$ from
\eqref{Mmat1} the former entails the velocity momentum relations
for $\pi^{\Delta}$, $\pi^{\psi}$. Finally, \eqref{JHthetaonshell3b} is
the evolution equation for the current, summarizing those of
$\pi^{\Delta}$, $\Delta$, $\pi^{\psi}$, $\psi$. In other words,
all the non-redundant velocity-momentum and evolution equations
are entailed solely by $e_0\bigl(L_1^H\bigr) - \partial_1 L_0^H +
\bigl[L_0^H, L_1^H\bigr] =0$.
The only equations missing are $\delta S^H/\delta \rho =0
= \delta S^H /\delta \pi^{\sigma}$, which are, however,
implied by the constraint equations.

In summary, the solubility of the pair of linear equations
$\partial_1 U = U L_1^H$ and $e_0(U) = U L_0^H$ is equivalent to
the full set of evolution equations~\eqref{JHthetaonshell3},
that is, all fields but $\sigma$ are on-shell. This is usually
combined with a~gauge fixing condition and the solution of the
constraints $\mathcal{H}_0 = \mathcal{H}_1 =0$, which allows one to express $\sigma$
in terms of the other fields. In this framework, the
main task in finding a~`general' solution to the evolution equations
lies in solving the pair of linear equations with `general' initial
data, see~\cite{Alekseev,Gravisolitons,Griffithsbook,Kleinbook} for detailed
accounts in (mostly) other 2-Killing vector reductions. As seen above,
neither fixing a~gauge nor imposing the constraints is central to
the basic fact: any solution of $\partial_1 U = U L_1^H$ and
$e_0(U) = U L_0^H$ in principle encodes a~solution of the
evolution equations~\eqref{JHthetaonshell3}. To do so, in addition to
initial conditions at some $x^0$ also a~spatial boundary condition
needs to be imposed.

For the on-shell aspects of Gowdy cosmologies, appropriate conditions
turn out to be
\begin{gather}
\mathbb{R} \times T^2\colon\
\lim_{x^1 \rightarrow -\infty} U\bigl(x^0,x^1;\theta\bigr) = \mathbbm{1},\nonumber\\
T^3\colon\ U_b\bigl(x^0;\theta\bigr) :=U\bigl(x^0,0;\theta\bigr) \mbox{ satisfies }
e_0(U_b) - U_b L_0^H|_{x^1=0} =0, \qquad \det U_b =1,
\label{Ubc}
\end{gather}
valid pointwise in $x^0$, $\theta$. For the $\mathbb{R} \times T^2$ topology,
$x^1$ ranges over $\mathbb{R}$ and $L_0^H$, $L_1^H$ vanish for~${\big|x^1\big| \rightarrow \infty}$,
so the condition is compatible with the differential equations for $U$.
For the~$T^3$ Gowdy cosmologies, the basic
fields and hence the currents $J_0^H$, $J_1^H$ are spatially periodic
and do not fall off. It is natural to single out a~solution of
$\partial_1 U = U L_1^H$ by prescribing its value at $x^1=0$. The specifics
are chosen such that the following two implications hold for both
the~${\mathbb{R}\times T^2}$ and the $T^3$ Gowdy systems.

First,
\begin{gather}
\det U\bigl(x^0,x^1;\theta\bigr) = 1.
\label{Udet}
\end{gather}
Indeed, since $L_1^H$ and $L_0^H$ are tracefree one has
$0 = \operatorname{tr}\bigl(U^{-1} \partial_1 U\bigr) = \partial_1 \operatorname{tr} \ln U =
\partial_1 \ln \det U$, and $0 = \operatorname{tr}\bigl(U^{-1} e_0(U)\bigr) = e_0( \operatorname{tr} \ln U)
= e_0(\ln \det U)$, so that $\det U$ can only be a~function of $\theta$.
Thus~\eqref{Udet} follows from~\eqref{Ubc}.
Second, subject to the boundary condition~\eqref{Ubc} one has
the implication
\begin{gather}
\partial_1 U = U L_1^H,\qquad
e_0\bigl(L_1^H\bigr) - \partial_1 L_0^H + \bigl[L_0^H, L_1^H\bigr] =0
 \Longrightarrow e_0(U) = U L_0^H.
\label{linsystemH1}
\end{gather}
For on-shell fields $U$ can therefore be defined by $\partial_1 U = U L_1^H$ only.
To see this, we set $A := e_0(U) - U L_0^H$. A~short computation using
the premise gives $\partial_1 A = A L_1^H$. Hence, if $A = A\bigl(x^0,x^1;\theta\bigr)$
vanishes for one $x^1$ and all $x^0$, $\theta$, it will be identically zero.
For the $\mathbb{R} \times T^2$ system, $A$~vanishes for $x^1 \rightarrow - \infty$.
For the $T^3$ systems, the condition on $U_b$ ensures $A=0$ for $x^1=0$.

\subsection{Conserved quantities from a~solution of the linear system}\label{section2.3}

Once a~solution of the linear system has been found one
can often construct one-parameter families of (on-shell) conserved
quantities from it. The principal ideas are drawn from the theory of
flat space integrable field theories~\cite{FT} and have been
applied to some 2-Killing vector reductions as well
\cite{BM,KSGowdyNC,KSYangian}. The adaptation to the $\mathbb{R} \times T^2$
Gowdy system is more-or-less straightforward, while that
to $T^3$ Gowdy cosmologies is not.

For the purpose of this subsection, we \textit{assume} that a~solution $U^\mathrm{on} = U^\mathrm{on}\bigl(x^0,x^1;\theta\bigr)$ of $\partial_1 U = U L_1^H$ and
$e_0(U) = U L_0^H$ has been found. Recall that this presupposes that
all but the $\sigma$ field are on-shell, see~\eqref{JHthetaonshell3}.
No gauge fixing enters but the boundary condition~\eqref{Ubc} is
taken to hold. As indicated,
we shall use a~superscript ``on'' to
stress that it is an on-shell object. Given such a~solution there are
three main construction principles for conserved quantities:

\textbf{Limit of} $\boldsymbol{U^\mathrm{on}}$. \textit{If} the limit $\mathcal{O}_T^\mathrm{on} =
\lim_{x^1 \rightarrow \infty} U^\mathrm{on}\bigl(x^0,x^1;\theta\bigr)$ exists
pointwise in $x^0$, $\theta$, it will be conserved: $e_0\bigl(\mathcal{O}_T^\mathrm{on}\bigr) =0$.

This follows by taking the limits of $e_0\bigl(U^\mathrm{on}\bigr) =
U^\mathrm{on} L_0^H$, $\partial_1 U^\mathrm{on} = U^\mathrm{on} L_1^H$,
and noting that $L_0^H$, $L_1^H$ vanish for $x^1 \rightarrow \infty$.
Since $e_0\bigl(U^\mathrm{on}\bigr) = (\partial_0 - s \partial_1) U^\mathrm{on}$, and $s$ stays
at least bounded for~${\big|x^1\big|\rightarrow \infty}$, this gives the
assertion. We anticipate that the assumption turns out to be satisfied for the~${\mathbb{R} \times T^2}$ Gowdy systems but not for the $T^3$ case.

\textbf{Monodromy matrix.} The on-shell monodromy matrix is schematically
of the form
\begin{equation*}
\mathcal{M}^\mathrm{on}(\theta) = U_+^\mathrm{on}\bigl(x^0,x^1;\theta\bigr) M\bigl(x^0,x^1\bigr)
U_-^\mathrm{on}\bigl(x^0,x^1;\theta\bigr)^{\mathsf T}, \qquad \operatorname{Im} \theta \neq 0,
\end{equation*}
where $U^\mathrm{on}_{\pm}$
are the solutions of the linear system \smash{$L_{\mu}^{H,\pm}(\theta)$}, respectively.
It can be seen to be constant in $x^1$ and conserved,
$e_0\bigl(\mathcal{M}^\mathrm{on}(\theta)\bigr)=0$. As such it also seems to generate nonlocal
conserved charges. A~detailed analysis, relegated to Appendix~\ref{appC},
shows, however, that for the~${\mathbb{R} \times T^2}$ systems $\mathcal{M}^\mathrm{on}(\theta) =\mathbbm{1}$
is just the unit matrix, while for $T^3$ cosmologies the quantity
is overconstrained by the requirements on $U_{\pm}$ from Appendix~\ref{appB}.

\textbf{Conserved currents.} Finally, a~solution $U^\mathrm{on}$
can be used to define a~one-parameter family of conserved currents that
are conserved without gauge fixing. We define $K_0^H(\theta) := 2 \partial_{\theta}
L_1^H(\theta)$, $K_1^H(\theta) := 2 \partial_{\theta} L_0^H(\theta)$, i.e.,
\begin{align}
K_0^H &= \frac{\pm \lambda_{\text{\sc n}}^2}{ \bigl[ \lambda_{\text{\sc n}}^2 (\theta + \tilde{\rho})^2 - \rho^2\bigr]^{3/2}}
 \left[ \lambda_{\text{\sc n}} ( \theta + \tilde{\rho}) J_0^H - \frac{\rho}{n} J_1^H \right],
\nonumber\\
K_1^H &= \frac{\pm \lambda_{\text{\sc n}}^2}{ \bigl[ \lambda_{\text{\sc n}}^2 (\theta + \tilde{\rho})^2 - \rho^2\bigr]^{3/2}}
 \big[ \lambda_{\text{\sc n}} ( \theta + \tilde{\rho}) J_1^H - \rho n J_0^H \big],
\label{KHdef}
\end{align}
where also the basic currents are on-shell, i.e.,~\eqref{JHthetaonshell3} holds.
Then set
\begin{gather}
\mathcal{J}_{\mu}^H(\theta) := U^\mathrm{on}(\theta) K_{\mu}^H(\theta)
U^\mathrm{on}(\theta)^{-1}, \qquad \mu =0,1, \qquad \operatorname{Im} \theta \neq 0.
\label{ls6}
\end{gather}
This construction has precursors in the integrable structure
of flat space nonlinear sigma-models~\cite{FT} and was originally
adapted to the Lax pair of the 2-Killing sectors by one of the authors
(M.N.). The rationale of the definition~\eqref{ls6} is the following.

\begin{Result} Let $U^\mathrm{on}(\theta)$, $\operatorname{Im}\theta \neq 0$, be a~solution of
\eqref{linsystemH}. Then the
one-parameter current~\eqref{ls6} is conserved without gauge fixing
\begin{gather}
e_0\bigl( \mathcal{J}_0^H(\theta) \bigr) = \partial_1 \mathcal{J}_1^H(\theta).
\label{ls5}
\end{gather}
Moreover, its conservation can be rendered manifest by the rewriting
\begin{gather}
\mathcal{J}_0^H(\theta) = 2 \partial_1 \bigl( \partial_{\theta} U^\mathrm{on} {U^\mathrm{on}}^{-1} \bigr),
\qquad
\mathcal{J}_1^H(\theta) = 2 e_0 \bigl( \partial_{\theta} U^\mathrm{on} {U^\mathrm{on}}^{-1} \bigr),
\label{ls7}
\end{gather}
where $\partial_{\theta} U^\mathrm{on} {U^\mathrm{on}}^{-1}$ is treated as a~spatial scalar.
\end{Result}

For the verification of the conservation equation, we proceed as follows.
Using~\eqref{linsystemH}, the conservation~\eqref{ls5} amounts to
\begin{gather}
e_0\bigl(K_0^H\bigr) = \partial_1 K_1^H + \bigl[L_1^H, K_1^H\bigr] -
\bigl[L_0^H, K_0^H\bigr].
\label{JHthetaonshell1}
\end{gather}
Inserting the definitions~\eqref{KHdef} one finds~\eqref{JHthetaonshell1}
implies the $\theta$-independence of $e_0\bigl(L_1^H\bigr) -
\partial_1 L_0^H + \bigl[L_0^H, L_1^H\bigr]$. Evaluating it for $\theta \rightarrow \infty$
shows its vanishing on account of~\eqref{JHthetaonshell3b}.
This in turn is necessary and sufficient for the local existence
of $U^\mathrm{on}$. Further, spelling out $\partial_{\theta} L_1^H$ using
$L_1^H = {U^\mathrm{on}}^{-1} \partial_1 U^\mathrm{on}$ and $\partial_{\theta} L_0^H$ using
$L_0^H = {U^\mathrm{on}}^{-1} e_0\bigl(U^\mathrm{on}\bigr)$, one finds~\eqref{ls7}
upon insertion into~\eqref{ls6}.

\section{Dirac observables: Strong, off-shell, and without gauge-fixing}\label{sec3}

The conserved quantities considered in Section~\ref{section2.3} were based on
an on-shell solution $U^\mathrm{on}$ of the linear system. The implication
\eqref{linsystemH1} shows that $U^\mathrm{on}$ is also characterized by
$\partial_1 U = U L_1^H$ and~${e_0\bigl(L_1^H\bigr) - \partial_1 L_0^H + \bigl[L_0^H, L_1^H\bigr] =0}$.
In this section, we drop the second condition (so that all fields are
off-shell) and seek appropriate solutions of $\partial_1 U = U L_1^H$ only.
With such a~solution at hand, we then explore whether one or more
of the construction principles in Section~\ref{section2.3} gives rise
to a~one-parameter family of Dirac observables $\mathcal{O}(\theta)$. If successful,
this construction principle has the ``off-shell'' and ``no gauge fixing''
properties aimed at built in. The strong vanishing of the Poisson brackets
$\{\mathcal{H}_0, \mathcal{O}(\theta) \} = 0 = \{\mathcal{H}_1, \mathcal{O}(\theta) \}$ is also a~feature
of this construction principle: the Lax pair
\eqref{LHdef} does not contain $\sigma$ and contains $\pi^{\sigma}$ only through $\tilde{\rho}$. Inspection
of the gauge variations~\eqref{gtransH} shows that the combinations $\pi^\sigma \pi^\rho$ and $\pi^\sigma \partial_1\sigma$ cannot occur
upon evaluation of the Poisson brackets. As the constraints
$\mathcal{H}_0$, $\mathcal{H}_1$ contain these combinations they can therefore also not occur
on the right-hand side, and $\{\mathcal{H}_0, \mathcal{O}(\theta) \}$, $\{\mathcal{H}_1, \mathcal{O}(\theta) \}$
will have to vanish strongly or not at all. In terms of the gauge variations~\eqref{gtransH}, we therefore take
\begin{gather}
\delta_{\epsilon}^H \mathcal{O}(\theta) =0,
\label{obs0}
\end{gather}
as the characterization of Dirac observables aimed at. Since~\eqref{gtransH}
also contains gauge variations of lapse and shift the condition
\eqref{obs0} is equivalent to $\mathcal{O}(\theta)$ having strongly vanishing Poisson
brackets with the extended constraints, where contributions
proportional to the primary constraints are added. In the end, we shall
check for our $\mathcal{O}(\theta)$'s, however, that the gauge variations of
lapse and shift do not
contribute, rendering~\eqref{obs0} equivalent to~\eqref{i2}, as claimed
in the introduction.
For clarity's sake, we add that we aim here at Dirac observables for
the Gowdy systems with two \textit{independent polarizations}.
The special case when the polarizations are aligned is much simpler
and Dirac observables have been known for some time~\cite{Torreobs}.

As noted earlier, the symmetry reductions with respect to
two spacelike Killing vectors also have a~global
$\mathrm{SL}(2,\mathbb{R})$ symmetry. Our observables $\mathcal{O}(\theta)$
will in general not be $\mathrm{SL}(2,\mathbb{R})$ invariant. This
can likewise be expressed canonically
using the $\mathrm{SL}(2,\mathbb{R})$ Noether current~\eqref{JHdef}.
The associated Noether charge $Q = \int {\rm d} x^1 J_0^H$
remains well-defined off-shell and generates infinitesimal $\mathrm{SL}(2,\mathbb{R})$
rotations of the fields $\Delta$, $\psi$. The finite transformations can be
implemented on phase space by the exponentiated Poisson action
of $Q$, schematically $F \mapsto \exp\{ \alpha \mathrm{ad} Q\} F$,
with~${\mathrm{ad Q}(F) = \{Q,F\}}$. The explicit formulas depend on
the situation considered and will be spelled out as needed.

The detailed constructions turn out to be very different for
the $\mathbb{R} \times T^2$ Gowdy systems (spatially noncompact)
and the $T^3$ Gowdy cosmologies (spatially compact). The
$\mathbb{R} \times T^2$ case is much simpler and is technically similar to
the situation for cylindrical gravitational waves
with appropriate spatial fall-off. For the latter, typically regularity on
the $r=0$ symmetry axis and a~suitable fall-off for
$r \rightarrow \infty$ is assumed~\cite{TorreRom}. The construction principles
for the on-shell conserved quantities in Section~\ref{section2.3} then apply
and have been studied in~\cite{KSYangian,Gerochreview}. The
existence of these conserved charges also constrains a~putative
S-matrix~\cite{PennaSmat}. The $r \rightarrow \infty$ limit of the
off-shell $U$ (solving the spatial part of the linear system
with regular data at $r=0$) also exists for cylindrical gravitational
waves and obeys a~closed quadratic Poisson algebra~\cite{KSYangian}.
In view of this rather satisfactory situation, we shall not
further consider cylindrical gravitational waves here.

An important auxiliary concept in our constructions for the Gowdy systems
is the so-called \textit{transition matrix}. This notion is
borrowed from the theory of flat space integrable field theories
\cite{FT}. In the present context, we need an off-shell Hamiltonian
version valid without gauge fixing. We define this off-shell
transition matrix as the
solution of the spatial initial value problem
\[
\partial_{x^1} T^H\bigl(x^0;x^1,y^1;\theta\bigr) =
T^H\bigl(x^0;x^1,y^1;\theta\bigr) L_1^H\bigl(x^0,x^1;\theta\bigr),\qquad T^H\bigl(x^0; y^1, y^1;\theta\bigr) =\mathbbm{1},
\]
where $L_1^H$ is given by~\eqref{LHdef}. This is solved by a~path
ordered exponential, which turns out to be a~convergent series
\begin{align}
T^H\bigl(x^0;x^1,y^1;\theta\bigr) ={}&
\exp_+ \left\{\int_{y^1}^{x^1} {\rm d}z L_1^H\bigl(x^0,z;\theta\bigr)\right\}
\label{THsol1}\\
:= {}& \mathbbm{1} + \sum_{n=1}^{\infty}
\int_{y^1}^{x^1} {\rm d}z_1 \int_{y^1}^{z_1} {\rm d}z_2
\cdots
\int_{y^1}^{z_{n-1}} {\rm d}z_n L_1^H\bigl(x^0,z_n;\theta\bigr) \cdots L_1^H\bigl(x^0,z_1;\theta\bigr),\nonumber
\end{align}
where the subscript $+$ is mnemonic for an ordering according
to increasing spatial arguments, $z_n < z_{n-1} < \ldots < z_1$,
and $\operatorname{Im} \theta \neq 0$ is assumed throughout.
For the inverse, one has
\begin{equation*}
T^H\bigl(x^0;x^1,y^1;\theta\bigr)^{-1} =T^H\bigl(x^0;y^1,x^1;\theta\bigr).
\end{equation*}
The convergence of the series holds under broad conditions.

For example, having $x \mapsto L_1^H\bigl(x^0,x;\theta\bigr)$ bounded
in some $2 \times 2$ matrix norm suffices to have the
series converge and be bounded in the same norm by an exponential
in $x^1 - y^1>0$, for fixed~$x^0$,~$\theta$. We will therefore
treat the series in~\eqref{THsol1} as convergent and to describe
the exact solution of $\partial_1 T^H = T^H L_1^H$ with initial
condition $T^H|_{x^1 = y^1} =\mathbbm{1}$. While $T^H$ is always well-defined,
the limits $y^1 \rightarrow - \infty$ and/or $x^1 \rightarrow + \infty$
are not necessarily. In fact, this is one of several aspects
where the $\mathbb{R} \times T^2$ and $T^3$ Gowdy systems differ significantly.

\subsection[R times T\^{}2 Gowdy]{$\boldsymbol{\mathbb{R} \times T^2}$ Gowdy}\label{sec3.1}

We take the transition matrix~\eqref{THsol1} as our starting point
and aim at determining its response under the Hamiltonian gauge variations
\eqref{gtransH}. Since $T^H$ is determined by $L_1^H$, which contains~$J_1^H$,~$J_0^H$ and $\tilde{\rho}$, we begin there.

\textbf{Gauge variations of $\boldsymbol{\mathbf{J_1^H}}$, $\boldsymbol{\mathbf{J_0^H}}$ and $\boldsymbol{\mathbf{\tilde{\rho}}}$.}
The gauge variation for $J_0^H$ was already found in~\eqref{HJvar}.
For $J_1^H$, it is useful to consider the gauge variation of $J_1^H/n$.
As it is a~spatial $+1$ density, one expects
\begin{gather}
\bigl\{ n^{-1} J_1^H, \mathcal{H}_1\bigl(\epsilon^1\bigr) \bigr\} = \partial_1 \bigl( \epsilon^1 n^{-1} J_1^H \bigr),
\label{JHvar2}
\end{gather}
which is indeed confirmed by a~direct componentwise computation.
The temporal gauge variation comes out as
\begin{align}
\bigl\{ n^{-1} \rho J_1^H, \mathcal{H}_0(\epsilon) \bigr\} &= - 2 \lambda_{\text{\sc n}} \epsilon \pi^{\sigma} n^{-1} J_1^H +
\rho^2 \partial_1\bigl( \epsilon \rho^{-1} J_0^H \bigr) - \lambda_{\text{\sc n}} \frac{\epsilon}{n}
\bigl[ J_1^H, J_0^H\bigr].
\label{JHvar3}
\end{align}
Here the explicit form of the commutator is used
\begin{equation*}
\frac{\epsilon}{n} \bigl[ J_1^H, J_0^H\bigr] =
\frac{2 \epsilon \rho}{\lambda_{\text{\sc n}} \Delta}
\bigl( \pi^{\Delta} \partial_1 \psi - \pi^{\psi} \partial_1 \Delta \bigr)
\begin{pmatrix} \psi & -\bigl(\Delta^2 + \psi^2\bigr) \\
 1 & - \psi \end{pmatrix},
\end{equation*}
and the computation proceeds again componentwise.

In~\eqref{rhotpdef},
we gave a~partial characterization of the field $\tilde{\rho}$
entering the linear system. A~complete characterization is
provided by the following three conditions: First,
$\tilde{\rho}$ is defined on $[t_i,t_f] \times \mathbb{R}$ and transforms according to diffeomorphisms whose gauge descriptors
$\bigl(\epsilon,\epsilon^1\bigr)$ are smooth with compact support, see
\eqref{epsnoncompact}. Second, $\tilde{\rho}$ is a~scalar
under compactly supported spatial diffeomorphisms. Third, it obeys
\begin{gather}
\partial_1 \tilde{\rho} = - \pi^{\sigma},\qquad
\delta_{\epsilon}^H \tilde{\rho} =
- \epsilon^1 \pi^{\sigma} + \frac{\epsilon}{\lambda_{\text{\sc n}}} \partial_1 \rho.
\label{rhotvar}
\end{gather}
The off-shell gauge variation is such that its on-shell specialization
to $\epsilon =n$, $\epsilon^1 =0$ reproduces the second relation in~\eqref{rhotpdef}.
The so-characterized $\tilde{\rho}$ can be viewed as an off-shell
Hamiltonian version of the conjugate $\tilde{\rho}$ of a~(pseudo-)harmonic
function $\rho$ entering the usual constructions~\mbox{\cite{Gravisolitons, Maison}}.
It remains to justify the gauge variation. Clearly, the general
solution of the first relation in~\eqref{rhotvar} is
\begin{gather}
\tilde{\rho}\bigl(x^0,x^1\bigr) := - \int_{y^1}^{x^1} {\rm d}x \pi^{\sigma}\bigl(x^0, x\bigr) +
\tilde{\rho}\bigl(x^0,y^1\bigr),
\label{rhotHdef}
\end{gather}
for some reference point $y^1$. Using $\delta_{\epsilon}^H \pi^{\sigma}$
from~\eqref{gtransH} to compute the gauge variation of the
integral, it reduces to boundary contributions, one at
$x^1$ and the other at $y^1$. By shifting the reference point
$y^1$ to $-\infty$, the associated boundary contribution vanishes,
because by assumption the descriptors of the underlying diffeomorphisms
obey~\eqref{epsnoncompact}. Hence one can consistently set
$\lim_{y^1 \rightarrow - \infty}\tilde{\rho}\bigl(x^0,y^1\bigr)=0$,
$\lim_{y^1 \rightarrow - \infty}\bigl(\delta_{\epsilon}^H\tilde{\rho}\bigr)\bigl(x^0,y^1\bigr)=0$.
The remaining terms in the gauge variation of~\eqref{rhotHdef} with
$y^1 \rightarrow -\infty$ are those in~\eqref{rhotvar}. For the $\mathbb{R} \times
T^2$ Gowdy system, this completes the derivation of the asserted
gauge variation for $\tilde{\rho}$.

\textbf{Gauge variations of $\boldsymbol{L_1^H(\theta)}$ and $\boldsymbol{T^H(\theta)}$.}
Using~\eqref{HJvar}, \eqref{JHvar2}, \eqref{JHvar3}, one finds
\begin{equation*}
\delta_{\epsilon}^H L_1^H(\theta) =
\partial_1\left( \frac{\epsilon}{n} L_0^H \right)
+ \frac{\epsilon}{n} \bigl[L_1^H, L_0^H\bigr] + \partial_1\bigl( \epsilon^1 L_1^H\bigr),
\end{equation*}
again valid off-shell and without gauge fixing. The spatial part
just reflects the fact that $L_1^H(\theta)$ is a~spatial $+1$ density.
For the behavior of $T^H$ under gauge transformations, this entails
\begin{gather}
\delta_{\epsilon}^H T^H\bigl(x^0;x^1,y^1;\theta\bigr) = T^H\bigl(x^0;x^1,y^1;\theta\bigr)
 C_{\epsilon}\bigl(x^1,\theta\bigr) -
 C_{\epsilon}\bigl(y^1,\theta\bigr)T^H\bigl(x^0;x^1,y^1;\theta\bigr),
\nonumber\\
C_{\epsilon}\bigl(x^1,\theta\bigr) = \left(\frac{\epsilon}{n} L_0^H + \epsilon^1 L_1^H\right)\bigl(x^1;\theta\bigr)
, \qquad \delta_{\epsilon} L_1^H = \partial_1 C_{\epsilon} + \bigl[L_1^H, C_{\epsilon}\bigr],
\label{THsol5}
\end{gather}
which is consistent with $T^H\bigl(x^0;y^1,y^1;\theta\bigr) =\mathbbm{1}$. In principle,
this can be verified from the series~\eqref{THsol1}. A~more economical line of reasoning is to temporarily define $ A\bigl(x^0;x^1,y^1;\theta\bigr)
:= \delta_{\epsilon}^H T^H\bigl(x^0;x^1,y^1;\theta\bigr) - T^H\bigl(x^0;x^1,y^1;\theta\bigr)
C_{\epsilon}\bigl(x^1,\theta\bigr)$. One verifies that $A$ satisfies the same first-order
equation as $T^{H}$, i.e.,~$\partial_{x^1} A = A L_1^H $, but with boundary
condition $A\bigl(x^0;y^1,y^1;\theta\bigr) = -C_{\epsilon}\bigl(y^1,\theta\bigr)$. Therefore, a~solution is $A\bigl(x^0;x^1,y^1;\theta\bigr) = -C_{\epsilon}\bigl(y^1,\theta\bigr)T^H\bigl(x^0;x^1,y^1;\theta\bigr)$
which yields the relation~\eqref{THsol5} for the gauge variation of $T^{H}$.
It remains to show that this $A$ is the unique solution. Suppose $A_1$
and $A_2$ both solve $\partial_{x^1} A = A L_1^H$ with the same boundary condition
$A\bigl(x^0;y^1,y^1;\theta\bigr) = -C_{\epsilon}\bigl(y^1,\theta\bigr)$, and consider
$\Delta A := A_1 - A_2$. Then $\partial_{x^1} \Delta A = \Delta A L^{H}_{1}$,
with $\Delta A\bigl(x^0;y^1,y^1;\theta\bigr) = 0$. Further,
\begin{align*}
\partial_{x^1}\bigl[ \Delta A \bigl(T^{H}\bigr)^{-1}\bigr]
&= \bigl(\partial_{x^1} \Delta A \bigr)\bigl(T^{H}\bigr)^{-1}
 + \Delta A \partial_{x^1}\bigl(T^{H}\bigr)^{-1}
\\
&= \Delta A L^{H}_{1}\bigl(T^{H}\bigr)^{-1} - \Delta A L^{H}_{1}\bigl(T^{H}\bigr)^{-1}
= 0.
\end{align*}
So, $\Delta A \bigl(T^{H}\bigr)^{-1}$ is constant in $x^1$. Evaluating the product
at $x^1=y^1$ gives $\Delta A\bigl(x^0;y^1,y^1;\theta\bigr)=0$, hence
$\Delta A\bigl(x^0;x^1,y^1;\theta\bigr) = 0$, for all $x^1$, and uniqueness follows.

\textbf{$\boldsymbol{U}$ and its gauge variation.}
To proceed, we seek to define $U$ by
\begin{gather}
U\bigl(x^0,x^1;\theta\bigr) := \lim_{y^1 \rightarrow -\infty} T^H\bigl(x^0,x^1,y^1;\theta\bigr).
\label{UdefNC}
\end{gather}
For the limit to exist $L_1^H$ must fall off sufficiently fast in $x^1$.
Specifically, we assume that the basic currents $J_0^H$, $J_1^H$ have
for $\big|x^1\big| \rightarrow \infty$ a~decay of order
\smash{$O\bigl(\big|x^1\big|^{-1-\delta}\bigr)$}, $\delta>0$. As explained earlier, we take
the gauge descriptors $\bigl(\epsilon,\epsilon^1\bigr)$ to be of the type
\eqref{epsnoncompact} and define $\tilde{\rho}$ by $\tilde{\rho}\bigl(x^0,x^1\bigr)
= \smash{- \int_{-\infty}^{x^1} {\rm d}x \pi^\sigma\bigl(x^0,x\bigr)}$. This assumes an appropriate
spatial fall-off of $\pi^\sigma$, see also~\cite{TorreRom}.
Under these fall-off assumptions, it follows that
\begin{equation*}
\sup_{z \in \mathbb{R}} \bigl\Vert (1 + |z|)^{1 + \delta} L_1^H\bigl(x^0, z;\theta\bigr) \bigr\Vert =
\varkappa\bigl(x^0,\theta\bigr) <\infty,
\end{equation*}
for some $2 \times 2$ matrix norm and $\delta>0$.
The series~\eqref{THsol1} then has a~convergent $y^1 \rightarrow -\infty$ limit
bounded in the same norm by
\begin{align}
\left\Vert U\bigl(x^0;x^1;\theta\bigr) \right\Vert
\leq{}& 1 + \sum_{n\geq 1} \varkappa\bigl(x^0,\theta\bigr)^n \int_{-\infty}^{x^1} {\rm d}z_1 \int_{-\infty}^{z_1} {\rm d}z_2
\cdots\nonumber\\
&
\times \int_{-\infty}^{z_{n-1}} {\rm d}z_n
(1 + |z_n|)^{-1 - \delta} \cdots (1 + |z_1|)^{-1 - \delta}
\leq \exp\bigl\{2 \varkappa\bigl(x^0,\theta\bigr)/\delta\bigr\}\label{UboundNC2}
\end{align}
for all $x^1 \in \mathbb{R}$. With~\eqref{UdefNC} well-defined, it also
obeys the boundary condition in~\eqref{Ubc}, i.e., $\lim_{x^1 \rightarrow - \infty} U\bigl(x^0,x^1;\theta\bigr) = \mathbbm{1}$.

Moreover, the so-defined $U$
obeys the following gauge variation:
\begin{gather}
\delta^H_\epsilon U\bigl(x^1,\theta\bigr) = U\bigl(x^1,\theta\bigr) C_\epsilon\bigl(x^1,\theta\bigr),
\label{Ugauge}
\end{gather}
where we omit the shared time argument $x^0$. This does not quite follow
by taking the $y^1 \rightarrow -\infty$ limit of~\eqref{THsol5}, as
on the left-hand side the $\delta^H_{\epsilon}$ gauge variation would have
to be interchanged with the limit. Inspection of~\eqref{THsol5}'s
derivation shows that the gauge variations of $n$, $s$ do not enter.
One can therefore strip off the descriptors $\epsilon$, $\epsilon^1$ (assumed
to obey~\eqref{epsnoncompact}) and split~\eqref{THsol5} into separate
equations for $\{\mathcal{H}_0, T^H\}$, $\{\mathcal{H}_1, T^H\}$. Here the
$y^1 \rightarrow - \infty$ limit can cleanly be taken and produces
$U L_0^H/n$, $U L_1^H$, respectively, on account of the fall-off
of $L_0^H\bigl(x^0,y^1;\theta\bigr)$, $L_1^H\bigl(x^0,y^1;\theta\bigr)$. From a~technical
viewpoint, the relation~\eqref{Ugauge} is the crucial ingredient for the
construction of Dirac observables in the $\mathbb{R} \times T^2$ Gowdy system.

\textbf{Relation $\boldsymbol{U_+}$ versus $\boldsymbol{U_-}$.} By direct computation using~\eqref{Mderiv3}, one verifies that $MU_-(\theta)^{\mathsf T}$ solves the same differential equation as $U_+(\theta)^{-1}$. By assumption~\eqref{Ubc}, both $U_+$ and $U_-$ approach $\mathbbm{1}$ as $x^1 \rightarrow -\infty$. The standard fall-off conditions on the fields $\Delta$, $\psi$ in the $\mathbb{R} \times T^2$ system are $\ln\Delta \rightarrow 0$, $\psi \rightarrow 0$ for $\big|x^1\big| \rightarrow \infty$, see~\cite{TorreRom}. This gives $M \rightarrow \mathbbm{1}$ as $x^1 \rightarrow - \infty$. Hence, $MU_-(\theta)^{\mathsf T}$ and $U_+(\theta)^-{1}$ solve the same differential equation with the same boundary conditions, and we conclude
\begin{gather}
U_-(\theta)
 =\big[U_+(\theta)^{\mathsf T}\big]^{-1}M^{-1}.
\label{Upm1}
\end{gather}
This relation is also consistent with the gauge variation~\eqref{Ugauge}. Indeed, writing $C_\epsilon^{\pm}(\theta)$ for the expression in~\eqref{THsol5} with $L_\mu^{H,\pm}(\theta)$, respectively, one has from~\eqref{Mderiv3}
\begin{gather}
M C_\epsilon^{\pm}(\theta)^{\mathsf T} M^{-1} = -
C_\epsilon^{\mp}(\theta) - \frac{\lambda_{\text{\sc n}}}{\rho} \left( \epsilon J_0^H + \epsilon^1 \frac{J_1^H}{n}\right).
\label{Upm2}
\end{gather}
Further, starting from~\eqref{Mmat1}, using the gauge variations
$\delta_{\epsilon}^H \Delta$, $\delta_{\epsilon}^H \psi$ from~\eqref{gtransH},
and expressing the result in terms of~\eqref{JHdef} produces
\begin{gather}
\delta_{\epsilon}^H M M^{-1} = \frac{\lambda_{\text{\sc n}}}{\rho} \left( \epsilon J_0^H +
\epsilon^1 \frac{J_1^H}{n} \right).
\label{Upm3}
\end{gather}
Combining~\eqref{Upm2} with~\eqref{Upm3} and~\eqref{Ugauge}, one can
verify that $MU_-(\theta)^{\mathsf T}$ obeys the same gauge variation as $U_+(\theta)^{-1}$.
Hence, the identification~\eqref{Upm1} is consistent with the respective
gauge variations $\delta_\epsilon U_{\pm} = U_{\pm} C_\epsilon^{\pm}$.

\textbf{Off-shell version of $\boldsymbol{U}$ limit.} With these preparations at hand, we
consider an off-shell version of the first construction principle from
Section~\ref{section2.3}.
Under the same assumptions as~\eqref{UdefNC},
the two-sided limit also exists. We set
\begin{gather}
\mathcal{O}_T\bigl(x^0,\theta\bigr) :=
\lim_{x^1 \rightarrow \infty} U\bigl(x^0,x^1;\theta\bigr) =
\lim_{x^1 \rightarrow \infty} \lim_{y^1 \rightarrow -\infty} T^H\bigl(x^0;x^1,y^1;\theta\bigr).
\label{OTdef}
\end{gather}
By a~reasoning analogous to the one leading to~\eqref{Ugauge},
one sees that the limit's gauge variation vanishes,
$\delta_{\epsilon}^H \mathcal{O}_T\bigl(x^0,\theta\bigr) =0$. This applies to both sign options
in~\eqref{LHdef}, whenever the distinction is relevant we write
$\mathcal{O}_T^{\pm}(\theta)$ for the limit deriving from $U_{\pm}$ solving
\smash{$\partial_1 U_{\pm} = U_{\pm} L_1^{H,\pm}$}.
For convenient reference, we summarize a~precise version as the following.

\begin{Result} \label{result0GowdyNC}
 On the phase space of the $\mathbb{R} \times T^2$ Gowdy system with generic
 polarizations consider the functions~\eqref{JHdef} with
 \smash{$O\bigl(\big|x^1\big|^{-1- \delta}\bigr)$}, $\delta>0$, type fall-off. In terms of them
 introduce the Lax pair~\eqref{LHdef},
where $\tilde{\rho}$ is a~solution of $\partial_1 \tilde{\rho}
= - \pi^{\sigma}$ with gauge variation $\delta_{\epsilon}^H \tilde{\rho} =
- \epsilon^1 \pi^{\sigma} + \epsilon \partial_1 \rho/\lambda_{\text{\sc n}}$.
\begin{itemize}
\itemsep=0pt
\item[$(a)$] Let $U(\theta)$, $\operatorname{Im} \theta \neq 0$, be given by~\eqref{UdefNC}
 with gauge variation~\eqref{Ugauge} and define
 $\mathcal{O}_T\bigl(x^0,\theta\bigr)$ by~\eqref{OTdef}, which refers to a~fixed time $x^0$.
 Then $\mathcal{O}_T\bigl(x^0,\theta\bigr)$ is a~one-parameter family of strong, off-shell
 Dirac observables, $\{\mathcal{H}_0,\mathcal{O}_T(\theta)\} =0=
 \{\mathcal{H}_1,\mathcal{O}_T(\theta)\}$.
\item[$(b)$] Part $(a)$ applies to both sign options $\mathcal{O}_T^{\pm}(\theta)$,
 which are related by
\begin{equation*}
\mathcal{O}_T^{-}\bigl(x^0,\theta\bigr) = \bigl[\mathcal{O}_T^{+}\bigl(x^0,\theta\bigr)^{\mathsf T}\bigr]^{-1}.
\end{equation*}
 They admit asymptotic expansions in $1/\theta$ whose coefficients are
 strong, off-shell Dirac observables.
\end{itemize}
\end{Result}

Part (a) is clear. The relation between $\mathcal{O}_T^-$ and $\mathcal{O}_T^+$
in (b) follows from~\eqref{Upm1}. Only the assertion about the expansions
remains to be justified; we relegate this to the end of this subsection.
\vspace{-2mm}

We add some remarks.

(i) A~similar (though technically somewhat different) construction of $\mathcal{O}_T$
and $\mathcal{M}$ in conformal gauge has been considered in
\cite{KSGowdyNC,KSYangian}. Their off-shell role as quantities Poisson commuting
with~${(\mathcal{H}_{0} \pm \mathcal{H}_1)_\mathrm{specialized}}$ (see before~\eqref{LHdef}) is
left implicit. The main difference, however, lies in the
boundary conditions for $\rho \rightarrow 0$, in that the fields and their
time derivatives are assumed to stay regular for $\rho \rightarrow 0$
(``regular sector''). As mentioned in~\cite{KSYangian}, this regularity
assumption is non-generic and entails nontrivial implicit relations
between the coordinate and momentum fields. It also has the
undesirable consequence that all the nonlocal conserved charges
obtained by expanding the on-shell versions in powers of
$1/\theta$ vanish identically~\cite{KSGowdyNC}. As a~workaround, a~variant
of the monodromy matrix~\cite{BM} was adopted in~\cite{KSYangian},
which is defined \textit{on} the branch cut in Figure~\ref{fig1}
and is intended to describe the before-mentioned regular sector
beyond the level of a~series expansion around $\theta =0$ or $\theta = \infty$.
In adapted coordinates $\rho = x^0$, $\tilde{\rho} = x^1$ this monodromy
matrix reduces to $M\bigl(x^0=0,x^1 = -\theta\bigr)$ (which is not even defined in
the present setting).

(ii) As detailed in Section~\ref{sec4}, the
AVD scenario provides a~physically more satisfactory account of
the behavior of the Gowdy systems near the big bang, $\rho \rightarrow 0$.
In brief, the fields may become singular near the big bang but
the basic current $J_0^H$ stays regular (with $J_1^H$ subleading).
Moreover, this behavior is expected to be generic and motivated our
search to find conserved charges and Dirac observables that are finite
and nonzero when backpropagated to the big bang.

(iii) In deriving the relation between $\mathcal{O}_T^-$ and $\mathcal{O}_T^+$,
we assumed for simplicity trivial asymptotics for $M$
as $x^1 \rightarrow \pm \infty$. This breaks the covariance under
global $\mathrm{SL}(2,\mathbb{R})$ transformations $M \mapsto C M C^{\mathsf T}$,
$C \in \mathrm{SL}(2,\mathbb{R})$; see before~\eqref{Mmat2}. The invariance
could be restored by assuming $\lim_{x^1 \rightarrow \pm \infty} M\bigl(x^0, x^1\bigr)
= M_{\pm}$ for constant, symmetric, unimodular $2\times 2$ matrices
that transform according to $M_{\pm} \rightarrow C M_{\pm} C^{\mathsf T}$. Then
$\mathbbm{1} = \mathcal{M}(\theta) = \mathcal{O}_T^+(\theta) \mathcal{O}_T^-(\theta)^{\mathsf T}$ in~\eqref{Mofftriv}
is replaced with
$M_- = \mathcal{M}(\theta) = \mathcal{O}_T^+(\theta) M_+ \mathcal{O}_T^-(\theta)^{\mathsf T}$, and
on account of $U( \theta) \mapsto C U(\theta) C^{-1}$,
i.e.,~$\mathcal{O}_T^{\pm} \mapsto C \mathcal{O}_T^{\pm}(\theta) C^{-1}$,
covariance is maintained.

(iv) The on-shell specializations of the $\mathcal{O}_T^{\pm}\bigl(x^0,\theta\bigr)$
reproduce the one-parameter families of conserved charges from Section~\ref{section2.3}.
Indeed, for $\epsilon = n, \epsilon^1 =0$ the gauge invariance
$\delta_{\epsilon}^H \mathcal{O}_T(\theta) =0$ reduces to
$0 = \bigl\{ \mathcal{O}_T^\mathrm{on}(\theta), \mathcal{H}_0(n) \bigr\} = e_0\bigl( \mathcal{O}_T^\mathrm{on}(\theta)\bigr)$.

(v) In the polarized subsystem, i.e.,~$\psi \equiv 0$, $\pi^{\psi}\equiv 0$,
the above assertions can be directly verified: the $L_0^H(\theta)$,
$L_1^H(\theta)$ become tracefree diagonal matrices determined by
their $(1,1)$ components $\ell^H_0(\theta)$, $\ell^H_1(\theta)$, respectively.
Hence $U(\theta)$ becomes diagonal unimodular with $(1,1)$ component $u(\theta)$.
The limit~\eqref{UdefNC} remains well-defined and gives
$\ln \bigl[\lim_{x^1 \rightarrow \infty} u\bigl(x^0,x^1;\theta\bigr)\bigr] =
\int_{-\infty}^{\infty} {\rm d}x \ell_1^H\bigl(x^0,x;\theta\bigr)$. This can directly be
verified to provide a~one-parameter family of strong, off-shell Dirac
observables for the polarized $\mathbb{R} \times T^2$ Gowdy system,
cf.\ \cite{GowdyPol}. Specializing $U_{\pm}$ to the polarized sector
leads to \smash{$\ln u_{\pm}\bigl(x^0,x^1;\theta\bigr) =
\int_{-\infty}^{x^1} {\rm d}x \ell_1^{H,\pm}\bigl(x^0,x;\theta\bigr)$}, with $\ell_1^{H,\pm}$
arising from the specializations of~\eqref{LHdef}. These are such that
\smash{$\ell_1^{H,+}(\theta) + \ell_1^{H,-}(\theta) = -\partial_1 \ln \Delta$}, from which one
recovers the specialization of~\eqref{Upm1}.

\textbf{Off-shell currents $\boldsymbol{\mathbf{\mathcal{J}_{\mu}^H(\theta)}}$.}
We proceed with the third construction principle in Section~\ref{section2.3}
and aim at promoting an
off-shell version \smash{$\mathcal{J}_{0}^H(\theta) := U(\theta) K_{0}^H(\theta) U(\theta)^{-1} $}
of the conserved current~\eqref{ls6} to a~one-parameter
family of Dirac observables. We begin with computing the gauge
variation of $K_0^H(\theta)$ as defined in~\eqref{KHdef}. With $\tilde{\rho}$
viewed as a~spatial scalar, $K_0^H$ is a~spatial $+1$ density. As such
one expects
\begin{gather}
\bigl\{ K_0^H, \mathcal{H}_1\bigl(\epsilon^1\bigr) \bigr\} = \partial_1\bigl( \epsilon^1 K_0^H\bigr),
\label{KHvar1}
\end{gather}
which is indeed confirmed using~\eqref{HJvar}, \eqref{JHvar2}. A~lengthy direct computation gives
\begin{align*}
\bigl\{ K_0^H, \mathcal{H}_0(\epsilon) \bigr\} = {}&\partial_1 \left( \frac{\epsilon}{n} K_1^H \right)
+ \frac{\pm \lambda_{\text{\sc n}}^2}{ \bigl[ \lambda_{\text{\sc n}}^2 (\theta + \tilde{\rho})^2 - \rho^2\bigr]^{3/2}}\\
& \times \left( 2 \frac{\epsilon \lambda_{\text{\sc n}}}{n} \partial_1 \tilde{\rho} J_1^H
+ \rho^2 \partial_1 \left( \frac{\epsilon}{\rho} J_0^H \right) -
\bigl\{ n^{-1}\rho J_1^H, \mathcal{H}_0(\epsilon)\bigr\} \right).
\end{align*}
Inserting~\eqref{JHvar3} and taking into account~\eqref{KHvar1},
one finds
\begin{gather}
\delta_{\epsilon}^H K_0^H =
\partial_1 \left( \frac{\epsilon}{n} K_1^H \right)
\pm \lambda_{\text{\sc n}}^3 \frac{\epsilon}{n} \frac{\big[J_1^H, J_0^H\big]}%
{ \bigl[ \lambda_{\text{\sc n}}^2 (\theta + \tilde{\rho})^2 - \rho^2\bigr]^{3/2}}
+ \partial_1\bigl( \epsilon^1 K_0^H\bigr).
\label{KHvar3}
\end{gather}
Combining~\eqref{KHvar3} and~\eqref{Ugauge} to evaluate $\mathcal{J}_0^H$'s
gauge variations, one arrives at
\begin{gather}
\delta_{\epsilon}^H \mathcal{J}_0^H(\theta) = U \bigl\{
\delta_{\epsilon}^H K_0^H + \big[ U^{-1} \delta_{\epsilon}^H U, K_0^H \big]
\bigr\} U^{-1}
= \partial_1\left( \frac{\epsilon}{n} \mathcal{J}_1^H(\theta) + \epsilon^1 \mathcal{J}_0^H(\theta) \right).
\label{JHthetaresult}
\end{gather}
Inspection of the computation shows that neither the shift
nor the lapse variation enters directly. This means, the left hand
side of~\eqref{JHthetaresult} can be identified with
$\bigl\{ \mathcal{J}_0^H(\theta), \mathcal{H}_0(\epsilon) + \mathcal{H}_1\bigl(\epsilon^1\bigr) \bigr\}$.
In summary, we have the following.

\begin{Result} \label{resultGowdyNC}
On the phase space of the $\mathbb{R} \times T^2$ Gowdy
system with generic polarizations consider the
functions~\eqref{JHdef} with $O\bigl(\big|x^1\big|^{-1- \delta}\bigr)$, $\delta>0$,
type fall-off. In terms of them introduce the Lax pair~\eqref{LHdef},
where $\tilde{\rho}$ is a~solution of $\partial_1 \tilde{\rho}
= - \pi^{\sigma}$ with gauge variation $\delta_{\epsilon}^H \tilde{\rho} =
- \epsilon^1 \pi^{\sigma} + \epsilon \partial_1 \rho/\lambda_{\text{\sc n}}$. Let $U(\theta)$, $\operatorname{Im} \theta \neq 0$,
be given by~\eqref{UdefNC} with gauge variation~\eqref{Ugauge}.
Finally, define
\begin{align}
\mathcal{J}_0^H(\theta) &:= \frac{\pm \lambda_{\text{\sc n}}^2}{ \bigl[ \lambda_{\text{\sc n}}^2(\theta + \tilde{\rho})^2 - \rho^2\bigr]^{3/2}}
U(\theta) \left[\lambda_{\text{\sc n}} ( \theta + \tilde{\rho}) J_0^H - \frac{\rho}{n} J_1^H \right]
U(\theta)^{-1},
\nonumber\\
\mathcal{J}_1^H(\theta) &:= \frac{\pm \lambda_{\text{\sc n}}^2}{ \bigl[\lambda_{\text{\sc n}}^2 (\theta + \tilde{\rho})^2 - \rho^2\bigr]^{3/2}}
U(\theta) \big[ \lambda_{\text{\sc n}}( \theta + \tilde{\rho}) J_1^H - \rho n J_0^H \big]
U(\theta)^{-1},
\label{Diracobs3}
\end{align}
which refer to a~fixed time $x^0$.
Then,
\begin{itemize}
\itemsep=0pt
\item[$(a)$] A~one-parameter family of strong, off-shell Dirac observables is given by
\begin{equation*}
\mathcal{O}(\theta) = \int_{-\infty}^{\infty} {\rm d}x \mathcal{J}_0^H\bigl(x^0,x;\theta\bigr).
\end{equation*}
Specifically,
\begin{gather}
\bigl\{ \mathcal{J}_0^H(\theta), \mathcal{H}_0(\epsilon) + \mathcal{H}_1\bigl(\epsilon^1\bigr) \bigr\} =
\partial_1\left( \frac{\epsilon}{n} \mathcal{J}_1^H(\theta) + \epsilon^1 \mathcal{J}_0^H(\theta) \right),
\label{Diracobs4}
\end{gather}
holds without gauge fixing and without use of any equations of motion.
\item[$(b)$] $\mathcal{O}(\theta)$ admits an asymptotic expansion in powers of
$1/\theta$ whose coefficients are strong, off-shell Dirac observables.
\end{itemize}
\end{Result}

We add some remarks.

(i) The on-shell specialization of~\eqref{Diracobs4} gives
rise to the conserved currents from~\eqref{ls6}.
Indeed, subject to the Hamiltonian equations of motion the (improper)
$\epsilon = n$,
$\epsilon^1= 0$ specialization of~\eqref{Diracobs4}'s left-hand side
equals $e_0\bigl(\mathcal{J}_0^H(\theta)\bigr)$. Hence, \eqref{Diracobs4} turns into
$e_0\bigl(\mathcal{J}^H_0(\theta)\bigr) = \partial_1 \mathcal{J}_1^H(\theta)$, consistent with~\eqref{ls5}.

(ii) Since $U(\theta) \mapsto C U(\theta) C^{-1}$ under global $\mathrm{SL}(2,\mathbb{R})$
transformations, the off-shell current $\mathcal{J}_{\mu}^H(\theta)$
transforms according to $\mathcal{J}_{\mu}^H(\theta) \mapsto C\mathcal{J}_{\mu}^H(\theta)C^{-1}$.
This gives $\mathcal{O}(\theta) \mapsto C \mathcal{O}(\theta) C^{-1}$ for the Dirac observables
in (a).

(iii) The above construction readily specializes to the polarized
$\mathbb{R} \times T^2$ Gowdy system. Since $\psi \equiv 0$, $\pi^{\psi}
\equiv 0$ there, the currents~\eqref{JHdef} become diagonal,
$J_0^H \leadsto \operatorname{diag}\bigl(\pi^{\phi}, - \pi^{\phi}\bigr)$,
$J_1^H \leadsto \rho n \operatorname{diag}(\partial_1 \phi, - \partial_1 \phi)/\lambda_{\text{\sc n}}$,
where $\phi = \ln \Delta$ and $\pi^{\phi}$ is its canonical momentum.
As a~consequence also $L_1^H(\theta)$ becomes
diagonal, and so does $U(\theta) \leadsto \operatorname{diag}\bigl(u(\theta), u(\theta)^{-1}\bigr)$.
In the definition~\eqref{Diracobs3}, the dependence on $u(\theta)$
therefore drops out and one obtains a~much simpler one-parameter
family of currents (see~\cite[Appendix A]{GowdyPol})
\begin{align}
\jmath_0^H(\theta) &:= \frac{\pm \lambda_{\text{\sc n}}^2}{\bigl[ \lambda_{\text{\sc n}}^2(\theta + \tilde{\rho})^2 - \rho^2\bigr]^{3/2}}
\left( \lambda_{\text{\sc n}} (\theta + \tilde{\rho}) \pi^{\phi}
- \frac{\rho^2}{\lambda_{\text{\sc n}}} \partial_1\phi \right),
\nonumber\\
\jmath_1^H(\theta) &:= \frac{ \pm\lambda_{\text{\sc n}}^2}{\bigl[ \lambda_{\text{\sc n}}^2 (\theta + \tilde{\rho})^2 - \rho^2\bigr]^{3/2}}
 \bigl( (\theta + \tilde{\rho}) \rho n \partial_1 \phi - \rho n \pi^{\phi} \bigr).
\label{ahjdef}
\end{align}
The counterpart of~\eqref{Diracobs4} can easily be verified directly,
\begin{gather}
\bigl\{ \jmath_0^H(\theta), \mathcal{H}_0(\epsilon) + \mathcal{H}_1\bigl(\epsilon^1\bigr) \bigr\} =
\partial_1\left( \frac{\epsilon}{n} \jmath_1^H(\theta) + \epsilon^1 \jmath_0^H(\theta)\right),
\label{ahdirac}
\end{gather}
again without using any equations of motion.

\textbf{Expansion in $\boldsymbol{\mathbf{1/\theta}}$.}
Finally, we consider the expansions anticipated in Result
\ref{result0GowdyNC}\,(b) and Result~\ref{resultGowdyNC}\,(b).
With $U_{\pm}$ defined by~\eqref{UdefNC} using $L_1^{H,\pm}$,
the inherited differential equations are
$\partial_1 U_{\pm}=U_{\pm}L_1^{H,\pm}$, with boundary conditions
$ \lim_{x^1\rightarrow-\infty}U_{\pm}\bigl(x^0,x^1;\theta\bigr)=\mathbbm{1}$ for all $x^0$.
Combining~\eqref{Tlimit2} and~\eqref{betapm} from Appendix~\ref{appB},
one has explicitly
\begin{align}
L_1^{\pm}(\theta)={}& - \frac{1}{2\theta} J_0^H +
\Bigg[\frac{\tilde{\rho}}{2\theta(\theta+\tilde{\rho})}-
 \frac{1}{2(\theta+\tilde{\rho})}
 \left(\frac{1}{\sqrt{1-\rho^2/\lambda_{\text{\sc n}}^2(\theta+\tilde{\rho})^2}} -1\right) \Bigg]J_0^H
\nonumber\\
&+ \left(\frac{1}{\sqrt{1-\rho^2/\lambda_{\text{\sc n}}^2(\theta+\tilde{\rho})^2}} -1\right)
\frac{\lambda_{\text{\sc n}} J_1^H}{2\rho n}, \qquad \pm\operatorname{Re}(\theta+\tilde{\rho})>0.
\label{LHpm0}
\end{align}
Expanding in the respective asymptotic regions results in an expansion
of the form
\begin{gather}
U_{\pm}(\theta)
 =\sum_{n\geq0}U_{n} \theta^{-n},\qquad
L_{1}^{H,\pm}(\theta)
 =\sum_{n\geq1}L_{1,n} \theta^{-n}, \qquad \pm\operatorname{Re}(\theta)>0,
\label{ls8}
\end{gather}
where the expansion coefficients are the same. To low orders, one
finds $\partial_1U_{0}=0$ and
\begin{equation*}
\partial_1U_{n}
 =\sum_{j=0}^{n-1}U_{j}L_{1,n-j},
\qquad n\geq1.
\end{equation*}
Taking into account the boundary conditions, this gives
$U_{0}=\mathbbm{1}$ and
\begin{align}
U_{n}\bigl(x^0,x^1\bigr)
&= \sum_{j=0}^{n-1}\int_{-\infty}^{x^1} {\rm d}x
\bigl(U_{j}L_{1,n-j}\bigr)\bigl(x^0,x\bigr),
\qquad \text{i.e.,} \qquad
U_{1}\bigl(x^0,x^1\bigr) = \int_{-\infty}^{x^1} {\rm d}x
L_{1,1}\bigl(x^0,x\bigr),
\nonumber\\
U_{2}\bigl(x^0,x^1\bigr)
&= \int_{-\infty}^{x^1} {\rm d}x
L_{1,2}\bigl(x^0,x\bigr)
+ \int_{-\infty}^{x^1} {\rm d}x
\int_{-\infty}^{x} {\rm d}y
L_{1,1}\bigl(x^0,y\bigr)L_{1,1}\bigl(x^0,x\bigr),
\label{ls10}
\end{align}
etc. Expanding the gauge variation formula~\eqref{Ugauge} for
each sign in powers of $1/\theta$, one finds
\begin{equation*}
\delta_{\epsilon}^{H}U_{n}
 =\sum_{m=0}^{n-1}U_{m}
 \left(
 \frac{\epsilon}{n}L_{0,n-m}
 +\epsilon^1L_{1,n-m}
 \right),
 \qquad n\in\mathbb{N}.
\end{equation*}
In order to take the $x^1\rightarrow+\infty$ limit, we proceed as in the
derivation of $\delta_{\epsilon}^{H}\mathcal{O}_T^{\pm}=0$ and
\eqref{Ugauge}. The descriptors $\bigl(\epsilon,\epsilon^1\bigr)$ are stripped off
to obtain separate equations for $\{\mathcal{H}_0,U_{n}\}$ and $\{\mathcal{H}_1,U_{n}\}$.
The limits of their right-hand sides vanish since $L_{0,m}$ and $L_{1,m}$
vanish for $x^1\rightarrow\infty$. Hence, in the expansions
$ \mathcal{O}_T^{\pm}(\theta) =\mathbbm{1}+\sum_{n\geq1}\mathcal{O}_{T,n}\theta^{-n}$,
$\pm \operatorname{Re}(\theta)>0 $, all the $\mathcal{O}_{T,n}$ are Dirac observables, as claimed.

For clarity's sake, we add that the expansions assume a~different form
if the \textit{same} large $\theta$ is used in both expansions. This is because
\eqref{LHpm1} relates $L_1^{H,+}$ and $L_1^{H,-}$. For example,
\begin{equation*}
L_{1}^{H,+}(\theta) = \sum_{n\geq1}L_{1,n} \theta^{-n},
\qquad
 L_{1}^{H,-}(\theta)
 = - \frac{\lambda_{\text{\sc n}} J_1^H}{n\rho} -\sum_{n\geq1}L_{1,n} \theta^{-n},
 \qquad \operatorname{Re}(\theta)>0.
\end{equation*}
Correspondingly, one can use the $U_+$ expansion in~\eqref{ls8} and~\eqref{Upm1} to infer the expansion of~$U_-$ for large $\theta$ with $\operatorname{Re}(\theta)>0$. To low orders, one finds
\begin{equation*}
U_-(\theta) =
\bigl[ \mathbbm{1}-U_{1}^{\mathsf T}\theta^{-1}
+\bigl(\bigl(U_{1}^{\mathsf T}\bigr)^2-U_{2}^{\mathsf T}\bigr)\theta^{-2} +O\bigl(\theta^{-3}\bigr)
\bigr]M^{-1}, \qquad \operatorname{Re}(\theta)>0.
\end{equation*}
To proceed, we expand \smash{$\mathcal{O}^\pm(\theta) = \int_{-\infty}^{\infty} {\rm d}x \mathcal{J}_0^{H,\pm}\bigl(x^0,x;\theta\bigr)$}. In the two sign cases $\smash{\mathcal{J}_0^{H,\pm}(\theta)}=\smash{U_\pm(\theta) K_0^{H,\pm}(\theta) U_\pm^{-1}(\theta)}$ is expanded in the respective asymptotic regimes
$\pm \operatorname{Re} \theta>0$. Since \smash{$K_0^{H,\pm}(\theta) =
2 \partial_{\theta} L_1^{H,\pm}(\theta)$}, the earlier relations suffice
to expand the master current in~\eqref{Diracobs3} as
\begin{equation*}
\mathcal{J}_0^{H,\pm}(\theta) = \sum_{n \geq 0} \mathcal{J}_{0,n} \theta^{-n -2}, \qquad \pm \operatorname{Re} \theta>0,
\end{equation*}
where the expansion coefficients are the same in both cases. To low orders,
\begin{align*}
\mathcal{J}_{0,0} &= 2 \bigl(-1 L_{1,1}\bigr),\qquad
\mathcal{J}_{0,1} = 2 \bigl( - 2 L_{1,2}
+ [L_{1,1}, U_1]\bigr),\\
\mathcal{J}_{0,2} &= 2 \bigl( - 3 L_{1,3}
+ [L_{1,1}, U_2]
+ 2[ L_{1,2}, U_1] - [L_{1,1}, U_1]U_1 \bigr).
\end{align*}
Here the explicit forms of the $L_{1,n}$'s have to be inserted.
Taking into account~\eqref{LHpm0}, one finds for
$\pm \operatorname{Re}(\theta + \tilde{\rho}) >0$
\begin{align}
L_{1,1} &= - \frac{1}{2} J^H_0,
\qquad
L_{1,2} = \frac{1}{4} \left(
\rho \frac{J^H_1}{ \lambda_{\text{\sc n}} n} + 2 \tilde{\rho} J^H_0 \right),
\nonumber\\
L_{1,3} &= -\frac{1}{4} \left( 2 \rho \tilde{\rho} \frac{J^H_1}{\lambda_{\text{\sc n}} n}
+ \left(\frac{\rho^2}{\lambda_{\text{\sc n}}^2} + 2 \tilde{\rho}^2\right)J^H_0 \right).
\label{ls13}
\end{align}
On account of Result~\ref{resultGowdyNC}, this yields a~countable set of Dirac
observables
\begin{gather}
\delta_{\epsilon}^H \mathcal{O}_n = 0, \qquad
\mathcal{O}_n := \int_{-\infty}^{\infty} {\rm d}x^1 \mathcal{J}_{0,n}\bigl(x^0,x^1\bigr),\qquad
n \in \mathbb{N}.
\label{Diracobs6}
\end{gather}
These are again valid strongly, off-shell, and without gauge fixing.
Their on-shell specialization must by the result
\eqref{ls7} be of the form $\mathcal{J}_{0,n}^H = 2\partial_1 \Omega_n$,
$\mathcal{J}_{1,n}^H = 2 e_0(\Omega_n)$, where $\partial_{\theta} U U^{-1} =
\sum_{n \geq 0} \Omega_n \theta^{-n -2}$. Explicitly,
$
\Omega_0 = -U_1$, $\Omega_1 = U_1^2 - 2 U_2$, $
\Omega_2 = -3 U_3 + 2 U_2 U_1 +U_1 U_2 - U_1^3$.
In relation to the earlier discussion of the results in
\cite{KSGowdyNC,KSYangian} it is worth stressing that these
charges are finite and nonzero when backpropagated to the big bang,
$\rho \rightarrow 0^+$. This is because, based on the AVD scenario, the
current $J_0^H$ is regular at $\rho =0$, while $J_1^H$ is subleading,
cf.\ Section~\ref{sec4.4}.
The expansion coefficients~\eqref{ls13} reduce to $L_{1,n} =
(-)^n \tilde{\rho}^{n-1} J_0^H/2$, and the~$U_n$'s in
\eqref{ls10} become very simple. The same holds for their
$x^1 \rightarrow \infty$ limits and the on-shell values~${\mathcal{O}_n =
2 \lim_{x^1 \rightarrow \infty} \Omega_n\bigl(x^0,x^1\bigr)}$ in~\eqref{Diracobs6}.

\subsection[T\^{}3 Gowdy]{$\boldsymbol{T^3}$ Gowdy}\label{sec3.2}

Spatially periodic systems often exhibit properties that are
qualitatively different from those of their spatially noncompact
counterparts. This also
holds for integrable field theories in their Hamiltonian formulation
\cite{FT}. For the $\mathbb{R} \times T^2$ versus $T^3$ Gowdy systems
very specific differences occur. The $T^3$ Gowdy cosmologies are
arguably more natural and more interesting than the~${\mathbb{R} \times T^2}$
versions. More natural, because three-tori treat all spatial
directions on the same level and are a~standard option for
cosmological spacetimes. More interesting, because $T^3$ Gowdy
cosmologies have the remarkable property of being asymptotically
velocity dominated (AVD). This has been rigorously proven
\cite{IsenMonc,Ringstroemproof} and places the $T^3$ systems into
the broader context of the Belinski--Khalatnikov--Lifshitz scenario
\cite{BelHenbook,RingstroemCR}.

Concerning our present quest for Dirac observables for
$T^3$ Gowdy cosmologies, there are
four major obstructions to the direct application of standard
techniques from spatially periodic flat space field theories or from other
2-Killing vector reductions:
\begin{itemize}\itemsep=0pt
\item[(i)] Although the basic fields are spatially periodic, the
 Lax pair~\eqref{LHdef} is not. This is because the field
 $\tilde{\rho}$ is only quasi-periodic, $\tilde{\rho}\bigl(x^0,x^1 + 2\pi\bigr)
 = \tilde{\rho}\bigl(x^0,x^1\bigr) - \pi_0^{\sigma}$, for some nonzero
 constant~$\pi_0^{\sigma}$.
\item[(ii)] There is no known recursive algorithm for (spatially nonlocal)
 conserved charges in the $T^3$ Gowdy systems.
\item[(iii)] Related to (i) the relevant limits in the path ordered
 exponential $T^H$ \eqref{THsol1} of $L_1^H$ are $y^1 \rightarrow -\infty$,
 $x^1 \rightarrow \infty$. However $L_1^H\bigl(x^0,x;\theta\bigr)$ falls off only like
 $O(1/|x|)$, too soft for the unbounded ${\rm d}x$ integral to exist.
 A~one-sided limit $y^1 \rightarrow -\infty$ can be constructed by a~renormalization
 procedure, but the resulting $U$ does not have an $x^1 \rightarrow \infty$ limit.
 No counterparts of $\mathcal{O}_T$ from~\eqref{OTdef} exist.
\item[(iv)] The off-shell $U$ from (iii) is only gauge invariant
 on certain parts of phase space. This can be overcome by
 considering the periodic extension of a~suitable trace of $\mathcal{J}_0^H(\theta)$,
 which will lead to a~one-parameter family of strong, off-shell Dirac
 observables.
\end{itemize}

Several authors considered the precursor problem of constructing nonlocal
conserved charges for $T^3$ Gowdy cosmologies~\cite{AHusain,Husain,ManoS}.
These approaches turn out to be blocked by properties~(i),~(ii).
For spatially periodic integrable systems, one usually considers the
trace of the transition matrix~\cite{FT}. In the case at hand,
taking the trace of equation~\eqref{THsol5} gives for $x^1,y^1 \in [0,2\pi]$
\begin{equation*}
\delta_{\epsilon}^H \operatorname{tr} T^H\bigl(x^0;x^1,y^1;\theta\bigr) = \operatorname{tr}
\bigl\{ T^H\bigl(x^0;x^1,y^1;\theta\bigr) \bigl[C_\epsilon\bigl(x^1,\theta\bigr) - C_\epsilon\bigl(y^1,\theta\bigr)\bigr] \bigr\}.
\end{equation*}
If the Lax pair was spatially periodic, $L_{1}^H\bigl(x^0, x^1 + 2\pi;\theta\bigr) =
L_{1}^H\bigl(x^0,x^1;\theta\bigr)$, so would be the gauge variation matrices
$C_{\epsilon}\bigl(x^1,\theta\bigr)$ as the descriptors $\epsilon$, $\epsilon^1$ are naturally
taken to be periodic as well. This is the strategy adopted
(for the conserved charges) in~\cite{ManoS}, see
their equations~(3.3),~(2.17),~(2.7). In our understanding,
this fails because $L_0^H$, $L_1^H$ are not spatially periodic,
see (i), as detailed below.

For flat space integrable systems, often a~recursive algorithm exists
that produces a~denumerable set of conserved charges. For
flat space nonlinear sigma-models,~\cite{Brezin} provides such
an algorithm. After a~complete gauge fixing the nontrivial evolution
equations of the $T^3$ system become that of a~$\mathrm{SL}(2,\mathbb{R})/\mathrm{SO}(2)$
sigma-model with an explicit coordinate dependence. This motivated
\cite{AHusain,Husain} to try to `correct' the recursively produced
flat space expressions by ad-hoc terms that ensure conservation and
periodicity of the resulting currents. The method was not shown to
be recursive. More importantly, the first non-Noether charge
found thereby is also nonlocal in time. This invalidates the contemplated
generation of further conserved charges via equal time Poisson brackets.
Generally, giving up temporal locality risks trivializing the
notion of a~conserved charge: if for some functional $Q$ of the
phase space variables $\partial_0 Q = R$ holds on account of the evolution
equations, should one regard \smash{$Q - \int^{x^0} {\rm d} t R$} as `conserved'?
In the absence of other attempts, we conclude that, while a~recursive
algorithm might exist, it can not easily be guessed.

Our strategy in the following is to accept (i), (ii) as properties of the
integrable structure of the $T^3$ Gowdy cosmologies and to
follow the route (iii), (iv), centered around a~`renormalized'~$U$.
The details of the renormalization procedure are relegated
to Appendix \ref{appB}. By assumption, the basic fields $(n,s)$, $\bigl(\rho,\pi^{\rho}\bigr)$,
$\bigl(\Delta, \pi^{\Delta}\bigr)$, $\bigl(\psi, \pi^{\psi}\bigr)$, are spatially periodic
in the $T^3$ Gowdy system. For definiteness we take the period to
be $2\pi$, so that $A\bigl(x^0, x^1 + 2\pi\bigr) = A\bigl(x^0,x^1\bigr)$ for all
phase space variables, including lapse and shift. The Lax pair,
however, also contains the auxiliary field $\tilde{\rho}$, which
turns out to be quasi-periodic only.

\textbf{Quasi-periodicity of $\boldsymbol{\tilde{\rho}}$}. We return to
the discussion~\eqref{rhotvar}, \eqref{rhotHdef} of the $\tilde{\rho}$
field in the $\mathbb{R} \times T^2$ Gowdy system.
In the $T^3$ Gowdy system, the additional point to note is that the
general solution~\eqref{rhotHdef} is not spatially periodic even
if $\pi^{\sigma}$ is. From~\eqref{rhotHdef}, one has
\begin{gather}
\tilde{\rho}\bigl(x^0,x^1 + 2\pi\bigr) - \tilde{\rho}\bigl(x^0,x^1\bigr)
= - \int_{x^1}^{x^1 + 2\pi} {\rm d}x \pi^{\sigma}\bigl(x^0,x\bigr)
= - \int_0^{2\pi} {\rm d}x \pi^{\sigma}\bigl(x^0,x\bigr) =: -\pi^{\sigma}_0,
\label{rhotHqperiodic}
\end{gather}
using the spatial periodicity of $\pi^{\sigma}$ in the second step.
Note that the reference point $y^1$ in~\eqref{rhotHdef} drops out.
Importantly, $\pi_0^{\sigma}$ itself is gauge invariant,
\begin{gather}
\delta^H_{\epsilon} \pi^{\sigma}_0 =0.
\label{rhotHshift1}
\end{gather}
This follows irrespective of assumptions about the descriptors
$\bigl(\epsilon,\epsilon^1\bigr)$ from
the fact that $\{\sigma_0, \pi_0^{\sigma} \}\allowbreak =1$, where \smash{$\sigma_0
= (2\pi)^{-1} \int_0^{2\pi} {\rm d}x \sigma(x)$} is the zero mode
of $\sigma$. This zero mode simply does not occur in $\mathcal{H}_0$, $\mathcal{H}_1$,
which accounts for~\eqref{rhotHshift1}. The zero modes are nevertheless
an integral part of the dynamics of the system~\cite{TorreRom} and cannot
just be set to zero.

On the one hand, this justifies the assumption to use as base
descriptors the same as in the~${\mathbb{R} \times T^2}$ system to obtain the
previous expression~\eqref{rhotvar}
for the gauge variation $\delta_{\epsilon}^H \tilde{\rho}$. On the other hand,
the quasi-periodicity allows one to restrict $\tilde{\rho}\bigl(x^0, \cdot \bigr)$
to $[0,2\pi]$ as fundamental domain and extend it via
$\tilde{\rho}\bigl(x^0, x^1 + l 2\pi\bigr) = \tilde{\rho}\bigl(x^0,x^1\bigr) - l \pi_0^{\sigma}$,
$l \in \mathbb{Z}$. By~\eqref{rhotHshift1}, this does not change the gauge variation
of the restricted field, which will still be given by~\eqref{rhotvar}.
For the restricted field, it makes sense to also restrict the
descriptors $\bigl(\epsilon,\epsilon^1\bigr)$ to $x^1 \in [0,2\pi]$. In particular, one
can view~\eqref{rhotHshift1} alternatively as a~consequence of
$\delta_{\epsilon}^H \pi^{\sigma}$ in~\eqref{gtransH} with periodic
descriptors~\eqref{epscompact}. Consistency then
requires that the restricted descriptors are spatially periodic.
This is because~$\pi^{\sigma}$,~$\partial_1 \rho$ on the right-hand side of
$\delta_{\epsilon}^H \tilde{\rho}$ are spatially periodic and the
gauge variation is (by the previous step) the same if the
interval $[l 2\pi, (l+1) 2\pi]$, $l \neq 0$ is considered instead.
This justifies the gauge variation~\eqref{rhotvar} also for the
$T^3$ Gowdy system, where for $\tilde{\rho}$ restricted to $[0,2\pi]$
the same descriptors~\eqref{epscompact} as for all other fields
should be used.

\textbf{The regularized $\boldsymbol{ T^{H,\pm}}$.} The Lax pair~\eqref{LHdef}
contains $\tilde{\rho}$ in addition to fields that are spatially periodic.
The quasi-periodicity~\eqref{rhotHqperiodic} can be used to transplant
the Lax pair's $x^1$ argument into the fundamental
domain $[0,2\pi]$, using
$L_{\mu}^{H,\pm}\bigl(x^0,x^1 + 2\pi;\theta\bigr) =
L_{\mu}^{H,\pm}\bigl(x^0,x^1;\theta- \pi^{\sigma}_0\bigr)$, $\mu=0,1$.
For the $T^{H,\pm}$ in~\eqref{THsol1}, this implies
$T^H\bigl(x^0;x^1 + 2\pi, y^1 + 2\pi;\theta\bigr) =
T^{H,\pm}\bigl(x^0;x^1, y^1;\theta - \pi_0^{\sigma}\bigr)$. However, the construction
of a~well-defined $U_{\pm}$ from $T^{H,\pm}$ obeying the
gauge variation of~\eqref{Ugauge} is now considerably more subtle, as
$L_1^{H,\pm}$ will not decay sufficiently fast to ensure the existence of
an unregularized limit as in~\eqref{UdefNC}. In Appendix \ref{appB}, we construct
appropriate `renormalized' $U_{\pm}$ as~${\text{\sc l}\rightarrow \infty}$ limits of a~family of regularized $T_{\text{\sc l}}^{\pm}$, $\text{\sc l}\in \mathbb{N}$ defined by
\begin{gather}
T_{\text{\sc l}}^{\pm}\bigl(x^0,x^1;\theta\bigr) = \exp_+\left\{
\frac{Q}{2} \sum_{l=1}^\text{\sc l}\frac{1}{\theta + l \pi_0^{\sigma}}
+ \int_{- 2\pi \text{\sc l}}^{x^1} {\rm d}x
L_1^{H,\pm}\bigl(x^0,x;\theta\bigr) \right\}, \qquad \text{\sc l} \in \mathbb{N},
\label{Uregdef}
\end{gather}
where $Q = \int_0^{2\pi} {\rm d}x^1 J_0^H$ is the $\mathfrak{sl}(2,\mathbb{R})$ Noether
charge, and $\exp_+$ denotes the path ordered exponential with
the factors containing larger integration variables to the right.
Importantly, the~${\text{\sc l} \rightarrow \infty}$ limit of~\eqref{Uregdef} defining $U_{\pm}$
only exists subject to $\pm \operatorname{Re}(\theta + \tilde{\rho}) >0$.
Further, the so-obtained $U_{\pm}$ are gauge covariant only
under additional conditions. The specifics are recorded in
Result~\ref{resultB1}.

With this in place, we consider again off-shell versions of the
construction principles from Section~\ref{section2.3}. As noted in Result~\ref{result0GowdyNC}, for the $\mathbb{R} \times T^2$ Gowdy system
Dirac observables could simply be obtained by a~double limit
$x^1 \rightarrow + \infty$, $y^1 \rightarrow - \infty$
of the transition matrix $T^H\bigl(x^0;x^1,y^1;\theta\bigr)$. For the $T^3$ Gowdy
system, one is led to consider a~double limit $x^1 = 2\pi L'$, $y^1 =
- 2\pi L$, with~${L, L' \in \mathbb{N}}$ sent to infinity. This limit will
in general \textit{not} exist and neither does the $L' \rightarrow \infty$
limit of the `renormalized' $U\bigl(x^0, 2\pi L';\theta\bigr)$ from Result
\ref{resultB1}. The latter can be seen by iterating~\eqref{quasiperB1}.
One might try a~correlated limit
$L= L' \rightarrow \infty$, in which case the combined conditions needed
for the existence of a~limit again appear to overconstrain it.
As outlined in Section~\ref{section2.3}, the second of the standard construction
principles for spatially periodic systems is the monodromy matrix.
For the $T^3$ systems, it is discussed in Appendix~\ref{appC} and we find that
the conditions for a~renormalized limit to exist clash with the
limit being nontrivial. This leaves only the construction principle
via the one parameter currents $\mathcal{J}_{\mu}^{H, \pm}(\theta)$.

\textbf{Off-shell currents $\boldsymbol{\mathcal{J}_{\mu}^{H,\pm}(\theta)}$.}
In terms of $T_{\text{\sc l}}^{\pm}$, we define
\begin{gather}
\mathcal{J}_{\mu}^{H,\pm}(\theta)_{\text{\sc l}} := T_{\text{\sc l}}^{\pm}(\theta) K_{\mu}^{H,\pm}(\theta)
T_{\text{\sc l}}^{\pm}(\theta)^{-1}, \qquad \mu =0,1, \qquad \operatorname{Im}(\theta) \neq 0.
\label{Jregdef}
\end{gather}
The instrumental result is the following.

\begin{Result}\mbox{} \label{resultJtheta}
\begin{itemize}
\itemsep=0pt
\item[$(a)$] The limits \smash{$\mathcal{J}_{\mu}^{H,\pm}\bigl(x^0,x^1;\theta\bigr):=
 \lim_{\text{\sc l} \rightarrow \infty} \mathcal{J}_{\mu}^{H,\pm}\bigl(x^0,x^1;\theta\bigr)_{\text{\sc l}}$}, $\mu =0,1$,
 exist pointwise for $x^1 \in [0,2\pi]$ and $x^0$ satisfying
 $\pm \operatorname{Re}(\theta + \tilde{\rho}) >0$. They can be extended
 to other spatial intervals via the quasi-periodicity properties
 \begin{gather}
\mathcal{J}_{\mu}^{H,\pm}\bigl(x^0,x^1 + 2\pi;\theta\bigr) = {\rm e}^{ - \frac{Q}{2\theta}}
\mathcal{J}_{\mu}^{H,\pm}\bigl(x^0,x^1;\theta- \pi^{\sigma}_0\bigr) {\rm e}^{\frac{Q}{2\theta}}.
\label{Jquasiper}
\end{gather}
\item[$(b)$] If in addition $\operatorname{tr} Q^2 < 2 \bigl(\pi_0^{\sigma}\bigr)^2$,
$\pm \pi_0^{\sigma} >0$ is assumed also the following holds:
\begin{equation*}
\delta_{\epsilon}^H \mathcal{J}_0^{H,\pm}(\theta) = \partial_1\left( \frac{\epsilon}{n}
\mathcal{J}_1^{H,\pm}(\theta)+ \epsilon^1 \mathcal{J}_0^{H,\pm}(\theta) \right).
\end{equation*}
\item[$(c)$] There exist separately gauge invariant $2 \times 2$ matrices
 $\nu_{\pm}$ such that for $\pm \operatorname{Re}(\theta + \tilde{\rho}) >0$,
 $\pm \pi_0^{\sigma} >0$, the limits
 $\operatorname{tr}\bigl\{\nu_{\pm} \mathcal{J}_{\mu}^{H,\pm}(\theta)\bigr\} := \lim_{\text{\sc l} \rightarrow \infty}
 \operatorname{tr}\bigl\{ \nu \mathcal{J}_{\mu}^{H,\pm}(\theta)_{\text{\sc l}}\bigr\}$, $\mu=0,1$,
exist and{\samepage
 \begin{gather}
\delta_{\epsilon}^{H} \operatorname{tr}\bigl\{ \nu_{\pm} \mathcal{J}_0^{H,\pm}(\theta)\bigr\} =
\partial_1\left( \frac{\epsilon}{n}
 \operatorname{tr}\{\nu_{\pm} \mathcal{J}_1^{H,\pm}(\theta) \}+ \epsilon^1
 \operatorname{tr} \bigl\{ \nu_{\pm} \mathcal{J}_0^{H,\pm}(\theta)\bigr\} \right),
\label{traceJvar}
\end{gather}
holds without condition on $\operatorname{tr} Q^2$.}
\end{itemize}
\end{Result}

The derivation is relegated to Appendix \ref{appB}. As detailed in Section~\ref{sec4.4},
the condition $\operatorname{tr} Q^2 < 2 \bigl(\pi_0^{\sigma}\bigr)^2$ for off-shell configurations is
not especially natural. We therefore shall not make use of
part (b) of the result but use (c) for the further construction.
It can be thought of as expressing in a~basis independent way the
fact that of the three independent matrix components
of $\mathcal{J}_0^{H,\pm}(\theta)$ one exists that has the desired gauge variation
without subsidiary condition on $\operatorname{tr} Q^2$.

Nevertheless, the integrated version of~\eqref{traceJvar} does not
directly give rise to Dirac observables. Rather,
\begin{align}
\delta_{\epsilon}^H \int_0^{2\pi} {\rm d}x
 \operatorname{tr}\{\nu \mathcal{J}_0^{H,\pm}(x;\theta)\}
={}& \left(\frac{\epsilon}{n}\right)(0)\operatorname{tr}\bigl\{\nu
 \bigl[ \mathcal{J}_1^{H,\pm}(2\pi;\theta) - \mathcal{J}_1^{H,\pm}(0;\theta) \bigr]\bigr\}
\nonumber\\
& + \epsilon^1(0)
 \operatorname{tr} \bigl\{\nu\bigl[ \mathcal{J}_0^{H,\pm}(2\pi;\theta) - \mathcal{J}_0^{H,\pm}(0;\theta)
 \bigr]\bigr\},
\label{JHthetaavg1}
\end{align}
where we suppress the shared time argument $x^0$, which also enters the
equal time Poisson brackets. On the right-hand side we used the spatial
periodicity of $\epsilon$, $\epsilon^1$.

The idea in the following is simply to consider the sum of all
$2\pi$ translates in~\eqref{JHthetaavg1}. Once shown to be convergent the
right-hand side becomes a~telescopic sum with vanishing end term
contributions. To establish convergence, the instrumental identity
is
\begin{gather}
 \sum_{n=-N}^N \operatorname{tr}\bigl\{ \nu_{\pm} \mathcal{J}_{\mu}^{H,\pm}\bigl(x^1 + 2 \pi n;\theta\bigr) \bigr\}\nonumber\\
 \qquad
= \operatorname{tr}\bigl\{ \nu_{\pm} \mathcal{J}_{\mu}^{H,\pm}\bigl(x^1;\theta\bigr) \bigr\} + \sum_{n=1}^N \operatorname{tr}\bigl\{ \check{\nu}_{\pm}(n,\theta)
\mathcal{J}_{\mu}^{H,\pm}\bigl(x^1;\theta - n \pi_0^{\sigma}\bigr) \bigr\}\nonumber\\
 \qquad\phantom{=}{}
+ \sum_{n=1}^N \operatorname{tr}\bigl\{ \nu_{\pm}(n,\theta)
\mathcal{J}_{\mu}^{H,\pm}\bigl(x^1;\theta + n \pi_0^{\sigma}\bigr) \bigr\},
\nonumber\\
\check{\nu}_{\pm}(n,\theta) = {\rm e}^{\pm \check{\varkappa}_n(\theta) (\operatorname{tr} Q^2/2)^{1/2}}
\nu_{\pm},\qquad
\nu_{\pm}(n,\theta) = {\rm e}^{\mp \varkappa_n(\theta) (\operatorname{tr} Q^2/2)^{1/2}} \nu_{\pm},
\label{Jthetaavg2}
\end{gather}
as shown in~\eqref{Jthetaper2}, \eqref{nun}. Assuming
$\pm \pi_0^{\sigma} >0$,
the terms in the sums turn out to fall off pointwise at least
like $O\bigl(n^{-(2 + \delta)}\bigr)$, $\delta >0$. The $N \rightarrow \infty$ limit
in~\eqref{Jthetaavg2} therefore
exists pointwise in $\bigl(x^0,x^1,\theta\bigr)$ and is by construction spatially
periodic. It remains to verify that the integral of this limit has
vanishing gauge variations. For the finite periodic extensions, one finds
\begin{gather}
 \delta_{\epsilon}^H \sum_{n=-N}^N \int_0^{2\pi} {\rm d}x
 \operatorname{tr} \bigl\{ \nu_{\pm} \mathcal{J}_0^{H,\pm}(x + 2\pi n;\theta) \bigr\}
\nonumber\\
 \qquad = \left(\frac{\epsilon}{n}\right)(0) \operatorname{tr} \bigl\{
\check{\nu}_{\pm}(N + 1,\theta)\mathcal{J}_1^{H,\pm}\bigl(0, \theta - (N + 1) \pi_0^{\sigma}\bigr)
-\nu_{\pm}(N,\theta)\mathcal{J}_1^{H,\pm}\bigl(0, \theta + N \pi_0^{\sigma}\bigr) \bigr\}
\nonumber\\
 \phantom{=}{}\qquad +\epsilon^1(0) \operatorname{tr} \bigl\{
\check{\nu}_{\pm}(N + 1,\theta)\mathcal{J}_0^{H,\pm}\bigl(0, \theta - (N + 1) \pi_0^{\sigma}\bigr)\nonumber\\
 \phantom{=}{}\qquad
-\nu_{\pm}(N,\theta)\mathcal{J}_0^{H,\pm}\bigl(0, \theta + N \pi_0^{\sigma}\bigr) \bigr\}.
\label{Jthetaavg3}
\end{gather}
The right-hand side falls off faster than $O\bigl(1/N^2\bigr)$ yielding our
main result for $T^3$.

\begin{Result} \label{resultT3obs}
The periodic extension \smash{$\sum_{n \in \mathbb{Z}}
\operatorname{tr}\bigl\{\nu_{\pm} \mathcal{J}_0^{H,\pm}(x + 2\pi n;\theta)\bigr\}$}
of \smash{$\operatorname{tr}\bigl\{\nu_{\pm} \mathcal{J}_0^{H,\pm}(x;\theta)\bigr\}$}
is for ${\pm} \operatorname{Re}(\theta + \tilde{\rho})>0$,
$\pm \pi_0^{\sigma} >0$ pointwise well-defined
and its integral defines a~one-parameter family of strong,
off-shell Dirac observables for the $T^3$ Gowdy system
\begin{gather}
\delta^H_{\epsilon} \mathcal{O}(\theta) =0, \qquad
\mathcal{O}(\theta) = \int_0^{2\pi} {\rm d}x^1 \sum_{n \in \mathbb{Z}}
 \operatorname{tr}\bigl\{\nu_{\pm} \mathcal{J}_0^{H,\pm}(x + 2\pi n;\theta)\bigr\}.
\label{DiracobsT3}
\end{gather}
Here, $\nu_{\pm}$ is the separately gauge invariant $2 \times 2$ matrix
constructed in Appendix~{\rm \ref{appB}}.
\end{Result}

We add several remarks.
\begin{enumerate}\itemsep=0pt
\item[(i)] To the best of our knowledge, this is the first successful construction
of Dirac observables for $T^3$ Gowdy cosmologies beyond the polarized sector.
\item[(ii)] Since $\mathcal{J}_0^H(\theta) \mapsto C \mathcal{J}_0^H(\theta) C^{-1}$ under global
$\mathrm{SL}(2,\mathbb{R})$ transformations, the
$\nu$ matrix transforms according to $\nu \mapsto C^{-1} \nu C$.
For $\nu$ in the range of $\mathrm{Nil}_{\pm}$ in~\eqref{nil1}, one needs to
take into account that the projectors $P_{\pm}$ are built from
the Noether charge $Q$, which itself transforms according to
$Q \mapsto C Q C^{-1}$. By~\eqref{sl2g}, this gives $P_{\pm} \mapsto
C P_{\pm} C^{-1}$, so that for $\nu_a = P_\pm a P_{\mp}$, $a \in \mathfrak{sl}(2,\mathbb{R})$,
in the range of $\mathrm{Nil}_{\pm}$ the net transformation law is $\nu_a
\mapsto \nu_{C^{-1} a C}$. The image remains in the range of $\mathrm{Nil}_{\pm}$,
and since the latter is one-dimensional this merely changes the description
of the Dirac observables~\eqref{DiracobsT3}.
\item[(iii) ]In the polarized case, a~sign-modified periodic extension
of~\eqref{ahjdef} can be constructed explicitly and leads to a~one-parameter family
of off-shell Dirac observables strongly commuting with the constraints
\cite{GowdyPol}. A~modulation of the plain periodic extension by
the generalized sign function from~\eqref{rootsign}
is useful there.
\end{enumerate}

\section{The velocity dominated Gowdy system}\label{sec4}

As noted in the introduction, $T^3$ Gowdy cosmologies have the
remarkable property of being asymptotically velocity dominated. This gives a~\textit{dynamical} rationale to consider the velocity dominated field
equations, as the ones governing the dynamics close to the big bang.
This viewpoint is, however, not suited for the investigation of
off-shell properties underlying our Dirac observables. We therefore
use the kinematical scaling from Appendix~\ref{appA} to isolate an action principle
for the symmetry reduced gravity theory governing the velocity dominated
(VD) limit. The result is $S^L_{\text{\sc 2kvd}} = S^L_{\text{\sc vd2k}}$
in Figure~\ref{2Kdiagram} of the introduction. The Hamiltonian version of this action
and its gauge symmetries are taken as the point of departure here.
The Dirac observables of this symmetry reduced VD gravity theory are
constructed in
Section~\ref{sec4.2}. These can be matched to the leading terms of
an anti-Newtonian expansion of those in Section~\ref{sec3}, thereby
providing an off-shell generalization of the AVD property.

\subsection{Hamiltonian action and its gauge symmetries}\label{sec4.1}

Fully explicitly, the Hamiltonian
action of the VD system takes the form
\begin{gather}
S^\text{\sc h}_{\text{\sc 2kvd}} = \int {\rm d}^2 x \bigl\{ \pi^{\rho} \partial_0 \rho
+ \pi^{\sigma} \partial_0 \sigma + \pi^{\psi} \partial_0 \psi
+ \pi^{\Delta} \partial_0 \Delta - n\mathcal{H}_0^{\text{\sc vd}} - s \mathcal{H}_1\bigr\},
\nonumber\\
\mathcal{H}_0^{\text{\sc vd}} = - \lambda_{\text{\sc n}} \pi^{\sigma} \pi^{\rho}
+ \frac{\lambda_{\text{\sc n}}}{2\rho} \Delta^2 \bigl(\bigl(\pi^{\psi}\bigr)^2+\bigl(\pi^{\Delta}\bigr)^2 \bigr),
\nonumber\\
\mathcal{H}_{1} = \pi^{\rho} \partial_{1} \rho
+ \pi^{\sigma} \partial_{1} \sigma
- 2 \partial_{1} \pi^{\sigma} + \pi^{\psi} \partial_{1} \psi
+ \pi^{\Delta} \partial_{1} \Delta.
\label{VDHaction}
\end{gather}
Strictly speaking, different symbols should be used for the phase space
variables of the VD system. In order not to clutter the notation, we shall
not do so systematically; however for the basic currents we write
$J_0^{\text{\sc vd}}$, $J_1^{\text{\sc vd}}$ for contradistinction.
The VD counterpart of Result~\ref{resultSHgaugeinv} is the following.

\begin{Result}
Let $S^\text{\sc h}_{\text{\sc 2kvd}}$ be the action~\eqref{VDHaction} with
the temporal integration restricted to $x^0 = t \in [t_i,t_f]$
and fall-off or boundary conditions in $x^1$ that ensure
the absence of spatial boundary terms. Then
\begin{gather}
\delta^{\text{\sc hvd}}_{\epsilon} S^\text{\sc h}_{\text{\sc 2kvd}} = 0,
 \qquad \epsilon|_{t_i} = 0 = \epsilon|_{t_f}.
\label{VDHginv}
\end{gather}
Here, all but the gauge variations of lapse and shift are canonically
generated via $\delta_{\epsilon}^{\text{\sc hvd}} F =
\bigl\{ F, \mathcal{H}_0^{\text{\sc vd}}(\epsilon) + \mathcal{H}_1\bigl(\epsilon^1\bigr)\bigr\}$.
The gauge variations of lapse and shift are designed such that
\eqref{VDHginv} holds and coincide with those of the Lagrangian formalism.
\end{Result}

Since these gauge variations govern the VD Dirac observables and
for contradistinction with the original ones, we spell them
out in full. For lapse and shift, one has
\begin{gather}
\delta^{\text{\sc hvd}}_{\epsilon} n = \partial_{0} \epsilon
- (s \partial_{1} \epsilon - \epsilon \partial_{1} s)
+ \epsilon^{1} \partial_{1} n - n\partial_{1} \epsilon^{1},
\nonumber\\
\delta^{\text{\sc hvd}}_{\epsilon} s = \partial_{0} \epsilon^{1}
+\bigl(\epsilon^{1} \partial_{1} s - s\partial_{1} \epsilon^{1}\bigr).
\label{VDHvar0}
\end{gather}
The gauge variations for the configuration space variables
are the same as in equation~\eqref{gtransH}
\begin{alignat}{3}
&\delta_{\epsilon}^{\text{\sc hvd}} \rho = -\lambda_{\text{\sc n}} \epsilon \pi^{\sigma}
+ \epsilon^1 \partial_1 \rho,
\qquad&&
\delta_{\epsilon}^{\text{\sc hvd}} \sigma = -\lambda_{\text{\sc n}} \epsilon \pi^\rho + \epsilon^1 \partial_1 \sigma + 2 \partial_1 \epsilon^1,&
\nonumber\\
&\delta_{\epsilon}^{\text{\sc hvd}} \Delta = \lambda_{\text{\sc n}} \frac{\epsilon}{\rho} \Delta^2 \pi^\Delta + \epsilon^1 \partial_1 \Delta,
\qquad&&
\delta_{\epsilon}^{\text{\sc hvd}} \psi = \lambda_{\text{\sc n}} \frac{\epsilon}{\rho} \Delta^2 \pi^\psi + \epsilon^1 \partial_1 \psi.&
\label{VDHvar1}
\end{alignat}
The canonical momenta however transform differently
\begin{alignat}{3}
&\delta_{\epsilon}^{\text{\sc hvd}} \pi^{\rho} = \lambda_{\text{\sc n}} \frac{\epsilon \Delta^2}{2 \rho^2}\bigl( \bigl(\pi^{\psi}\bigr)^2 + \bigl(\pi^{\Delta}\bigr)^2 \bigr) + \partial_1\bigl(\epsilon^1 \pi^{\rho}\bigr),
\qquad&&
\delta_{\epsilon}^{\text{\sc hvd}} \pi^{\sigma} = \partial_1 \bigl(\epsilon^1 \pi^{\sigma}\bigr),&
\nonumber\\
&\delta_{\epsilon}^{\text{\sc hvd}} \pi^{\Delta} = - \lambda_{\text{\sc n}} \frac{\epsilon}{\rho} \Delta \bigl( \bigl(\pi^{\psi}\bigr)^2 + \bigl(\pi^{\Delta}\bigr)^2 \bigr) + \partial_1\bigl(\epsilon^1 \pi^{\Delta}\bigr),
\qquad&&
\delta_{\epsilon}^{\text{\sc hvd}} \pi^{\psi} = \partial_1 \bigl(\epsilon^1 \pi^\psi\bigr).&
\label{VDHvar2}
\end{alignat}
Note that some of the right hand sides are independent of the temporal
descriptor $\epsilon$. This can occur because the VD Hamiltonian constraint
no longer contains the conjugate spatial derivative terms. In particular,
this happens for the additional field $\tilde{\rho}$ entering the
earlier constructions. It is still defined by~\eqref{rhotHdef}
but its gauge variation is now
\begin{gather}
\delta_{\epsilon}^{\text{\sc hvd}} \tilde{\rho}
= \epsilon^1 \partial_1 \tilde{\rho}.
\label{VDHvar3}
\end{gather}

In addition to these local gauge symmetries, there is still
the global $\mathrm{SL}(2,\mathbb{R})$ symmetry that acts on the matrix $M(x)$
in~\eqref{Mmat1} via $M(x) \mapsto C M(x) C^{\mathsf T}$, $C \in \mathrm{SL}(2,\mathbb{R})$.
The associated Noether charge is $Q^{\text{\sc vd}} = \int
{\rm d} x^1 J_0^{\text{\sc vd}}\bigl(x^0,x^1\bigr)$, with
\begin{gather}
J_0^{\text{\sc vd}} :=
\begin{pmatrix} \Delta \pi^{\Delta} + \psi \pi^{\psi} &
 \bigl(\Delta^2 - \psi^2\bigr) \pi^{\psi} - 2 \psi \Delta \pi^{\Delta}\\
\pi^{\psi} &- \Delta \pi^{\Delta} - \psi \pi^{\psi}
\end{pmatrix}.
\label{sl2charge2}
\end{gather}
In the VD system, this will be the preeminent building block
for Dirac observables.

The gauge variations $\delta_{\epsilon}^{\text{\sc hvd}}$ in
\eqref{VDHvar0}, \eqref{VDHvar1}, \eqref{VDHvar2} can be
verified to leave the action $S^{\text{\sc hvd}}$ invariant
by direct computation. Alternatively, they can be obtained by subjecting the original gauge variations
\eqref{gtransH} to the scale transformation~\eqref{scale1}
and keeping only the leading terms. We omit the details.

\subsection{Dirac observables for VD Gowdy}\label{sec4.2}

In the following, we treat the VD system with action~\eqref{VDHaction}
as a~gravity theory in its own right and seek to identify Dirac
observables that strongly commute with the constraints
\begin{equation*}
\bigl\{ \mathcal{H}_{\text{\sc 0vd}}, \mathcal{O}^\text{\sc vd}\bigr\} =0=
\bigl\{ \mathcal{H}_1, \mathcal{O}^\text{\sc vd}\bigr\},
\end{equation*}
without using equations of motion or gauge fixing. Here, the
basic Poisson brackets from~\eqref{PB} remain in place for the
velocity dominated system. The first candidate is the integrand
\eqref{sl2charge2} of the Noether charge, which obeys
$\delta_{\epsilon}^{\text{\sc hvd}} J_0^{\text{\sc vd}} =
\bigl\{J_0^{\text{\sc vd}}, \mathcal{H}_0^{\text{\sc vd}}(\epsilon) + \mathcal{H}_1\bigl(\epsilon^1\bigr)
\bigr\} = \partial_1 \bigl(\epsilon^1 J_0^{\text{\sc vd}}\bigr)$.
Note the difference to~\eqref{HJvar}: there is no spatial component
of the current, the temporal gauge variation vanishes identically.
By~\eqref{VDHvar3}. the $\tilde{\rho}$ field separately has the same
property, and so will have any differentiable function of $\tilde{\rho}$.
In view of the earlier constructions inverse powers of $\theta + \tilde{\rho}$
are of special interest and we note
\begin{gather}
\delta_{\epsilon}^{\text{\sc hvd}} \left( \frac{J_0^{\text{\sc vd}}}%
 {(\theta + \tilde{\rho})^m} \right)
 = \partial_1 \left( \epsilon^1 \frac{J_0^{\text{\sc vd}}}%
 {(\theta + \tilde{\rho})^m} \right), \qquad m \in \mathbb{N}.
\label{sl2charge4}
\end{gather}
In fact, such combinations occur in the small $\rho$ limit
of $L_1^H$ and $K_0^H$ with suitable sign choices. Indeed, from~\eqref{LHpm2} one has
\begin{gather}
L_1^{H,\pm}(\theta) = - \frac{J_0^H}{2 (\theta + \tilde{\rho})} +
O(\rho),\qquad
K_0^{H,\pm}(\theta) = \frac{J_0^H}{(\theta + \tilde{\rho})^2} +
O(\rho), \nonumber\\
 \pm\operatorname{Re}(\theta + \tilde{\rho})>0.
\label{LKlimit}
\end{gather}
Reading $\rho$ as time in the adapted foliation, the $K_0^{H,\pm}(\theta)$
limit should enter the behavior of \smash{$\mathcal{J}_0^{H,\pm}(\theta) = U_\pm(\theta) K_0^{H,\pm}(\theta)
U_\pm(\theta)^{-1}$} near the big bang. In other words, the `dressing' with a~solution of some linear system as in Result~\ref{resultGowdyNC} is not
necessary to produce a~pre-Dirac observable in the VD system.

It is however indispensable if one seeks to make contact to the Dirac
observables in the full Gowdy system. Guided by the above
$L_1^H(\theta)$ limit, we define for $\operatorname{Im}\theta \neq 0$
\begin{align}
T^{\text{\sc vd}}\bigl(x^0;x^1,y^1; \theta\bigr) ={}&
\exp_+\left\{ - \frac{1}{2} \int_{y^1}^{x^1} {\rm d}x
\left(\frac{J_0^{\text{\sc vd}}}{\theta + \tilde{\rho}} \right)\bigl(x^0,x\bigr)
\right\}
\nonumber\\
={}&
\mathbbm{1} + \sum_{n \geq 1}
\frac{(-)^n}{2^n}
\int_{y^1}^{x^1} {\rm d}z_1
\int_{y^1}^{z^1} {\rm d}z_2 \cdots
\int_{y^1}^{z_{n-1}} {\rm d}z_n
\nonumber\\
& \times
\left(\frac{J_0^{\text{\sc vd}}}{\theta + \tilde{\rho}}\right)\bigl(x^0, z_n\bigr)
\left(\frac{J_0^{\text{\sc vd}}}{\theta + \tilde{\rho}}\right)\bigl(x^0, z_{n-1}\bigr)
\cdots \left(\frac{J_0^{\text{\sc vd}}}{\theta + \tilde{\rho}}\right)\bigl(x^0, z_1\bigr)
,
\label{THVDsol1}
\end{align}
in parallel to~\eqref{THsol1}. Again, the series is convergent under
broad conditions, for example, when $\mathbb{R} \ni x \mapsto
\bigl(J_0^{\text{\sc vd}}/(\theta + \tilde{\rho})\bigr)\bigl(x^0,x\bigr)$ is
bounded in some $2\times 2$ matrix norm. For the $y^1 \rightarrow -\infty$
limit to exist stronger fall-off conditions are needed, as detailed below.

Next, we consider the gauge transformations. Proceeding as in the derivation
of~\eqref{THsol5} and using the $m=1$ instance of~\eqref{sl2charge4},
one finds
\begin{align*}
\delta_{\epsilon}^{\text{\sc hvd}} T^{\text{\sc vd}}\bigl(x^0;x^1,y^1; \theta\bigr)
 ={}& T^{\text{\sc vd}}\bigl(x^0;x^1,y^1; \theta\bigr)
 \left( - \frac{\epsilon^1}{2}
 \frac{J_0^{\text{\sc vd}}}{\theta + \tilde{\rho}} \right)\bigl(x^0,x^1\bigr)\\
&-
\left( - \frac{\epsilon^1}{2}
\frac{J_0^{\text{\sc vd}}}{\theta + \tilde{\rho}} \right)\bigl(x^0,y^1\bigr)
T^{\text{\sc vd}}\bigl(x^0;x^1,y^1; \theta\bigr).
\end{align*}
The next steps depend on the boundary conditions.

\textbf{$\boldsymbol{\mathbb{R} \times T^2}$ topology.} We may assume that
$J_0^{\text{\sc vd}}\bigl(x^0,x^1\bigr)$ itself falls off as $O\bigl(\big|x^1\big|^{-1-\delta}\bigr)$,
$\delta >0$, for~${\big|x^1\big|\rightarrow \infty}$. In this case, the integrals of~\eqref{sl2charge4}
directly yield Dirac observables,
\begin{equation*}
\delta_{\epsilon}^{\text{\sc hvd}} \int_{-\infty}^{\infty} {\rm d}x
\left( \frac{J_0^{\text{\sc vd}}}{(\theta + \tilde{\rho})^m} \right)\bigl(x^0,x\bigr)
=0, \qquad m \in \mathbb{N}.
\end{equation*}
Concerning the `dressing' with some $U^{\text{\sc vd}}$, we
note that under the above assumption one has for some
$2 \times 2$ matrix norm and $\delta >0$
\begin{equation*}
\sup_{z \in \mathbb{R}} \Big\{ (1 + |z|)^{1 + \delta}
\Big\Vert \left(\frac{J_0^{\text{\sc vd}}}{\theta + \tilde{\rho}} \right)\bigl(x^0,z\bigr)
\Big\Vert \Big\} = \varkappa^{\text{\sc vd}}\bigl(x^0,\theta\bigr) < \infty.
\end{equation*}
Proceeding as in~\eqref{UboundNC2}, one sees that the series
\eqref{THVDsol1} has a~convergent $y^1 \rightarrow - \infty$ limit,
which can be bounded in the same $2 \times 2$ norm by
$\exp\bigl\{ 2 \varkappa^{\text{\sc vd}}\bigl(x^0,\theta\bigr)/\delta\bigr\}$.
Hence,
\begin{equation*}
U^{\text{\sc vd}}\bigl(x^0,x^1;\theta\bigr) =
\exp_+\left\{ - \frac{1}{2} \int_{-\infty}^{x^1} {\rm d}x
\left(\frac{J_0^{\text{\sc vd}}}{\theta + \tilde{\rho}} \right)\bigl(x^0,x\bigr)
\right\},
\end{equation*}
is well-defined and enjoys the instrumental relations
\begin{gather}
\partial_1 U^{\text{\sc vd}} = - \frac{1}{2(\theta + \tilde{\rho})} U^{\text{\sc vd}}
J_0^{\text{\sc vd}}, \qquad
\delta_{\epsilon}^{\text{\sc hvd}} U^{\text{\sc vd}} =
- \frac{\epsilon^1}{2(\theta + \tilde{\rho})} U^{\text{\sc vd}}
J_0^{\text{\sc vd}}.
\label{UVDdef2}
\end{gather}
This allows one to `dress' any of the quantities~\eqref{sl2charge4}
to obtain new ones that are gauge invariant modulo a~total spatial derivative.
In view of~\eqref{LKlimit}, we focus on the $m=2$ case and
set
\begin{gather}
\mathcal{J}_0^{\text{\sc vd}}(\theta) := U^{\text{\sc vd}}(\theta)
\frac{J_0^{\text{\sc vd}}}{(\theta + \tilde{\rho})^2}
 U^{\text{\sc vd}}(\theta)^{-1}.
\label{VDobs1}
\end{gather} A~short computation in parallel to~\eqref{JHthetaresult} then gives
\begin{gather}
\delta_{\epsilon}^{\text{\sc hvd}} \mathcal{J}_0^{\text{\sc vd}}(\theta) =
\bigl\{ \mathcal{J}_0^{\text{\sc vd}}(\theta), \mathcal{H}^{\text{\sc vd}}_0(\epsilon)
+ \mathcal{H}_1\bigl(\epsilon^1\bigr) \bigr\} = \partial_1
\bigl( \epsilon^1 \mathcal{J}_0^{\text{\sc vd}}(\theta)\bigr).
\label{VDobs2}
\end{gather}
This yields the following result.

\begin{Result} \label{resultVDNC}
 Assume $J_0^{\text{\sc vd}}\bigl(x^0,x^1\bigr)$ to fall off as
 $O\bigl(\big|x^1\big|^{-1 - \delta}\bigr)$, $\delta >0$. Then the quantities
\begin{equation*}
\mathcal{O}^{\text{\sc vd}}_m(\theta) := \int_{-\infty}^{\infty} {\rm d}x
\left( \frac{J_0^{\text{\sc vd}}}{(\theta + \tilde{\rho})^m} \right)\bigl(x^0,x\bigr),\quad
 m \in \mathbb{N},\qquad
\mathcal{O}^{\text{\sc vd}}(\theta) = \int_{-\infty}^{\infty} {\rm d}x
\mathcal{J}_0^{\text{\sc vd}}\bigl(x^0,x;\theta\bigr),
\end{equation*}
provide one-parameter families of strong, off-shell Dirac observables for
the $\mathbb{R} \times T^2$
VD Gowdy system, i.e.,~$\delta_{\epsilon}^{\text{\sc hvd}}
\mathcal{O}_m^{\text{\sc vd}}(\theta) = 0$, $\delta_{\epsilon}^{\text{\sc hvd}}
\mathcal{O}^{\text{\sc vd}}(\theta) = 0$ hold without using equations of motion
and without invoking a~gauge fixing.
\end{Result}

Before proceeding, we check that~\eqref{VDobs2} has the correct
on-shell specialization. As usual, for $\epsilon = n$, $\epsilon^1=0$ the
Poisson bracket in~\eqref{VDobs2} should generate $\mathcal{J}_0^{\text{\sc vd}}(\theta)$'s
time evolution. This amounts to
$
e_0\bigl( \mathcal{J}_0^{\text{\sc vd}}(\theta)\bigr) =0$, (on-shell),
which indeed holds: The evolution equations derived from the action
\eqref{VDHaction} combine to $e_0\bigl(J_0^{\text{\sc vd}}\bigr) =0$.
Further, $e_0(\tilde{\rho})$ we interpret as $\delta^\text{\sc hvd}_{\epsilon} \tilde{\rho}$
for $\epsilon =n$, $\epsilon^1 =0$. By~\eqref{VDHvar3} this vanishes, so
$e_0\bigl( J_0^{\text{\sc vd}}/(\theta + \tilde{\rho})^m\bigr) =0$ follows.
Finally, we need $e_0\bigl(U^{\text{\sc vd}}(\theta)\bigr)=0$, which we obtain
from the $e_0$ derivative of the first relation in~\eqref{UVDdef2}
by an argument analogous to the one after equation~\eqref{linsystemH1}.

\textbf{$\boldsymbol{T^3}$ topology.} The periodic extension of
\eqref{sl2charge4} gives
\begin{align*}
& \delta_{\epsilon}^\text{\sc hvd} \int_0^{2\pi} {\rm d}x J_0^{\text{\sc vd}}\bigl(x^0,x\bigr)
\check{s}_m\bigl(x^0,x;\theta\bigr) =0, \qquad 2 \leq m \in \mathbb{N},\\
& \check{s}_m\bigl(x^0,x^1;\theta\bigr) := \sum_{l \in \mathbb{Z}}
\frac{1}{\bigl( \theta + \tilde{\rho} + l \pi_0^{\sigma}\bigr)^m},
\end{align*}
where the sums can be evaluated explicitly and result in trigonometric
Laurent polynomials
\begin{equation*}
\check{s}_m = \left(\frac{\pi}{\pi_0^{\sigma}}\right)^2
\frac{(-)^m}{(m - 1)!} \left( \frac{\partial}{\partial \theta} \right)^{m-2}
\frac{1}{\sin^2 \frac{\pi}{\pi_0^{\sigma}}(\theta + \tilde{\rho})}.
\end{equation*}
For the `dressing' with some $U^{\text{\sc vd}}$, we
proceed as in Appendix \ref{appB}. We define
\begin{equation*}
T^{\text{\sc vd}}_\text{\sc l}\bigl(x^0,x^1;\theta\bigr) =
\exp_+\left\{ \frac{Q}{2} \varkappa_{\text{\sc l}}(\theta)
- \frac{1}{2} \int_{-2\pi \text{\sc l}}^{x^1} {\rm d}x
\left(\frac{J_0^{\text{\sc vd}}}{\theta + \tilde{\rho}} \right)\bigl(x^0,x\bigr)
\right\},
\end{equation*}
with the $\varkappa_{\text{\sc l}}(\theta)$ from~\eqref{aux2}, \eqref{aux4}.
Convergence of the series is ensured under broad conditions, for example, if $\sup_{ x \in [0,2\pi]} \Vert J_0^{\text{\sc vd}}/
(\theta + \tilde{\rho})\Vert < \varkappa^{\text{\sc vd}}\bigl(x^0,\theta\bigr)$,
for some matrix norm. A~precise counterpart of the Result~\ref{quasiperB1}
holds for the sequence $T^{\text{\sc vd}}_\text{\sc l}$, which we call

\textbf{Result~\ref{quasiperB1} VD}. The sequence $T^{\text{\sc vd}}_\text{\sc l}$, $\text{\sc l} \in
\mathbb{N}$, has properties (a), (b), (c) paralleling those of~$T_\text{\sc l}$,~${\text{\sc l} \in
\mathbb{N}}$, in Result~\ref{quasiperB1}.

We omit restating the assertions, but quickly run through
the main steps leading to parts (a), (b), (c). First, the existence of the
$\text{\sc l} \rightarrow \infty$ limit is ensured by
\begin{align*}
& \frac{Q}{2} \varkappa_{\text{\sc l}}(\theta)
- \frac{1}{2} \int_{-2\pi \text{\sc l}}^{x^1} {\rm d}x
\left(\frac{J_0^{\text{\sc vd}}}{\theta + \tilde{\rho}} \right)\bigl(x^0,x\bigr)
\\
& \qquad =- \frac{1}{2} \int_{0}^{x^1} {\rm d}x
\left(\frac{J_0^{\text{\sc vd}}}{\theta + \tilde{\rho}} \right)\bigl(x^0,x\bigr)
- \frac{1}{2} \int_0^{2\pi} {\rm d}x
[\varkappa_{\text{\sc l}}(\tilde{\rho} + \theta) -\varkappa_{\text{\sc l}}(\theta)]
J_0^{\text{\sc vd}}\bigl(x^0,x\bigr),
\end{align*}
with $\varkappa_{\text{\sc l}}(\theta)$ from~\eqref{aux2}. By~\eqref{aux4}, the
difference $\varkappa_{\text{\sc l}}(\tilde{\rho} + \theta) -\varkappa_{\text{\sc l}}(\theta)$
has a~finite $\text{\sc l} \rightarrow \infty$ limit given by $-2 c_{\infty}(\theta)$
from~\eqref{aux3}. In order to avoid the poles of the DiGamma
function, we assume~$\pi_0^{\sigma} \operatorname{Re}(\theta) >0$ throughout.
This gives part (a): The limit
\[
U^{\text{\sc vd}}\bigl(x^0,x^1;\theta\bigr) :=
\lim_{\text{\sc l} \rightarrow \infty} T^{\text{\sc vd}}_\text{\sc l}\bigl(x^0,x^1;\theta\bigr)
\]
exists for $\pi_0^{\sigma} \operatorname{Re}(\theta) >0$, pointwise in
$x^0$ and $x^1 \in [0,2\pi]$.
It is differentiable in $x^1$, solves $\partial_1 U^{\text{\sc vd}} =
-U^{\text{\sc vd}} J_0^{\text{\sc vd}}/(2(\theta + \tilde{\rho}))$,
and has the pointwise fall-off $U^{\text{\sc vd}}\bigl(x^0,x^1;\theta\bigr) =
\mathbbm{1} + O(1/|\theta|)$, for~${\pi_0^{\sigma} \operatorname{Re}(\theta) >0}$.
Next, the quasi-periodicity
\[
T^{\text{\sc vd}}_\text{\sc l}\bigl(x^0,x^1 + 2\pi;\theta\bigr) =
\exp\{- Q/(2\theta)\}T^{\text{\sc vd}}_\text{\sc l}\bigl(x^0,x^1;\theta- \pi_0^{\sigma}\bigr),
\]
and its generalizations are derived in parallel to~\eqref{TLquasiper}.
In particular, this can be used to justify the boundary
condition $T^{\text{\sc vd}}_\text{\sc l}\bigl(x^0, -2\pi \text{\sc l};\theta\bigr) =
\exp\{ Q\varkappa_\text{\sc l}(\theta)/2\}$ as in~\eqref{TLbc}.
Further, it implies part~(b): $U^{\text{\sc vd}}$ can be extended to other
spatial intervals using the quasi-periodicity
\begin{equation*}
U^{\text{\sc vd}}\bigl(x^0,x^1 + 2\pi;\theta\bigr) =
{\rm e}^{ -\frac{Q}{2\theta}}
U^{\text{\sc vd}}\bigl(x^0,x^1;\theta- \pi_0^{\sigma}\bigr).
\end{equation*}
Finally, for the gauge variation we proceed as in
the derivation of~\eqref{Tlimit8}. With $\delta_{\epsilon}^H$, $T_{\text{\sc l}}$,
$C_{\epsilon}$, replaced by \smash{$\delta_{\epsilon}^{\text{\sc hvd}},
T_{\text{\sc l}}^{\text{\sc vd}}, C_{\epsilon}^{\text{\sc vd}}
:= - \bigl(\epsilon^1/2\bigr) J_0^{\text{\sc vd}}/(\theta + \tilde{\rho})$},
respectively, the same formula holds. In order for it to have a~well-defined $\text{\sc l} \rightarrow \infty$ limit again, the condition $\operatorname{tr} Q^2 <
2 \bigl(\pi_0^{\sigma}\bigr)^2$ is needed. This gives the VD counterpart of Result
\ref{quasiperB1}\,(c): assume \smash{$\operatorname{tr} Q^2 <
2 \bigl(\pi_0^{\sigma}\bigr)^2$} and $\pi_0^{\sigma} \operatorname{Re}(\theta) >0$. Then
$\delta_{\epsilon}^{\text{\sc hvd}} U^{\text{\sc vd}} =
U^{\text{\sc vd}} C_{\epsilon}^{\text{\sc vd}}$ holds.

Next, we turn to the one-parametric current and obtain VD
counterparts of Result~\ref{resultJtheta}. Defining
$\mathcal{J}_0^{\text{\sc vd}}(\theta)_{\text{\sc l}} :=
T_{\text{\sc l}}^{\text{\sc vd}}(\theta) J_0^{\text{\sc vd}}
T_{\text{\sc l}}^{\text{\sc vd}}(\theta)^{-1}/(\theta + \tilde{\rho})^2$,
the limit
\begin{equation*}
\mathcal{J}_0^{\text{\sc vd}}(\theta) :=
\lim_{\text{\sc l} \rightarrow \infty} \mathcal{J}_0^{\text{\sc vd}}(\theta)_{\text{\sc l}} =
U^{\text{\sc vd}}(\theta)\frac{ J_0^{\text{\sc vd}}}{(\theta + \tilde{\rho})^2}
U^{\text{\sc vd}}(\theta)^{-1}, \qquad \pi_0^{\sigma}\operatorname{Re}(\theta) >0,
\end{equation*}
exists pointwise in $x^0$ and $x^1 \in [0,2\pi]$. It can be
extended to other spatial intervals using the quasi-periodicity
\begin{equation*}
\mathcal{J}_0^{\text{\sc vd}}\bigl(x^0,x^1 + 2\pi;\theta\bigr) = {\rm e}^{- \frac{Q}{2\theta}}
\mathcal{J}_0^{\text{\sc vd}}\bigl(x^0,x^1;\theta - \pi_0^{\sigma}\bigr)
{\rm e}^{\frac{Q}{2\theta}}.
\end{equation*}
However, $\mathcal{J}_0^{\text{\sc vd}}(\theta)$ has the desired gauge variation
only if $\operatorname{tr} Q^2 < 2 \bigl(\pi_0^{\sigma}\bigr)^2$ is imposed in addition to
$\pi_0^{\sigma} \operatorname{Re}(\theta) >0$. This changes
upon taking the trace with either one of the $\nu = \nu_{\pm} =
\mathrm{Nil}_{\pm}(a)$ matrices in~\eqref{sim6}.
Proceeding along the same lines leading to~\eqref{traceJ2} one finds
\begin{equation*}
\delta_{\epsilon}^{\text{\sc hvd}} \operatorname{tr}\bigl\{\nu \mathcal{J}_0^{\text{\sc vd}}(\theta)
\bigr\} =
\partial_1 \bigl( \epsilon^1 \operatorname{tr}
\bigl\{ \nu \mathcal{J}_0^{\text{\sc vd}}(\theta)\bigr)\bigr\}, \qquad
\pi_0^{\sigma} \operatorname{Re}(\theta) >0,
\end{equation*}
without condition on $\operatorname{tr} Q^2$. This is the VD counterpart of the
Result~\ref{resultJtheta}\,(c).

For the periodic extension $\sum_{n=-N}^N \operatorname{tr}\bigl\{\nu
\mathcal{J}_0^{\text{\sc vd}}\bigl(x^0,x^1 + 2 \pi n;\theta\bigr)\bigr\}$ we follow the same
steps as in~\eqref{Jthetaavg2}, \eqref{Jthetaavg3} to arrive at the

\begin{Result} \label{resultVDT3}
Let $\pi_0^{\sigma} \operatorname{Re}(\theta) >0$. Then, the quantities
\begin{align}
\mathcal{O}^{\text{\sc vd}}_m(\theta) &:= \sum_{l \in \mathbb{Z}} \int_0^{2\pi} {\rm d}x
\frac{J_0^{\text{\sc vd}}\bigl(x^0,x\bigr)}{\bigl( \theta + \tilde{\rho}\bigl(x^0,x\bigr)
 + l \pi_0^{\sigma}\bigr)^m},\qquad
 2 \leq m \in \mathbb{N},
\nonumber\\
\mathcal{O}^{\text{\sc vd}}(\theta) &:= \sum_{n \in \mathbb{Z}} \int_0^{2\pi} {\rm d}x
\operatorname{tr}\bigl\{\nu \mathcal{J}_0^{\text{\sc vd}}\bigl(x^0,x + 2\pi n;\theta\bigr)\bigr\},
\label{VDobs6}
\end{align}
provide one-parameter families of strong, off-shell Dirac observables for
the $T^3$ VD Gowdy system, i.e.,~$\delta_{\epsilon}^{\text{\sc hvd}}
\mathcal{O}_m^{\text{\sc vd}}(\theta) = 0$, $\delta_{\epsilon}^{\text{\sc hvd}}
\mathcal{O}^{\text{\sc vd}}(\theta) = 0$ hold without using equations of motion
and without invoking a~gauge fixing.
\end{Result}

Next, we seek to relate the Dirac observables
in the VD Gowdy systems to those in the full Gowdy cosmologies.

\subsection{Anti-Newtonian expansion of Dirac observables}\label{sec4.3}

In the following, we use the scale transformation~\eqref{scale1}
to obtain a~scaling decomposition of the Dirac observables
found in Section \ref{sec3}. The leading order terms will turn out to
correspond to Dirac observables in the VD system. To this end,
we consider, respectively, the scaling
of basic currents $J_{\mu}^H$, the quantities $K_{\mu}^H(\theta)$
in $\mathcal{J}_{\mu}^H(\theta) = U(\theta) K_{\mu}^H(\theta) U(\theta)^{-1}$,
as well as $L_{\mu}^H(\theta)$ and $U(\theta)$.

In a~first step, we recall from~\eqref{JHdef} the definitions of $J_0^H$, $J_1^H$.
Using~\eqref{scale1}, one has
$
J_0^H \mapsto J_0^H$, \smash{$
\frac{J_1^H}{n} \mapsto \frac{1}{\ell} \frac{J_1^H}{n}$}.
In $L_{\mu}^H(\theta)$, $K_{\mu}^H(\theta)$, we keep the definition~\eqref{rhotHdef}
of $\tilde{\rho}$ and view it as scale invariant, $\tilde{\rho}
\mapsto \tilde{\rho}$. Augmented thereby the scale transformation
\eqref{scale1} induces an $\ell$ dependence on \smash{$L_{\mu}^H(\theta)$},
\smash{$K_{\mu}^H(\theta)$}, which we seek to expand in inverse powers of $\ell$.
In doing so one encounters complex roots of the form
\smash{$\bigl[z^2 - \rho^2/\ell^2\bigr]^{-p/2}$}, for $p =1,3$ and $z = \theta + \tilde{\rho}$,
$\operatorname{Im}\theta \neq 0$. In a~small $\rho/\ell$ expansion the branch cut in Figure~\ref{fig1} shrinks to the point $z=0$. The expansion takes the form
\begin{align}
& \bigl[ z^2 - \rho^2/\ell^2\bigr]^{-p/2} = s(z)
 \frac{1}{z^p} \sum_{j \geq 0} \binom{- p/2}{j} (-)^j
 \left( \frac{\rho}{z \ell} \right)^{2j}.
\nonumber\\
& s(z) := {\rm e}^{{\rm i} p \operatorname{Arg}(z)}\bigl( {\rm e}^{ 2{\rm i} \operatorname{Arg}(z)} \bigr)^{-p/2}
= \operatorname{sign}\bigl(\operatorname{Re}(z)\bigr), \qquad \operatorname{Im}(z) \neq 0,\qquad
p \mathrm{ odd}.
\label{rootsign}
\end{align}
This will be applied to $z = \theta + \tilde{\rho}$ and eventually
we set $\ell =1$. A~sufficient condition for convergent series
to arise then is
\begin{gather}
\inf |\theta + \tilde{\rho}| > \delta, \qquad
\sup |\rho/\lambda_{\text{\sc n}}| < \delta,
\label{ANconv}
\end{gather}
for some $\delta >0$. Here the infimum/supremum is taken
for fixed $x^0$ over either $x^1 \in \mathbb{R}$ or $x^1 \in [0,2\pi]$,
depending on the Gowdy system considered. By adjusting $\theta$ and $\lambda_{\text{\sc n}}$
both conditions are easily met.

Using~\eqref{rootsign}, the $K_0^H$, $K_1^H$ currents~\eqref{KHdef}
have a~series expansion in order of scaling weights as follows
\begin{gather}
K_0^H(\theta) = \pm s(\theta + \tilde{\rho}) \bigg\{
 \underbrace{\frac{J_0^H}{(\theta + \tilde{\rho})^2}}_{\ell^{0}}\nonumber
\\
\phantom{K_0^H(\theta) =}{}+ \sum_{j \geq 1} \binom{-3/2}{j} (-)^{j}
\underbrace{\frac{\lambda_{\text{\sc n}}^2 \rho^{2 j}}{\big[\lambda_{\text{\sc n}}(\theta + \tilde{\rho})\big]^{2 j+2}}}_{\ell^{-2 j}}
\underbrace{\left[ J_0^H - \frac{2j}{2j+1}
 \frac{\lambda_{\text{\sc n}}(\theta + \tilde{\rho})}{\rho}
 \frac{J_1^H}{n} \right]}_{\ell^0}\bigg\},
\nonumber\\
K_1^H(\theta) = \pm s(\theta + \tilde{\rho}) \Bigg\{
\sum_{j \geq 0} \binom{-3/2}{j} (-)^{j}
\underbrace{\frac{\lambda_{\text{\sc n}}^2\rho^{2 j}}{\big[\lambda_{\text{\sc n}}(\theta + \tilde{\rho})\big]^{2 j+3}}}_{\ell^{-3 -2 j}}
\underbrace{\big[ \lambda_{\text{\sc n}}(\theta + \tilde{\rho}) J_1^H - \rho n J_0^H\big]}_{\ell^1}
\Bigg\}.
\label{Hscaling3}
\end{gather}
Here the scaling is not performed but the $\ell$-powers carried by
individual terms upon rescaling are indicated by an underbrace.
Observe that by splitting up the terms in square brackets
the expansion can be re-interpreted as one in powers of $\rho/\lambda_{\text{\sc n}}$.
Performing the rescaling, we write this as
\begin{equation*}
K_0^H(\theta) = \sum_{j \geq 0} \ell^{-2j} K_0^H(\theta)_{2j},
\qquad
K_1^H(\theta) = \frac{1}{\ell^2} \sum_{j \geq 0} \ell^{-2j} K_1^H(\theta)_{2j},
\end{equation*}
with coefficients that can be read off from~\eqref{Hscaling3}.
Note that the leading term $K_0^H(\theta)_0$ has (up to a~sign) the same form
as the $J_0^\text{\sc vd}/(\theta + \tilde{\rho})^2$ entering
\eqref{VDobs1}, just with the original Gowdy fields replaced by
their VD counterparts. For later reference, we indicate this as
\begin{gather}
K_0^H(\theta)_0 \big|_{\text{\sc vd}} = s(\theta + \tilde{\rho})
\frac{J_0^\text{\sc vd}}%
{(\theta + \tilde{\rho})^2}.
\label{Hscaling5}
\end{gather}
In order to obtain a~scaling decomposition of $U(\theta)$, the one
for $L_{\mu}^H(\theta)$ is needed. Using again~\eqref{rootsign}
and the expansions of \smash{$L_0^H(\theta)$}, \smash{$L_1^H(\theta)$} turns out to
contain an unwanted term, depending on the sign of $\operatorname{Re}(\theta+
\tilde{\rho})$. Specifically, using~\eqref{rootsign} one finds
\begin{align}
 L_0^H(\theta) ={}& \bigl( \pm s(\theta + \tilde{\rho}) -1 \bigr)
\underbrace{\frac{\lambda_{\text{\sc n}} n J_0^H}{2 \rho }}_{\ell^{0}}\nonumber
\\
&\pm s(\theta + \tilde{\rho})
\frac{1}{2}\sum_{j \geq 0} \binom{-1/2}{j} (-)^{j+1}
\underbrace{\frac{\lambda_{\text{\sc n}} \rho^{2 j}}{\big[\lambda_{\text{\sc n}}(\theta
 + \tilde{\rho})\big]^{2 j+1}}}_{\ell^{-2 j}}
\underbrace{\left[ J_1^H - \frac{1 + 2 j}{2 + 2 j}
 \frac{\rho}{\lambda_{\text{\sc n}}(\theta + \tilde{\rho})}
 n J_0^H \right]}_{\ell^{-2}},
\nonumber\\
 L_1^H(\theta) ={}&
\bigl( \pm s(\theta + \tilde{\rho}) -1 \bigr)
\underbrace{\frac{\lambda_{\text{\sc n}} J_1^H}{2 n \rho }}_{\ell^{0}}
\mp s(\theta + \tilde{\rho})
\underbrace{\frac{J_0^H}{2(\theta + \tilde{\rho})}}_{\ell^{0}}
\nonumber\\
&\pm
s(\theta + \tilde{\rho}) \frac{1}{2}\sum_{j \geq 1} \binom{-1/2}{j} (-)^{j+1}
\underbrace{\frac{\lambda_{\text{\sc n}} \rho^{2 j}}{\big[\lambda_{\text{\sc n}}(\theta + \tilde{\rho})\big]^{2 j+1}}}_{\ell^{-2 j}}
\underbrace{\left[ J_0^H - \frac{\lambda_{\text{\sc n}}(\theta + \tilde{\rho})}{\rho}
 \frac{J_1^H}{n} \right]}_{\ell^0}.\label{Hscaling8}
\end{align}
The $O(1/\rho)$ terms on the right hand sides are unwanted. In
order to eliminate them, we use the upper sign in~\eqref{LHdef}
for $\operatorname{Re}(\theta + \tilde{\rho}) >0$ and the lower sign
for $\operatorname{Re}(\theta + \tilde{\rho}) <0$. This, of course, reflects
the structure~\eqref{LHpm2} where the two sign options were
distinguished by $\pm$ superscripts. Doing this here as well,
we write
\begin{equation*}
K_0^{H,\pm}(\theta) = 2 \partial_{\theta} L_1^{H,\pm}(\theta), \qquad
K_1^{H,\pm}(\theta) = 2 \partial_{\theta} L_0^{H,\pm}(\theta),
\end{equation*}
and apply the previous expansions for $\pm \operatorname{Re}(\theta + \tilde{\rho})>0$,
respectively. As $\pm s(\theta + \tilde{\rho}) =1$ then, no explicit
signs occur and we may omit the $\pm$ superscripts in the
expansion coefficients. Restoring the powers of $1/\ell$, we write
\begin{gather}
L_0^{H,\pm}(\theta) = \frac{1}{\ell^2} \sum_{j \geq 0} \ell^{-2j} L_0^H(\theta)_{2j},
\qquad
L_1^{H,\pm}(\theta) = \sum_{j \geq 0} \ell^{-2j} L_1^H(\theta)_{2j}, \nonumber\\
\pm \operatorname{Re}(\theta+ \tilde{\rho}) >0.
\label{Hscaling9}
\end{gather}
Inspection of~\eqref{Hscaling8} shows that with an overall
$1/(\theta + \tilde{\rho})$ pulled out the coefficients only
involve positive powers of $\rho/(\theta + \tilde{\rho})$.

Next, we expand the defining differential equation for $U$,
i.e.,~$\partial_1 U_{\pm} = U_{\pm} L_1^{H,\pm}$, using an ansatz of the form
\begin{gather}
U_{\pm}(\theta) = \sum_{j \geq 0} \ell^{-2j} U(\theta)_{2j}, \qquad
\pm \operatorname{Re}(\theta + \tilde{\rho}) >0.
\label{Hscaling10}
\end{gather}
This gives the recursion
\begin{align}
& \partial_1 U(\theta)_0 = U(\theta)_0 L_1^H(\theta)_0,
\nonumber\\
& \partial_1 U(\theta)_{2j} - U(\theta)_{2j} L_1^H(\theta)_0 =
\sum_{k=0}^{j-1} U(\theta)_{2k} L_1^H(\theta)_{2(j-k)}, \qquad j \geq 1.
\label{Hscaling17}
\end{align}
For the lowest orders, one has in parallel to~\eqref{Hscaling5}
the relations
\begin{equation*}
L_1^H(\theta)_0 \big|_{\text{\sc vd}} =
- \frac{J_0^\text{\sc vd}}{2(\theta + \tilde{\rho})},\qquad
U(\theta)_0 \big|_{\text{\sc vd}} = U^{\text{\sc vd}}(\theta).
\end{equation*}
In particular, $U(\theta)_0$ is given by a~convergent path ordered
exponential under the same conditions as $U^{\text{\sc vd}}(\theta)$
in Section~\ref{sec4.2}. In the second relation~\eqref{Hscaling17},
the right-hand side is known from the preceding orders. Once $U(\theta)_0$ is
known the second equation integrates to the recursion
\begin{equation*}
U\bigl(x^0,x^1;\theta\bigr)_{2j} = \sum_{k=0}^{j-1} \int_{-\infty}^{x^1} {\rm d}z
\bigl(U(\theta)_{2k} L_1^H(\theta)_{2(j-k)} U(\theta)_0^{-1} \bigr)
\bigl(x^0,z;\theta\bigr) U\bigl(x^0,x^1;\theta\bigr)_0,
\end{equation*}
for $j \geq 1$. Here, we built in the boundary condition~\eqref{Ubc}
by using $-\infty$ as the lower integration boundary. This produces
convergent integrals because since $L_1^H(\theta)_{2j}$ is of
order $(\theta + \tilde{\rho})^{-2j}$ the integrand can be
assumed to vanish at least like $|z|^{-2}$ for large $|z|$.

Next, we address the gauge variations of the $U(\theta)_{2j}$ components.
The expected behavior can be inferred from a~scaling decomposition of
the original gauge variations, i.e., 
\begin{equation*}
\left\{ U_{\pm}(\theta), \mathcal{H}^{\text{\sc vd}}_0(\epsilon) + \mathcal{H}_1\bigl(\epsilon^1\bigr) +
\frac{1}{\ell^2} \mathcal{V}(\epsilon) \right\} = U_{\pm}(\theta)
\left[\frac{\epsilon}{n} L_0^{H,\pm}(\theta) + \epsilon^1 L_1^{H,\pm}(\theta) \right].
\end{equation*}
One finds
\begin{gather}
\bigl\{ U(\theta)_0, \mathcal{H}_{0\text{\sc vd}}(\epsilon) + \mathcal{H}_1\bigl(\epsilon^1\bigr) \bigr\} =
U(\theta)_0 \epsilon^1 L_1^H(\theta)_0,
\nonumber\\
\bigl\{ U(\theta)_2, \mathcal{H}_{0\text{\sc vd}}(\epsilon) + \mathcal{H}_1\bigl(\epsilon^1\bigr) \bigr\}
= - \{ U(\theta)_0, \mathcal{V}(\epsilon) \} + U(\theta)_2 \epsilon^1 L_1^H(\theta)_0
\nonumber\\
\phantom{\bigl\{ U(\theta)_2, \mathcal{H}_{0\text{\sc vd}}(\epsilon) + \mathcal{H}_1\bigl(\epsilon^1\bigr) \bigr\}
=}{}
+ U(\theta)_0 \left[\frac{\epsilon}{n} L_0^H(\theta)_0 +
 \epsilon^1 L_1^H(\theta)_2 \right],\label{Hscaling21}
\end{gather}
etc. Notably, on the left-hand side only the gauge variations of the VD
system enter.

To proceed, we expand $\mathcal{J}_{\mu}^{H,\pm}(\theta) = U_{\pm}(\theta)
K_{\mu}^{H,\pm}(\theta) U_{\pm}(\theta)^{-1}$
in powers of $1/\ell^2$,
\begin{gather}
\mathcal{J}_0^{H,\pm}(\theta) = \sum_{j \geq 0} \ell^{-2 j}
\mathcal{J}_0^H(\theta)_{2 j},\qquad
\mathcal{J}_1^{H,\pm}(\theta) = \frac{1}{\ell^2} \sum_{j \geq 0} \ell^{-2 j}
\mathcal{J}_1^H(\theta)_{2 j}, \nonumber\\
 \pm \operatorname{Re}(\theta + \tilde{\rho}) >0.
\label{Hscaling22}
\end{gather}
To low orders,
\begin{align}
& \mathcal{J}_0^H(\theta)_0 = U(\theta)_0 K^H_0(\theta)_0 U(\theta)_0^{-1},
\qquad
 \mathcal{J}_1^H(\theta)_0 = U(\theta)_0 K_1^H(\theta)_0 U(\theta)_0^{-1},
\nonumber\\
&\mathcal{J}_0^H(\theta)_2 = U(\theta)_0 K^H_0(\theta)_2 U(\theta)_0^{-1}
+ \big[ U(\theta)_2 U(\theta)_0^{-1}, \mathcal{J}_0^H(\theta)_0\big],
\nonumber\\
&\mathcal{J}_1^H(\theta)_2 = U(\theta)_0 K^H_1(\theta)_2 U(\theta)_0^{-1}
+ \big[ U(\theta)_2 U(\theta)_0^{-1}, \mathcal{J}_1^H(\theta)_0\big].
\label{Hscaling23}
\end{align}
On the other hand, we can expand the key equation~\eqref{Diracobs4}
which results in the expected gauge variations
\begin{align}
& \bigl\{ \mathcal{J}_0^H(\theta)_0, \mathcal{H}^{\text{\sc vd}}_0(\epsilon) + \mathcal{H}_1\bigl(\epsilon^1\bigr) \bigr\}
= \partial_1 \bigl( \epsilon^1 \mathcal{J}_0^H(\theta)_0\bigr),
\label{Hscaling24}\\
& \bigl\{ \mathcal{J}_0^H(\theta)_2, \mathcal{H}^{\text{\sc vd}}_0(\epsilon) + \mathcal{H}_1\bigl(\epsilon^1\bigr) \bigr\}
= - \bigl\{ \mathcal{J}_0^H(\theta)_0, \mathcal{V}(\epsilon) \bigr\}
+ \partial_1 \left( \frac{\epsilon}{n} \mathcal{J}_1^H(\theta)_0
+ \epsilon^1 \mathcal{J}_0^H(\theta)_2 \right).
\nonumber
\end{align}
Consistency requires that these come out from the gauge variations
of the $U(\theta)_{2j}$ and $K_{\mu}^H(\theta)_{2j}$. The latter
can be inferred from the decomposition of the ones in~\eqref{KHvar3}.
To low orders,
\begin{gather}
 \{ K_0^H(\theta)_0, \mathcal{H}_{0\text{\sc vd}}(\epsilon) + \mathcal{H}_1\bigl(\epsilon^1\bigr) \} = \partial_1 \bigl( \epsilon^1 K_0^H(\theta)_0\bigr),
\nonumber\\
 \{ K_0^H(\theta)_2, \mathcal{H}_{0\text{\sc vd}}(\epsilon) + \mathcal{H}_1\bigl(\epsilon^1\bigr) \}
= - \{ K_0^H(\theta)_0, \mathcal{V}(\epsilon) \} + \partial_1 \left( \frac{\epsilon}{n} K_1^H(\theta)_0 + \epsilon^1 K_0^H(\theta)_2 \right)
\nonumber\\
\phantom{ \{ K_0^H(\theta)_2, \mathcal{H}_{0\text{\sc vd}}(\epsilon) + \mathcal{H}_1\bigl(\epsilon^1\bigr) \}
=}{} \pm\frac{\epsilon}{n} \frac{1}{ (\theta +\tilde{\rho})^3} \big[J_1^H, J_0^H].
\label{Hscaling25}
\end{gather}
Using~\eqref{Hscaling21} and~\eqref{Hscaling25}, one can
verify that the currents~\eqref{Hscaling23} indeed satisfy~\eqref{Hscaling24}.

Under the assumption~\eqref{ANconv}, all expansions~\eqref{Hscaling9},
\eqref{Hscaling10}, \eqref{Hscaling3}, \eqref{Hscaling22} can for $\ell=1$
be reinterpreted as convergent series in powers of $\rho/\lambda_{\text{\sc n}}$.
This gives the following result.

\begin{Result} Assume~\eqref{ANconv}. Then,
 the currents $\mathcal{J}_{\mu}^{H,\pm}\bigl(x^0,x^1;\theta\bigr)$, $\mu=0,1$, in the $\mathbb{R}\times T^2$ and $T^3$ Gowdy systems admit
for $\pm \operatorname{Re}(\theta + \tilde{\rho}) >0$, $\pm \pi_0^{\sigma} >0$, a~convergent expansion in powers of $\rho/\lambda_{\text{\sc n}}$. This implies a~concomitant expansion in inverse powers of $\lambda_{\text{\sc n}}$ of
the off-shell Dirac observables
\begin{equation*}
\mathcal{O}(\theta) = \int_{-\infty}^{\infty} {\rm d}x \mathcal{J}_0^{H,\pm}\bigl(x^0,x;\theta\bigr),
\qquad
\mathcal{O}(\theta) := \int_0^{2\pi} {\rm d}x \sum_{n \in \mathbb{Z}}
\operatorname{tr}\bigl\{\nu \mathcal{J}_0^{H,\pm}\bigl(x^0,x + 2\pi n;\theta\bigr)\bigr\},
\end{equation*}
from Results~{\rm\ref{resultGowdyNC}} and~{\rm\ref{resultT3obs}}
in the $\mathbb{R} \times T^2$ and $T^3$ Gowdy systems, respectively.
The leading terms~$\mathcal{O}(\theta)_0$ in the expansions coincide,
upon replacement of the original with the VD phase space variables,
with the Dirac observables $\mathcal{O}^{\text{\sc vd}}(\theta)$ in the
VD system as summarized in Results~{\rm\ref{resultVDNC}} and~{\rm\ref{resultVDT3}},
respectively.
\end{Result}

\subsection{On-shell specialization and relation to AVD}\label{sec4.4}

So far, we performed on-shell specializations without
also fixing a~gauge. In all discussions of asymptotic velocity
domination (AVD) only the velocity dominated field equations
enter, moreover they do so for a~standard coordinate
choice inherited from the full gauge fixed Gowdy system,
see~\cite{RingstroemLR} for a~review. The
interplay between gauge fixing and this standard coordinate
choice is detailed in~\cite{GowdyPol,TorreRom}. Here we only mention
that once the proper time gauge $s =0 = \partial_0 n$ is imposed,
$e_0$ acts like $\partial_t$, and a~spatial coordinate can be introduced by
\smash{$\zeta = \int^{x^1} {\rm d} x n(x)^{-1}$}.
The stability group of the $s=0= \partial_0 n$
condition with respect to the $\delta^{\text{\sc vd}}_{\epsilon}$
gauge transformations from~\eqref{scale4} is still
$\epsilon \equiv 0$, $\epsilon^1 = \epsilon^1\bigl(x^1\bigr)$ and the subsequent steps
can proceed as in~\cite{GowdyPol}. For the following discussion,
we simply adopt the resulting net identifications
\begin{gather}
\rho = x^0 = t>0, \qquad \tilde{\rho} = \zeta \in \mathbb{R},
\label{coords}
\end{gather}
where in the $T^3$ Gowdy systems the basic fields are
$2\pi$ periodic in $\zeta$. Commonly $-\ln t$ is used
as time variable; in order to have the big bang located at
$\tau \rightarrow - \infty$, we take $t = {\rm e}^{\tau}$ here.
The VD field equations in terms of $\Delta$, $\psi$ then read
\begin{align*}
& \partial_{\tau}\left( \frac{\partial_{\tau} \Delta}{\Delta} \right) +
\left(\frac{\partial_{\tau}\psi }{\Delta}\right)^2 =0,
\qquad
\partial_{\tau} \left( \frac{\partial_{\tau} \psi}{\Delta^2} \right) =0,
\qquad
\partial_{\tau}^2 \sigma + \frac{ (\partial_{\tau} \Delta)^2 +
 (\partial_{\tau} \psi)^2}{ 2 \Delta^2} =0,\\
& \partial_{\tau} \sigma - \frac{ (\partial_{\tau} \Delta)^2
+ (\partial_{\tau} \psi)^2}{ 2 \Delta^2}=0,
 \qquad \partial_{\zeta}( \sigma +2 \ln n) -
\frac{\partial_{\tau} \Delta \partial_{\zeta} \Delta +
\partial_{\tau} \psi \partial_{\zeta} \psi}{\Delta^2}=0.
\end{align*}
Using the Hamiltonian constraint, the $\sigma$ evolution
equation simplifies to $\partial_{\tau}^2 \sigma=0$.
The $\Delta$, $\psi$ evolution equations can be
recognized as the geodesic equations in the Poincar\'{e} upper
half plane $\{ (\psi,\Delta) \in \mathbb{R}^2 | \Delta >0\}$,
with Riemannian metric ${\rm d}s^2 = \Delta^{-2} \bigl( {\rm d}\Delta^2 + {\rm d} \psi^2\bigr)$.

\textbf{Solution of the VD field equations.}
Using the two conserved quantities $ \partial_{\tau}\psi/\Delta^2 =:
- 2 v b/a$ and $\bigl[ (\partial_{\tau}\Delta)^2 + (\partial_{\tau} \psi)^2\bigr]
/\Delta^2 =: v^2$, one finds as the general solution
of the $\Delta$, $\psi$ evolution equations
\begin{gather}
\Delta(\tau,\zeta) = \frac{a}{{\rm e}^{ v \tau} + b^2 {\rm e}^{- v \tau}},
\qquad
\psi(\tau,\zeta) = \frac{ab}{b^2 + {\rm e}^{2 v \tau}}+ c,
 \qquad a > 0,
\label{temo2}
\end{gather}
where $b$, $c$ and $a,v>0$ are arbitrary (real-valued, differentiable,
$2\pi$-periodic) functions of $\zeta$. For $M(\tau,\zeta)$ from
\eqref{Mdef} this implies $M(\tau,\zeta) = {\rm e}^{- \tau v} M_0(\zeta) + O(1)$
for $\tau \rightarrow - \infty$, where $M_0(\zeta)$ has rank $1$.
Re-parameterizing the solution in terms of initial
data at $\tau =0$:
$\Delta_0( \cdot ) = \Delta(0, \cdot )$,
$\Delta_1( \cdot ) = (\partial_{\tau}\Delta)(0, \cdot )$,
$\psi_0( \cdot ) = \psi(0, \cdot )$,
$\psi_1( \cdot ) = (\partial_{\tau}\psi)(0, \cdot )$,
gives
\begin{gather}
\Delta(\tau,\zeta) = \frac{\Delta_1^2 + \psi_1^2}{v} \frac{ 2 {\rm e}^{v \tau}
 \bigl(\Delta_1 + \sqrt{\Delta_1^2 + \psi_1^2} \bigr)}%
 {\bigl(1 + {\rm e}^{2 v \tau}\bigr) \psi_1^2
 + 2 \Delta_1 \bigl(\Delta_1 + \sqrt{\Delta_1^2 + \psi_1^2} \bigr)},
\nonumber\\
\psi(\tau,\zeta) = \psi_0 +
 \frac{(-1 + {\rm e}^{2 v \tau})\psi_1}{v}
 \frac{ \psi_1^2 + \Delta_1
 \bigl(\Delta_1 + \sqrt{\Delta_1^2 + \psi_1^2} \bigr)}%
 {\bigl(1 + {\rm e}^{2 v \tau}\bigr) \psi_1^2
 + 2 \Delta_1 \bigl(\Delta_1 + \sqrt{\Delta_1^2 + \psi_1^2} \bigr)},
\label{temoinit}
\end{gather}
with the $\zeta$ dependence suppressed on the right-hand side.
Here $\Delta_0$ has been replaced with $\sqrt{\Delta_1^2 + \psi_1^2}/v$
since $v>0$ governs the time dependence. From~\eqref{temoinit}, one may
check that
\begin{equation*}
\left( \psi(\tau, \cdot ) - \left( \psi_0 +
\frac{\Delta_1\Delta_0}{\psi_1} \right) \right)^2
+ \Delta(\tau, \cdot )^2 =
\left(\frac{v\Delta_0^2}{\psi_1}\right)^2.
\end{equation*}
Pointwise in $\zeta$ this recovers the well-known description of
geodesics on the Poincar\'{e} upper half plane: they are semi-circles
with centers on the $\psi$-axis which touch it orthogonally as
$\tau \rightarrow \pm \infty$. The polarized case corresponds
to $\psi_0 = \psi_1 \equiv 0$, so that $\Delta = \Delta_0 {\rm e}^{v \tau}$,
with $\Delta_1 = v \Delta_0 >0$. The semi-circles then degenerate into
straight vertical lines.

Using~\eqref{temoinit}, the expressions entering the constraints read
\begin{gather*}
\frac{ (\partial_{\tau} \Delta)^2 + (\partial_{\tau} \psi)^2}{ 2 \Delta^2}
= \frac{v^2}{2},\\
\frac{\partial_{\tau} \Delta \partial_{\zeta} \Delta +
 \partial_{\tau} \psi \partial_{\zeta} \psi}{ \Delta^2}
= \tau \frac{1}{2} \partial_{\zeta} v^2 - \Delta_1
\partial_{\zeta} \left( \frac{v}{\sqrt{\Delta_1^2 + \psi_1^2}} \right)
+ v^2 \frac{\psi_1 \partial_{\zeta} \psi_0}{ \Delta_1^2 + \psi_1^2}.
\end{gather*}
In particular, the integrability condition: $\partial_{\zeta}$(first equation)
= $\partial_{\tau}$(second equation) manifestly holds. The Hamiltonian constraint
thus integrates to
$\sigma(\tau, \zeta) = v(\zeta)^2 (\tau - \tau_0)/2
+ \sigma(\tau_0, \zeta)$. Combined with the diffeomorphism
constraint this fixes
\begin{align*}
\sigma(\tau, \zeta) ={}& -2 \ln n(\zeta) +
\frac{v(\zeta)^2}{2} \tau\\
&+ \int_0^{\zeta} {\rm d}\zeta' \left\{ - \Delta_1
\partial_{\zeta'} \bigg( \frac{v}{\sqrt{\Delta_1^2 + \psi_1^2}} \bigg)
+ v^2 \frac{\psi_1 \partial_{\zeta'} \psi_0}{ \Delta_1^2 + \psi_1^2}
\right\}(\zeta'),
\end{align*}
up to an additive constant.
In order for $\sigma$ to be spatially periodic, the integral
in the second line over one period needs to vanish. After integration
by parts this amounts to
\begin{equation*}
\int_{0}^{2\pi} {\rm d}\zeta \bigg\{
\frac{\partial_{\zeta}\Delta_1}{\Delta_0}
+ \frac{\psi_1 \partial_{\zeta} \psi_0}{\Delta_0^2}
\bigg\} =0.
\end{equation*}
For completeness' sake, we also note the expression for the Noether
current evaluated on~\eqref{temoinit},
\begin{gather}
J_0^{\text{\sc vd}} = \frac{1}{\Delta_0} \begin{pmatrix}
 \Delta_1 & \Delta_0 \psi_1 \\
 \psi_1/\Delta_0 & - \Delta_1
 \end{pmatrix}
+ \frac{\psi_0}{\Delta_0^2} \begin{pmatrix}
 \psi_1 &
 - \psi_0 \psi_1 - 2 \Delta_0 \Delta_1 \\
 0 & - \psi_1
\end{pmatrix}.
\label{J0init}
\end{gather}
It comes out $\tau$ independent as required and obeys
$\operatorname{tr} \bigl(J_0^{\text{\sc vd}}\bigr)^2 = 2 \bigl(\Delta_1^2 + \psi_1^2\bigr)/\Delta_0^2
= 2 v^2$.

\textbf{Asymptotic velocity domination.}
In the general relativity literature commonly used notations are
$
\Delta = {\rm e}^{-P}$, $ \psi = Q$,
with $\tau_{\text{\sc gr}} = - \tau$ and $\theta_{\text{\sc gr}} = \zeta$
as coordinates. The evolution equations for $P$, $Q$ derived from
\eqref{2KLaction} in coordinates~\eqref{coords} then read
\begin{align}
\partial_{\tau}^2 P - {\rm e}^{2 \tau} \partial_{\zeta}^2 P -
 {\rm e}^{2 P}\big[ (\partial_{\tau}Q)^2 - {\rm e}^{2 \tau} (\partial_{\zeta}Q)^2\big]&= 0,
\nonumber\\
\partial_{\tau}^2 Q - {\rm e}^{2 \tau} \partial_{\zeta}^2 Q +2
\big[ \partial_{\tau}Q \partial_{\tau} P - {\rm e}^{2 \tau} \partial_{\zeta} Q \partial_{\zeta} P\big]
&= 0.
\label{temo9}
\end{align}
The evolution equations of the velocity dominated system are
obtained by dropping the terms with $\partial_{\zeta}$ derivatives.
Motivated by the large $-\tau$ behavior of the exact velocity
dominated solution~\eqref{temo2} (rewritten in terms of $P$, $Q$
and $v >0$), one considers an ansatz of the form
\begin{align}
P(\tau,\zeta) &= -v(\zeta) \tau + p_0(\zeta) + u(\tau,\zeta),
\qquad
Q(\tau,\zeta) = q_0(\zeta) + {\rm e}^{ 2 v(\zeta) \tau} [ q_1(\zeta)
 + w(\tau, \zeta)].
\label{temo10}
\end{align}
From~\eqref{temo2}, one can find simple ${\rm e}^{ 2 v(\zeta) \tau}$
dependent choices for $u$, $w$ that vanish rapidly
for $\tau \rightarrow -\infty$ and that render~\eqref{temo10} an exact
solution of the velocity dominated system. Using~\eqref{temo10}
as an ansatz for a~solution of the full evolution equations~\eqref{temo9}, one sees that
\begin{gather}
0 < v(\zeta) < 1,
\label{temo11}
\end{gather}
is needed for consistency: the ${\rm e}^{2 P} {\rm e}^{2 \tau} (\partial_\zeta Q)^2$
term in the first equation behaves like ${\rm e}^{2 (1-v(\zeta)) \tau}\allowbreak
(\partial_{\zeta} q_0)^2$, so that~\eqref{temo11} is needed for all
$\zeta$ for which $\partial_{\zeta} q_0$ is nonzero.
All other terms in~\eqref{temo9} decay without further assumptions.
In the polarized case $Q \equiv 0$, the constraint~\eqref{temo11}
is absent and the existence of generic solutions
$P(\tau,\zeta)$ of the form~\eqref{temo10} goes back to
\cite{IsenMonc}. In the non-polarized system the most interesting
case is when~\eqref{temo11} with $q_0(\zeta) \neq 0$
holds for all $\zeta \in [0, 2\pi]$.%
\footnote{Isolated points $\zeta_0$ at which $q_0(\zeta_0) =0$
may give rise to solutions with spatial spikes, see
\cite[Section 9]{RingstroemLR}.}
The ansatz~\eqref{temo10} then contains $4$ arbitrarily specifiable
functions $v$, $p_0$, $q_0$, $q_1$ of $\zeta$, which is the correct
number for a~generic solution of a~pair of second-order PDEs
in $\tau$ and $\zeta$. The existence of solutions of the
form~\eqref{temo10} was originally shown by Ringstr\"{o}m
\cite{Ringstroemproof}. This includes a~proof of the existence of $v^2$ as the $\tau \rightarrow -\infty$ limit of $(\partial_\tau P)^2+ {\rm e}^{2P} (\partial_\tau Q)^2$ \cite{RingstroemVelocity}. An earlier variant
was based on the insight that~\eqref{temo9},
\eqref{temo10} transcribes into a~so-called Fuchsian system for~$u$,~$w$~\cite{KRend}. If~$v$,~$p_0$,~$q_0$, $q_1$ are assumed to be real
analytic and~\eqref{temo11} holds for all $\zeta \in [0,2\pi]$,
this entails that there exists a~unique solution of~\eqref{temo9}
of the form~\eqref{temo10}, where $u$, $w$ tend to zero as~${\tau \rightarrow -\infty}$; see \cite[Theorem~1]{KRend}. This result also
justifies an earlier formal expansion~\cite{GMonc} and
has been used to establish valid asymptotic expansions
under weaker assumptions, see~\cite{RingstroemLR} for a~review.
Solutions of the form~\eqref{temo10}, \eqref{temo11} turn out
to always have a~curvature singularity as $\tau \rightarrow -\infty$.
The data~$v$,~$p_0$,~$q_0$, $q_1$ can thus be viewed as `initial data
at the big bang', while the relation to actual initial data
$(P,Q, \partial_{\tau} P, \partial_{\tau} Q)|_{\tau = \tau_0}$ needs to be
established separately, see~\cite{Li,RingstroemInitial}
in relation to `BKL bounces'.
For the present purposes, we may loosely summarize the situation
by saying that
`almost all' solutions of the full Gowdy fields equations have
an `inevitable' spacelike big bang singularity and behave for
$\tau \rightarrow -\infty$ asymptotically like~\eqref{temo10}.

{\bf Bounds on $\boldsymbol{{\operatorname{tr}Q^2}}$.} In view of its
significance in Results~\ref{resultJtheta} and~\ref{resultT3obs},
it is worth examining the significance of the condition
$\operatorname{tr}Q^2 < 2 \bigl(\pi_0^{\sigma}\bigr)^2$. Recall for the off-shell
generator
\begin{equation*}
Q\bigl(x^0\bigr) = \int_0^{2\pi} {\rm d}x^1 J_0^H\bigl(x^0,x^1\bigr),
 \qquad \bigl(J_0^H\bigr)^2 = \Delta^2 \bigl[ \bigl(\pi^{\Delta}\bigr)^2 + \bigl(\pi^{\psi}\bigr)^2\bigr] \mathbbm{1},
\end{equation*}
where the prefactor of the unit matrix is the kinetic energy density
on the reduced phase space. On-shell this quantity is
the previously encountered $(\partial_{\tau} P)^2 + {\rm e}^{2 P} (\partial_{\tau} Q)^2$
which controls the behavior of solutions backpropagated to the big bang.
As mentioned, in $x^0 = {\rm e}^{\tau}$ time it always has a~finite $\tau \rightarrow - \infty$ limit~\cite{RingstroemInitial}
\begin{gather}
v\bigl(x^1\bigr) = \bigl(\lim_{\tau \rightarrow - \infty}
\Delta^2 \bigl[ \bigl(\pi^{\Delta}\bigr)^2 + \bigl(\pi^{\psi}\bigr)^2\bigr]\bigr)^{1/2},
\label{qprop2}
\end{gather}
and the low velocity condition $0 < v\bigl(x^1\bigr) < 1$ is sufficient for
AVD behavior. Ideally, the condition $0< v\bigl(x^1\bigr) < 1$ would
relate to the decay condition \smash{$\operatorname{tr}Q^2 < 2 \bigl(\pi_0^{\sigma}\bigr)^2$}
(with $\pi_0^{\sigma} = -2\pi$ on-shell). However, this is the case
only for restricted initial data:
The double integral over $\operatorname{tr}\bigl[ J_0^H(x)J_0^H(x')\bigr]$ can
be related to that over \smash{$\operatorname{tr}\bigl[ J_0^H(x)^2\bigr]^{1/2}
\operatorname{tr}\bigl[J_0^H(x')^2\bigr]^{1/2}$} on account of the
Cauchy--Schwarz inequality for the Frobenius inner product
only if the transpose~${\operatorname{tr}\bigl[ J_0^H(x)^{\mathsf T} J_0^H(x')\bigr]}$ enters
in the first position. For a~symmetric
$J_0^H(x) = J_0^H(x)^{\mathsf T}$ one has indeed
\begin{equation*}
\operatorname{tr} Q^2 =
 \int_0^{2 \pi} {\rm d}x \int_0^{2 \pi} {\rm d}x '
\operatorname{tr}\bigl[ J_0^H(x)^{\mathsf T} J_0^H(x')\bigr] \leq
2 \int_0^{2 \pi} {\rm d}x v(x) \int_0^{2 \pi} {\rm d}x ' v(x')
< 2 (2 \pi)^2,
\end{equation*}
using $0< v(x) < 1$ in the last two steps. Initial data for
which $J_0^H$ is symmetric are not generic. From~\eqref{JHdef},
one sees that the symmetry condition
$\pi^{\psi} \bigl(\Delta^2 -1 - \psi^2\bigr) = 2 \pi^{\Delta} \psi \Delta$
amounts to one constraint for the four functions
$\bigl(\Delta,\pi^{\Delta}\bigr)$, $\bigl(\psi,\pi^{\psi}\bigr)$, at some $\tau = \tau_0$.
Assuming that AVD holds, one can also push $\tau_0$ back to $- \infty$,
where the parameterization~\eqref{J0init} applies. Then
$\psi_1\bigl[ \Delta_0^2 -1 - \psi_0^2\bigr] = 2 \Delta_1 \psi_0 \Delta_0$,
is the symmetry condition.

\textbf{The value of the conserved charges.} By construction, the on-shell
specialization of the Dirac observables~\eqref{Diracobs6} gives rise to a~one-parameter family of conserved charges. This remains true
if in addition the coordinate choice~\eqref{coords} is made.
The original field equations can then be described by
currents $J_0^H|^\mathrm{on}$, $J_1^H|^\mathrm{on}$ that solve the appropriate
specialization of~\eqref{JHthetaonshell3b} and~\eqref{JHthetaonshell3c}. The one-parameter
family of conserved charges~\eqref{DiracobsT3} can be viewed as a~functional of
$J_0^H|^\mathrm{on}$, $J_1^H|^\mathrm{on}$ for which we write $\mathcal{O}(\theta)|^\mathrm{on}$.
Whenever AVD holds the integrand of $\mathcal{O}(\theta)|^\mathrm{on}$ can
be backpropagated to large negative $\tau = \ln t$, where the
underlying solution should approach the time independent $J_0^{\text{\sc vd}}$,
with the spatial component subleading. On the other hand, being a~conserved charge, the value of $\mathcal{O}(\theta)|^\mathrm{on}$
will be unaffected by this. The conclusion is that
\begin{gather}
\mathcal{O}(\theta)|^\mathrm{on} = \mathcal{O}^{\text{\sc vd}}(\theta)|^\mathrm{on},
\label{Diraconshell}
\end{gather}
where $\mathcal{O}^{\text{\sc vd}}(\theta)|^\mathrm{on}$ is the on-shell
and coordinate specialization of the VD Dirac observables
from~\eqref{VDobs6}. The $J_0^{\text{\sc vd}}$ occurring in
the definitions can be replaced with the time independent
matrix~\eqref{J0init} and one obtains a~fairly explicit parameterization
of $\mathcal{O}(\theta)|^\mathrm{on}$ in terms of initial data. In the
polarized case the identity~\eqref{Diraconshell} can be verified
explicitly, see~\cite[Appendix A]{GowdyPol}.

\section{Conclusions}

Dirac observables are a~conceptual cornerstone of gravitational
theories that are, however, technically rather elusive in general
\cite{Giddings,Hoehn,Pons,Tambornino}.
Here we constructed one-parameter families of exact Dirac observables
for a~specific infinite-dimensional subsector of Einstein gravity,
the Gowdy cosmologies.
These observables are spatially nonlocal functionals on the phase
space of the
system which have the properties (a)--(d) highlighted in the
introduction: they strongly commute with the constraints,
are valid off-shell and without resorting to gauge-fixing,
refer to a~fixed time, and when backpropagated to the
big bang remain regular. The last feature is especially relevant in
the context of the broader asymptotic velocity domination (AVD) scenario~\cite{RingstroemCR},
which posits that generic cosmological solutions when
backpropagated to the big bang are often governed by velocity dominated
field equations where dynamical spatial gradients are absent.
This feature has been rigorously proven for solutions of the
$T^3$ Gowdy type~\cite{Ringstroemproof}. The anti-Newtonian expansion
\eqref{i3} of our Dirac observables can therefore be viewed
as an off-shell generalization of the AVD property.

There are several follow-up questions. By referring to a~fixed
time, our Dirac observables in principle have well-defined equal-time Poisson brackets. In a~different setting a~closed quadratic
Poisson algebra was found~\cite{KSGowdyNC} in the $\mathbb{R} \times T^2$
system for a~generating functional of nonlocal conserved charges.
It should be interesting to see whether such a~result persists for the present $\mathbb{R} \times T^2$ Dirac observables and for $T^3$ cosmologies. A~second open question concerns the feasibility
of a~general `lift' of Dirac observables in the VD Gowdy
system to those of the full Gowdy cosmologies. Some of
our VD Gowdy observables have been `reverse engineered' to allow
for such a~lift, but the systematic feasibility remains to be explored.
Having presented several one-parameter families of Dirac observables
one may also ask in which sense they are independent and what precisely
would constitute a~`complete' set.

Finally, one might aim at a~quantum version of AVD. Such a~quantum version, formulated on the level of integrands of Dirac
observables, has been established in the polarized Gowdy system
\cite{GowdyPol}. The extension to the non-polarized case considered
here will be much harder, as the action~\eqref{2KLaction} contains
self-interactions of exponential type in $\phi = \ln \Delta$.
Taking advantage of a~reformulation as a~Riemannian sigma-model with a~four-dimensional
target space, an all-order perturbative (weak coupling) quantization
has been accomplished in~\cite{2Kren1,2Kren2}, see also~\cite{reducedquant}.
When aiming at a~quantum AVD property, however, a~strong coupling
expansion in inverse powers of~$\lambda_{\text{\sc n}}$ is relevant and requires a~different methodology.

\appendix

\section{Two Killing vector reduction and its VD counterpart}\label{appA}

 A~convenient starting point is the Gibbons--Hawking action on a~foliated geometry $M = [t_i,t_f] \times \Sigma$. For metrics parameterized
in terms of ADM fields $\bigl(N,N^i,\mathrm{g}_{ij}\bigr)$ (called lapse, shift, spatial
metric) as
${\rm d}s^2 = - N^2 {\rm d}t^2 + \mathrm{g}_{ij} \bigl(N^i {\rm d}t + {\rm d}x^i\bigr) \bigl(N^j {\rm d}t + {\rm d}x^j\bigr)$, the action
is $S^L_{1+d}$ below. Simply evaluating~\smash{$S^L_{1 + d}$} on the subclass of
metrics~\eqref{2Kmetric} turns out to produce a~valid action principle
governing the associated ``midi-superspace'' subsector of Einstein
gravity. In this appendix, we introduce the reduced Lagrangian action,
its velocity dominated (VD) counterpart and discuss their
respective gauge symmetries. Our metric and curvature conventions
coincide with those in~\cite{Straumannbook}.\looseness=-1

The $1 + d$ form of the pure gravity action reads
\begin{gather}
S_{1+d}^L = \frac{1}{2 \kappa} \int_{t_i}^{t_f} {\rm d}t \int_{\Sigma} {\rm d}^dx
\left\{ \frac{1}{4 n} e_0(\mathrm{g}_{ij}) G(\mathrm{g})^{ij,kl} e_0(\mathrm{g}_{kl})
+ n \mathrm{g} R(\mathrm{g}) \right\},
\label{1daction}
\end{gather}
where
\[
2 G^{ij,kl}(\mathrm{g}) = \mathrm{g}^{ik} \mathrm{g}^{jl} + \mathrm{g}^{il} \mathrm{g}^{jk}
-2 \mathrm{g}^{ij} \mathrm{g}^{kl}
\]
 is the deWitt metric and $R(\mathrm{g})$ is the
Ricci scalar of the spatial metric $\mathrm{g}_{ij}$.
Further, we take the lapse
anti-density $n = N/\sqrt{\mathrm{g}}$ as basic, while the shift $N^i$
is hidden in $e_0 = \partial_t - \mathcal{L}_{\vec{N}}$, with~$\mathcal{L}_{\vec{N}}$
the spatial Lie derivative. An action principle for the
associated velocity dominated (VD) gravity theory can be
extracted from~\eqref{1daction} via scaling limit
$(\ell \rightarrow \infty)$ that primarily enhances the
spatial metric, $\mathrm{g}_{ij} \mapsto \ell^2 \mathrm{g}_{ij}$
\cite{SvsCtensor}. The result is
\begin{gather}
S_{\text{\sc vd}}^L = \frac{1}{2\kappa} \int_{t_i}^{t_f} {\rm d}t \int_{\Sigma}
{\rm d}^dx \frac{1}{4n} e_0(q_{ij}) G(q)^{ij,kl} e_0(q_{kl}).
\label{VDLaction}
\end{gather}
Here we renamed the fields $\bigl(N, N^i, \mathrm{g}_{ij}\bigr)$ of
\eqref{1daction} into $\bigl(\nu,\nu^i,q_{ij}\bigr)$ to avoid confusion
with the original ADM fields. The densitized lapse $n$ is interpreted
in~\eqref{1daction} as $n = N/\sqrt{\mathrm{g}}$ and in~\eqref{VDLaction}
as $n = \nu/\sqrt{q}$, however. Similarly, $e_0$ is read as
$\partial_t - \mathcal{L}_{\vec{N}}$ in~\eqref{1daction} and as $\partial_t - \mathcal{L}_{\vec{\nu}}$
in~\eqref{VDLaction}. The action~\eqref{VDLaction} is equivalent
to the one obtained as the ``zero signature limit'' by
Henneaux~\cite{Henn1} and the system is now often referred to as
``electric Carroll gravity'' \cite{Carrollgrav1}. The equivalence to~\eqref{VDLaction} can be understood~\cite{SvsCtensor} via the redundancy
of the Ehresmann connection.

Evaluating~\eqref{1daction} on the metrics of the form~\eqref{2Kmetric}
results in the action $S^L_{\text{\sc 2k}}$ of the two-Killing vector
reduction in equation~\eqref{2KLaction} below. This action too can
be subjected to a~scaling limit~${(\ell \rightarrow \infty})$ that
models the enhancement of the spatial metric for geometries
of the form~\eqref{2Kmetric}. The scale transformation
is described in~\eqref{scale1}, \eqref{scale2} below and
results in the limiting action~$S^L_{\text{\sc 2kvd}}$ in
equation~\eqref{scale3} below. On the other hand, one can also
directly consider a~two-Killing vector reduction of $S^L_{\text{\sc vd}}$.
This requires the notion of a~Carroll Killing vector which
we introduce shortly. Performing a~symmetry reduction of
$S^L_{\text{\sc vd}}$ with respect to two commuting Carroll
Killing vectors results in the same action as $S^L_{\text{\sc 2kvd}}$,
schematically $S^L_{\text{\sc 2kvd}} = S^L_{\text{\sc vd2k}}$. The
interplay between the VD limit and the two-Killing vector reduction
has been summarized in the commutative diagram in Figure~\ref{2Kdiagram}
of the introduction.

Referring to the figure, only the right vertical arrow requires further justification. To this end,
we recall the definition: a~\textit{Carroll structure} consists
of a~$(1+d)$-dimensional smooth manifold equipped with a~degenerate metric
$Q_{IJ} {\rm d}X^I {\rm d}X^J$ of rank $d$ and a~vector field $n^I \partial/\partial X^I$,
such that~${Q_{IJ} n^J =0}$. In this setting we call $K= K^I \partial /\partial X^I$ a~Killing vector for the Carroll structure if
$
\mathcal{L}_K Q_{IJ} = 0$, $
\mathcal{L}_K n^I =0$.
This is a~fully covariant definition and results in a~symmetry
reduced Carroll structure. For a~somewhat different notion of a~Carroll Killing vector, see~\cite{CarrollKilling}. In the
application to VD gravity, we can draw on the results of
\cite{SvsCtensor} to identify
\begin{gather}
Q_{IJ} = \begin{pmatrix} \nu^k q_{kl} \nu^l & q_{jk} \nu^k
 \\ q_{ik} \nu^k & q_{ij} \end{pmatrix}, \qquad
n^I n^J = \begin{pmatrix} \nu^{-2} & - \nu^{-2} \nu^j
 \\ - \nu^{-2} \nu^i & \nu^{-2}\nu^i \nu^j \end{pmatrix}.
\label{VDCarroll}
\end{gather}
Here the rows and columns refer to coordinates
$\bigl(X^0, X^1, \ldots, X^d\bigr) = \bigl(t, x^1, \ldots, x^d\bigr)$ and their
differentials or vector fields, respectively.
Consistency with the non-standard action of the full diffeomorphism
group on $\bigl(\nu,\nu^i, q_{ij}\bigr)$ requires that
$q_{ij}\bigl({\rm d}t \nu^i + {\rm d}x^i\bigr)\bigl({\rm d}t \nu^j + {\rm d}x^j\bigr)$ and
$\nu^{-2} \bigl(\partial_t - \nu^i \partial_i\bigr)^2$ are separately invariant under it.
This is indeed the case, see \cite[equation~(21)]{SvsCtensor}.
In other words, \eqref{VDCarroll} provides a~Carroll structure
for VD gravity. To this VD Carroll structure, we now apply the
reduction with respect to two commuting spacelike Killing vectors.
The purely differential geometrical line of reasoning
(see, e.g.,~\cite[Appendix~C]{Straumannbook}) implying a~block diagonal
form of the metric, here $Q_{IJ} {\rm d}X^I {\rm d}X^J$, still applies
and leads to the identification
\begin{equation*}
q_{ij} = \begin{pmatrix} {\rm e}^{\tilde{\sigma}} &
 \begin{matrix} 0 & 0 \end{matrix} \\
 \begin{matrix} 0 \\ 0 \end{matrix} &
 \rho M_{ab} \end{pmatrix}, \qquad
\nu^i= \begin{pmatrix} s \\ 0 \\0 \end{pmatrix},
\end{equation*}
where now $d=3$, $\bigl(x^1,x^2,x^3\bigr) = \bigl(x^1, y^1,y^2\bigr)$, and the
matrix entries are independent of the Killing coordinates $y^1$, $y^2$,
but carry a~tacit $t$ dependence as well. A~simple computation then gives
\begin{gather}
e_0(q_{ik} ) q^{kj} =
\begin{pmatrix} e_0(\tilde{\sigma}) &
 \begin{matrix} 0 & 0 \end{matrix} \\
 \begin{matrix} 0 \\ 0 \end{matrix} &
 e_0(\rho M_{ac}) \rho^{-1} M^{cb}\end{pmatrix},
\label{VD2Kcurrent}
\end{gather}
where automatically the correct one-dimensional Lie derivatives
contained in $e_0$ arise: $e_0(\tilde{\sigma}) = \partial_t \tilde{\sigma}
- s \partial_1 \tilde{\sigma} -2 \partial_1 s$, and $e_0( \rho M_{ab}) =
\partial_t (\rho M_{ab}) - s \partial_1 ( \rho M_{ab})$. Using~\eqref{VD2Kcurrent}
to evaluate~\eqref{VDLaction}, one recovers equation~\eqref{scale3},
i.e.,~$S^L_{\text{\sc vd2k}} = S^L_{\text{\sc 2kvd}}$.
This justifies the vertical arrow
in Figure~\ref{2Kdiagram} and turns it into a~commutative diagram.

\subsection{Reduced Lagrangian action and its symmetries}

Inserting~\eqref{2Kmetric} into the $1+3$ action~\eqref{1daction} and
interpreting ${\rm d}^2 y/(2\kappa)$ as $\mbox{(unit volume)}/\lambda_{\text{\sc n}}$, one finds the
following reduced action:
\begin{align}
S^L_{\text{\sc 2k}}={} & \frac{1}{\lambda_{\text{\sc n}}} \int {\rm d}^2 x \left\{ - \frac{1}{n} e_0(\rho) e_0(\sigma)
+ n \bigl( \partial_1 \rho \partial_1 \sigma - 2 \partial_1^2 \rho\bigr)\right.
\nonumber\\
&\left.+
\frac{\rho}{2n} \frac{e_0(\Delta)^2 + e_0(\psi)^2}{\Delta^2}
-\frac{\rho n}{2} \frac{(\partial_1\Delta)^2 + (\partial_1\psi)^2}{\Delta^2} \right\}.
\label{2KLaction}
\end{align}
Here $\lambda_{\text{\sc n}} >0$ is the dimensionless reduced Newton constant and
$e_0 = \partial_0 - \mathcal{L}_{s}$ is the derivative transversal to
the leaves of the $x^0 = \mathrm{const}$ foliation with metric
\eqref{2Dmetric}. It acts on two-dimensional spatial tensor
densities according to their tensor and density type. In
particular, $e_0(\varphi) = \partial_0 \varphi - s \partial_1 \varphi$, on scalars
like $\rho$, $\Delta$, $\psi$. On the other hand,
$\sigma := \tilde{\sigma} + \frac{1}{2} \ln \rho$ is the logarithm
of a~spatial $+2$ density and $e_0$ acts according to
$e_0(\sigma) = \partial_0 \sigma - s \partial_1 \sigma -2 \partial_1 s$.
The action~\eqref{2KLaction} can be seen to produce the correctly
specialized Einstein field equation. That is, ``reduction of the action''
and ``variation to obtain the field equations'' are commuting
operations. Alternatively, one can appeal to the ``principle of
symmetric criticality'' \cite{Torre} to infer that the latter must hold on
general grounds.

\textbf{Local gauge symmetries.}
The above form of the reduced action masks its residual local
gauge symmetries. Indeed, from~\eqref{2Kmetric} one expects that
two-dimensional diffeomorphism invariance is preserved. This is indeed
the case and could be rendered manifest by rewriting~\eqref{2KLaction} in
terms of $\rho \sqrt{-\gamma} R^{(2)}(\gamma)$, where $R^{(2)}(\gamma)$
is the Ricci scalar of the metric~\eqref{2Dmetric}. The linearized
gauge transformations of the reduced action can be found by a~careful
specialization of the gauge variations of the $1+d$ action
\eqref{1daction}. Omitting the details, one obtains the following
result, which was verified by direct computation.

\begin{Result} \label{resultSLgaugeinv}
Let $S^L_{\text{\sc 2k}}$ denote the action
\eqref{2KLaction} with the
temporal integration restricted to $x^0= t \in [t_i, t_f]$ and fall-off
or boundary conditions in $x^1$ that ensure the absence of spatial boundary
terms. Then
\begin{gather}
\delta_{\epsilon} S^L_{\text{\sc 2k}} = 0, \qquad
\epsilon|_{t_i} = 0 = \epsilon|_{t_f},
\label{ginv}
\end{gather}
with the following local gauge variations:
\begin{align}
\delta_{\epsilon} \sigma &= \frac{\epsilon}{n} (\partial_0\sigma -
s \partial_1 \sigma - 2 \partial_1 s) +
\epsilon^{1} \partial_{1} \sigma + 2\partial_{1} \epsilon^{1}
\nonumber\\
\delta_{\epsilon} n &= \partial_{0} \epsilon
- (s \partial_{1} \epsilon - \epsilon \partial_{1} s)
+ \epsilon^{1} \partial_{1} n - n\partial_{1} \epsilon^{1},
\nonumber\\
\delta_{\epsilon} s &= \partial_{0} \epsilon^{1}
+\bigl(\epsilon^{1} \partial_{1} s - s\partial_{1} \epsilon^{1}\bigr)
+ \epsilon \partial_{1} n - n \partial_{1} \epsilon,
\qquad
\delta_{\epsilon} \varphi = \frac{\epsilon}{n} (\partial_0 \varphi - s\partial_1\varphi) +
\epsilon^{1} \partial_{1} \varphi.
\label{gtrans1}
\end{align}
Here $\varphi$ can represent any scalar field such as $\rho$, $\psi$, or
$\Delta$. Further, $\epsilon$, $\epsilon^1$ are in one-to-one correspondence
with a~linearized diffeomorphism, $t' = t - \xi^0(t,x) + O\bigl(\xi^2\bigr)$,
${x'}^1 = x^1 - \xi^1(t,x) + O\bigl(\xi^2\bigr)$, via the following relations
$\xi^{0} = \epsilon/n$, $\xi^{1} = \epsilon^{1} - s \epsilon/n$.
For spatially noncompact geometries, the descriptors $\bigl(\epsilon,\epsilon^1\bigr)$
are smooth with compact support; for spatially periodic boundary,
conditions the descriptors are spatially periodic as well.
\end{Result}

For all, but $\sigma$ the spatial variations are all of the form
$\mathcal{L}_{\epsilon^1} d = \epsilon^1 \partial_1 d + pd \partial_1 \epsilon^1$, for a~spatial density
$d$ of weight $p$. For ${\rm e}^{\sigma}$, this holds as well with $p=2$,
the difference arises from taking the logarithm.

For later use, we also note the finite version of the gauge
transformations~\eqref{gtrans1}. These arise from the usual tensorial
transformation law for $\gamma_{\mu\nu}$ in~\eqref{2Dmetric} under a~generic diffeomorphism, but are transcribed to the ${\rm e}^{\tilde{\sigma}}$,
$s$, $n$, component fields. The result can be obtained by specialization
of the general result for the $(1 + d)$-dimensional ADM fields, see
\cite{SvsCtensor}. To unclutter the notation, we momentarily
write $t = x^0$, $x = x^1$, without this indicating a~preferred choice
of coordinates in~\eqref{2Dmetric}. Let then $t' = \chi^0(t,x)$,
$x' = \chi^1(t,x)$ be a~generic smooth diffeomorphism of gauge-type
(i.e.,~of compact support or spatially periodic, respectively.)
As usual, we denote the transformed fields by a~`prime' and take them
to be evaluated
at the image point with the `primed' coordinates. For a~scalar $\varphi$,
the transformation law thus reads $\varphi'(t',x') = \varphi(t,x)$. In this
notation, one has
\begin{gather}
n' = \dfrac{ n \Big| \dfrac{\partial t'}{\partial t} \dfrac{\partial x'}{\partial x} \Big|}%
{ \Big|
 - n^2 \left( \dfrac{\partial t'}{\partial x} \right)^2 +
 \left( \dfrac{\partial t'}{\partial t} - s
 \dfrac{ \partial t'}{\partial x} \right)^2 \Big|},
\nonumber\\
s' = \dfrac{ n^2 \dfrac{\partial t'}{\partial x} \dfrac{\partial x'}{\partial x}
 - \left( \dfrac{\partial x'}{\partial t} - s \dfrac{ \partial x'}{\partial x} \right)
 \left( \dfrac{\partial t'}{\partial t} - s \dfrac{ \partial t'}{\partial x} \right)}%
{ n^2 \left( \dfrac{\partial t'}{\partial x} \right)^2 - \left( \dfrac{\partial t'}{\partial t} - s
 \dfrac{ \partial t'}{\partial x} \right)^2 },
\nonumber\\
{\rm e}^{\tilde{\sigma}'} = {\rm e}^{\tilde{\sigma}} \left[ - n^2
 \left( \dfrac{\partial t}{\partial x'} \right)^2 +
\left( \dfrac{\partial x}{\partial x'} + s
 \dfrac{ \partial t}{\partial x'} \right)^2 \right].
\label{gtrans3}
\end{gather}
 One may check that upon linearization $t' = t - \xi^0(t,x) +
 O\bigl(\bigl(\xi^0\bigr)^2\bigr)$, $x' = x - \xi^1(t,x) + O\bigl(\bigl(\xi^1\bigr)^2\bigr)$ and use of the
 reparameterization~\eqref{gtrans2} the relations~\eqref{gtrans3} induce
 \eqref{gtrans1} (where $\tilde{\sigma}$ transforms like $\sigma$
 in the first line of~\eqref{gtrans1}, in particular $\mathcal{L}_{\epsilon^1}
 \tilde{\sigma} = \epsilon^1 \partial_1 \tilde{\sigma} +2 \partial_1 \epsilon^1$).

\subsection{VD limit of two Killing vector reduction}

We now return to Figure~\ref{2Kdiagram} and introduce the before-mentioned
scale transformation
that enhances spacelike distances compared to timelike ones. Denoting
the scale parameter by $\ell\geq 1$, consider
\begin{alignat}{5}
& n \mapsto \ell^{-1} n, \qquad && s \mapsto s,
 \qquad&&  \lambda_{\text{\sc n}} \mapsto \ell^3 \lambda_{\text{\sc n}}, && &
\nonumber\\
&\rho \mapsto \ell^2 \rho, \qquad&&  \sigma \mapsto
\sigma + 3 \ln \ell, \qquad&&
\Delta \mapsto \Delta, \qquad&& \psi \mapsto \psi,&
\nonumber\\
&\pi^{\rho} \mapsto \ell^{-2} \pi^{\rho},\qquad&&
\pi^{\sigma} \mapsto \pi^{\sigma}, \qquad&&
\pi^{\Delta} \mapsto \pi^{\Delta}, \qquad&&
\pi^{\psi} \mapsto \pi^{\psi}.&
\label{scale1}
\end{alignat}
Since $\sigma = \tilde{\sigma} + \frac{1}{2} \ln \rho$,
this is such that spacelike distance elements are enhanced
\begin{align}
 g^\mathrm{2K}_{IJ}(X) {\rm d}X^I {\rm d}X^J \mapsto{}& - {\rm e}^{\tilde{\sigma}} n^2 \bigl({\rm d}x^0\bigr)^2+
\ell^2 \left\{ {\rm e}^{\tilde{\sigma}}\bigl({\rm d}x^1+ s {\rm d}x^0\bigr)^2\right.
\nonumber\\
& \left. +
\frac{\rho(x)}{\Delta(x)} \bigl({\rm d}y^2 + \psi(x) {\rm d}y^1\bigr)^2 +
\rho(x) \Delta(x) \bigl({\rm d}y^1\bigr)^2 \right\}.
\label{scale2}
\end{align}
The rescaling of Newton's constant $\lambda_{\text{\sc n}}$ is included in order for the action
\eqref{2KLaction} to have an invariant leading piece. Explicitly,
\begin{gather}
 S^L_{\text{\sc 2k}} \mapsto S^L_{\text{\sc 2kvd}} -
\frac{1}{\ell^2} \mathcal{V}_{\text{\sc 2k}},
\qquad
 S^L_{\text{\sc 2kvd}} = \frac{1}{\lambda_{\text{\sc n}}} \int {\rm d}^2 x
\left\{ - \frac{1}{n} e_0(\rho) e_0(\sigma)
+ \frac{\rho}{2n} \frac{e_0(\Delta)^2 + e_0(\psi)^2}{\Delta^2} \right\},
\nonumber\\
 \mathcal{V}_{\text{\sc 2k}} = \frac{1}{\lambda_{\text{\sc n}}} \int_{t_i}^{t_f} {\rm d} x^0 \int_{0}^{2\pi} {\rm d} x^1
n \left\{
\frac{\rho}{2} \frac{(\partial_1\Delta)^2 + (\partial_1\psi)^2}{\Delta^2}
- \partial_1 \rho \partial_1 \sigma + 2 \partial_1^2 \rho \right\}.
\label{scale3}
\end{gather}
The decomposition is designed such that for large $\ell \gg 1$ the kinetic
term in~\eqref{scale3} dominates and thus lends itself
to the exploration of the velocity dominated regime.

The Hamiltonian action~\eqref{Haction} has an analogous decomposition
\begin{equation*}
S^H_{\text{\sc 2k}} \mapsto S^H_{\text{\sc 2kvd}} -
\frac{1}{\ell^2} \mathcal{V}_{\text{\sc 2k}},
\end{equation*}
with the same $\mathcal{V}_{\text{\sc 2k}}$ as in~\eqref{scale3} and
$S^H_{\text{\sc 2kvd}}$ of the same form as
\eqref{Haction}, but with a~simplified Hamiltonian constraint
\begin{equation*}
\mathcal{H}_{\text{\sc 0vd}} = - \lambda_{\text{\sc n}} \pi^{\sigma} \pi^{\rho}
+ \frac{\lambda_{\text{\sc n}}}{2\rho} \Delta^2 \bigl(
\bigl(\pi^{\psi}\bigr)^2+\bigl(\pi^{\Delta}\bigr)^2 \bigr).
\end{equation*}
Since the shift is still exclusively carried by the $e_0$ derivatives,
the diffeomorphism constraint retains its form, $\mathcal{H}_{\text{\sc 1vd}}
= \mathcal{H}_1$. For the generator of gauge variations, this gives
\begin{equation*}
\mathcal{H}_0(\epsilon) + \mathcal{H}_1\bigl(\epsilon^1\bigr) \mapsto
\mathcal{H}_{\text{\sc 0vd}}(\epsilon) + \mathcal{H}_1\bigl(\epsilon^1\bigr) +
\frac{1}{\ell^2} V(\epsilon).
\end{equation*}
Here $V(\epsilon)$ is the spatial subintegral in $\mathcal{V}_{\text{\sc 2k}}$ with
$n$ replaced by $\epsilon$. Further, we assume that $\bigl(\epsilon,\epsilon^1\bigr)$
scale like $(n,s)$, i.e., $\epsilon \mapsto \ell^{-1} \epsilon$,
$\epsilon^1 \mapsto \epsilon^1$.

Remarkably, the leading term $S^L_{\text{\sc 2kvd}}$ in~\eqref{scale3}
still has a~local gauge invariance parameterized by two arbitrary
functions of $\bigl(x^0, x^1\bigr)$.\footnote{Strictly speaking, when viewing $S^L_{\text{\sc 2kvd}}$ as a~stand-alone gravity action its dynamical variables should be given
different names. For readability's sake, we retain here the same symbols,
$\Delta$, $\psi$, $\rho$, $\sigma$, $n$, $s$, the distinction will be clear
from the context.}
Its form is easiest to infer by scaling from~\eqref{gtrans1} by
noting that $n \mapsto \ell^{-1} n$ needs to be accompanied by
$\epsilon \mapsto \ell^{-1} \epsilon$, $\epsilon^1 \mapsto \epsilon^1$,
in order to have nondegenerate linearized diffeomorphisms in
\eqref{gtrans2}. Writing $\delta^{\text{\sc vd}}_{\epsilon}$ for the
limiting gauge variation, one finds
\begin{gather}
\delta^{\text{\sc vd}}_{\epsilon} S^L_{\text{\sc 2kvd}} = 0, \qquad
 \epsilon|_{t_i} = 0 = \epsilon|_{t_f}.
\label{scale4}
\end{gather}
Explicitly, the $\delta_{\epsilon}^{\text{\sc vd}}$ gauge transformations
coincide with the
ones in~\eqref{gtrans1} for all but the shift field. The latter's variation
reads $\delta_{\epsilon}^{\text{\sc vd}} s =
\partial_0 \epsilon^1 + \epsilon^1 \partial_1 s
- s \partial_1 \epsilon^1$, and is independent of the temporal descriptor
$\epsilon$. As before, the invariance~\eqref{scale4} can also be verified
by direct computation.

The finite gauge transformations for the velocity dominated system can
be found by taking the scaling limit of those in~\eqref{gtrans3}.
Subjecting both the `primed' and the `unprimed' fields in~\eqref{gtrans3}
to~\eqref{scale1} and taking the limit $\ell \rightarrow \infty$, one finds
\begin{gather}
n' =\dfrac{ n \Big| \dfrac{\partial t'}{\partial t} \dfrac{\partial x'}{\partial x} \Big|}%
{ \left( \dfrac{\partial t'}{\partial t} - s
 \dfrac{ \partial t'}{\partial x} \right)^2},
\qquad
s' = -\dfrac{
\dfrac{\partial x'}{\partial t} - s \dfrac{ \partial x'}{\partial x}}%
{\dfrac{\partial t'}{\partial t} - s\dfrac{ \partial t'}{\partial x}},
\qquad
{\rm e}^{\tilde{\sigma}'} = {\rm e}^{\tilde{\sigma}}
\left( \dfrac{\partial x}{\partial x'} + s
 \dfrac{ \partial t}{\partial x'} \right)^2,
\label{gtransVD}
\end{gather}
together with $\varphi' = \varphi$ for the scalars. Note that the
transformation law for $s$ is decoupled from the others.
Upon linearization
the transformations~\eqref{gtransVD} induce the gauge
variation $\delta^{\text{\sc vd}}_{\epsilon}$ described before. Since the latter
coincide with those $\delta_{\epsilon}$ in~\eqref{gtrans1} for all but $s$,
while the finite versions differ, this also highlights that the
linearized gauge transformations contain strictly less information
than the finite ones~\eqref{gtrans3} or~\eqref{gtransVD}. For
later use, we also note the following implications~\cite{SvsCtensor}:
under~\eqref{gtransVD} one has ${n'}^{-1} e_0'(\varphi') =
n^{-1} e_0(\varphi)$ for any scalar, where $e'_0 = \partial/\partial t' -
s' \partial/\partial x'$. Further, $n^{-1} e_0\bigl( {\rm e}^{\tilde{\sigma}}\bigr)$
transforms like ${\rm e}^{\tilde{\sigma}}$ under~\eqref{gtransVD}.

\section[Construction of U and J\^{}H\_0 for T\^{}3 Gowdy]{Construction of $\boldsymbol{U}$ and $\boldsymbol{\mathcal{J}^H_0}$ for $\boldsymbol{T^3}$ Gowdy}\label{appB}

Here we present the derivation of Results~\ref{resultJtheta}
and~\ref{resultT3obs}.
In the $T^3$ Gowdy system, the direct limit definition~\eqref{UdefNC}
of $U$ is no longer applicable, because $L_1^H$ only
falls off like $O\bigl(1/\big|x^1\big|\bigr)$, which produces a~logarithmic divergence in
the integral~\eqref{THsol1}. In order to construct a~(finite) $U$ with
the required gauge variation~\eqref{Ugauge}, a~renormalization is
needed that removes the logarithmic divergence. With such a~$U$ in place,
the current $\mathcal{J}_0^H(\theta)$ can be defined but, due to
boundary terms, its integral does not give rise to Dirac observables
yet. A~$2\pi$-periodic extension accomplishes this but only
subject to a~subsidiary condition on $\operatorname{tr}Q^2$. In the
last step, this condition is removed by considering
$\operatorname{tr}\bigl\{ \nu \mathcal{J}_0^H(\theta)\bigr\}$, with a~suitable matrix $\nu$.
For simplicity, we set $\lambda_{\text{\sc n}} =1$ in this appendix.

\subsection[The renormalized U]{The renormalized $\boldsymbol{U}$}

The starting point is the family of regularized transition matrices
$T^{\pm}_{\text{\sc l}}$, $\text{\sc l} \in \mathbb{N}$, introduced in~\eqref{Uregdef}. We shall
refer to its limit as the `renormalized' $U_{\pm}$ and aim at the
following result.

\begin{Result} \label{resultB1}\quad
\begin{itemize}\itemsep=0pt
\item[$(a)$] The limit \smash{$U_{\pm}\bigl(x^0,x^1;\theta\bigr) :=
 \lim_{\text{\sc l} \rightarrow \infty} T_{\text{\sc l}}^{\pm}\bigl(x^0,x^1;\theta\bigr)$}
 exists pointwise for all $x^0$ and $x^1 \in [0,2\pi]$ such that
 $\pm \operatorname{Re}(\theta + \tilde{\rho})>0$, $\rho>0$. It is
 differentiable in $x^1$, solves \smash{$\partial_1 U_{\pm} = U_{\pm} L_1^{H,\pm}$},
 and has the fall-off \smash{$U_{\pm}\bigl(x^0,x^1;\theta\bigr) = \mathbbm{1} + O(1/\theta)$},
 $\pm\operatorname{Re}\theta >0$,
pointwise in $x^0$ and $x^1 \in [0,2\pi]$.
\item[$(b)$] $U_{\pm}$ can be extended to other intervals
 $[2\pi (l - 1),2\pi l], l \in \mathbb{Z}$, using the quasi-periodicity
 property
\begin{gather}
U_{\pm}\bigl(x^0,x^1 + 2\pi;\theta\bigr) = {\rm e}^{ - \frac{Q}{2\theta}}
U_{\pm}\bigl(x^0,x^1;\theta - \pi_0^{\sigma}\bigr).
\label{quasiperB1}
\end{gather}
\item[$(c)$] Assume \smash{$\operatorname{tr} Q^2 < 2 \bigl(\pi_0^{\sigma}\bigr)^2$} and
 $\pm \pi_0^{\sigma} >0$, respectively, in addition to the conditions
 in $(a)$. Then the gauge variation $\delta_{\epsilon}^H U_{\pm}
 = U_{\pm} C_{\epsilon}^{\pm}$ holds, with $C_{\epsilon}^{\pm}$
 from~\eqref{THsol5}.
\end{itemize}
\end{Result}

In order to elucidate the origin of the regularization
and the ensued properties, we do not present the derivation
in a~strict ``proof of Result~\ref{resultB1}'' format.
Also, for part (c) some group theoretical results are needed
which we relegate to Appendix~\ref{appb2}. Aiming at the
first statement in part (a) we focus on the second term in the
exponent of~\eqref{Uregdef}.

To study its $\text{\sc l} \rightarrow \infty$
limit, we prepare the behavior of \smash{$L_1^{H,\pm}$}'s integral.
Since \smash{$J_0^H$}, \smash{$J_1^H$} and $\rho$ in~\eqref{LHdef} are periodic, the
large $x^1$ behavior of \smash{$L_1^{H,\pm}$} is governed by that of
$\tilde{\rho} +\theta$.
By~\eqref{rhotHqperiodic}, one has $\tilde{\rho}\bigl(x^0, x^1 - 2\pi l\bigr)
= \tilde{\rho}\bigl(x^0, x^1\bigr) + l\pi_0^{\sigma}$. Taking $x^1 \in [0,2\pi]$
we write
\begin{gather}
\int_{-2\pi \text{\sc l}}^{x^1} {\rm d}x f(x) = \int_0^{x^1} {\rm d}x f(x)
+ \sum_{l=1}^{\text{\sc l}} \int_{0}^{2\pi} {\rm d}x f(x -l 2\pi).
\label{Tlimit1}
\end{gather}
The integrand \smash{$f = L_1^{H,\pm}$} in question we decompose according to
\begin{gather}
L_1^{H,\pm}(\theta) = - \frac{1}{2 \theta} J_0^H + \alpha_{\pm}(\theta) J_0^H
+ \beta_{\pm}(\theta) \frac{J_1^H}{2 \rho n},\nonumber
\\
\alpha_{\pm}(\theta) = \frac{1}{2 (\theta + \tilde{\rho})}
\left( \frac{\tilde{\rho}}{\theta} - \beta_{\pm}(\theta) \right) =
\frac{1}{2 \theta} \mp \frac{1}{2\bigl[(\theta + \tilde{\rho})^2 - \rho^2\bigr]^{1/2}},
\nonumber\\
\beta_{\pm}(\theta) =
\pm \frac{\theta + \tilde{\rho}}%
 {\bigl[(\theta + \tilde{\rho})^2 - \rho^2\bigr]^{1/2}} -1,
\label{Tlimit2}
\end{gather}
where we suppress the shared $\bigl(x^0,x^1\bigr)$ arguments.
Note that $\beta_{\pm}(\theta)$ and $\alpha_{\pm}(\theta) - 1/(2\theta)$
are quasi-periodic under \smash{$x^1 \mapsto x^1 + 2\pi$}, but
$\alpha_{\pm}(\theta)$ separately is not.
Inserting~\eqref{Tlimit2} into~\eqref{Tlimit1} with~$\smash{f = L_1^{H,\pm}}$, one has
\begin{gather}
\sum_{l=1}^\text{\sc l} \int_0^{2\pi} {\rm d}x L_1^{H,\pm}(x - l 2\pi;\theta) =
- \frac{Q}{2} \varkappa_{\text{\sc l}}(\theta) +
\int_0^{2\pi} {\rm d}x \left\{ a_{\text{\sc l}}(\theta) J_0^H
+ b_{\text{\sc l}}(\theta) \frac{J^H_1}{2 \rho n} \right\}.
\label{aux1}
\end{gather}
Here \smash{$Q = \int_0^{2\pi} {\rm d}x J_0^H$} is the $\mathfrak{sl}(2,\mathbb{R})$
Noether charge, which is gauge invariant by itself, $\delta_{\epsilon}^H Q =0$.
The partial sums encountered are
\begin{gather}
\varkappa_{\text{\sc l}}(\theta) := \sum_{l=1}^{\text{\sc l}}
\frac{1}{\theta + l \pi_0^{\sigma}},\qquad
\check{\varkappa}_{\text{\sc l}}(\theta) = \sum_{l=0}^{\text{\sc l}-1}
\frac{1}{\theta - l\pi_0^{\sigma}},\qquad
a_{\text{\sc l}}^{\pm}(\theta) := \sum_{l=1}^{\text{\sc l}}
\alpha_{\pm}\bigl(\theta + l \pi_0^{\sigma}\bigr),\nonumber\\
b_{\text{\sc l}}^{\pm}(\theta) := \sum_{l=1}^{\text{\sc l}}
\beta_{\pm}\bigl(\theta + l \pi_0^{\sigma}\bigr),
\label{aux2}
\end{gather}
where we prepare $\check{\varkappa}_{\text{\sc l}}(\theta)$ for later use.
Moving the $Q$ term in~\eqref{aux1} to the left-hand side, we obtain
for $L \in \mathbb{N}$
\begin{gather}
T_{\text{\sc l}}^{\pm}\bigl(x^0,x^1;\theta\bigr) := \exp_+\left\{
\frac{Q}{2} \varkappa_{\text{\sc l}}(\theta) + \int_{- 2\pi \text{\sc l}}^{x^1} {\rm d}x
L_1^{H,\pm}\bigl(x^0,x;\theta\bigr) \right\}
\nonumber\\
\phantom{T_{\text{\sc l}}^{\pm}\bigl(x^0,x^1;\theta\bigr) :}{}= {\rm e}^{R_L^{\pm}\bigl(x^0,\theta\bigr)}\exp_+\left\{
\int_0^{x^1} {\rm d}x L_1^{H,\pm}\bigl(x^0,x;\theta\bigr)\right\}, \qquad x^1 \in[0,2\pi],
\nonumber\\
R_{\text{\sc l}}^{\pm}\bigl(x^0,\theta\bigr) := \int_0^{2\pi} {\rm d}x \left\{
a_{\text{\sc l}}^{\pm}(\theta)J_0^H + b_{\text{\sc l}}^{\pm}(\theta) \frac{J_1^H}{2n\rho} \right\}.
\label{Tlimit5}
\end{gather}
For $L=0$, we identify $T_0^{\pm}\bigl(x^0,x^1;\theta\bigr)$ with just the $\exp_+$ term.
The convergence of the path ordered exponential is ensured under broad
conditions. For example, a~bound
\begin{equation*}
\sup_{x \in [0,2\pi]}
\Vert L_1^{H,\pm}\bigl(x^0,x,\theta\bigr) \Vert \leq c\bigl(x^0,\theta\bigr)
\end{equation*}
for some matrix norm
gives $\exp\bigl\{ x^1 c\bigl(x^0,\theta\bigr)\bigr\}$
as a~bound on the exponential in the same norm. The extension to
other spatial intervals $x^1 \in [l 2\pi, (l+1) 2\pi]$, $l\in \mathbb{Z}$,
can be done using the quasi-periodicity property of $L_1^{H,\pm}$.
One finds slightly different expressions for the right and
the left translates
\begin{align}
T_{\text{\sc l}}^{\pm}\bigl(x^0;x^1+n 2\pi;\theta\bigr) &= {\rm e}^{- \frac{Q}{2} \check{\varkappa}_n(\theta)}
T^{\pm}_{L + n} \bigl(x^0, x^1; \theta - n \pi_0^{\sigma}\bigr), \qquad n \in \mathbb{N},
\nonumber\\
T_{\text{\sc l}}^{\pm}\bigl(x^0;x^1- n 2\pi;\theta\bigr) &= {\rm e}^{\frac{Q}{2} \varkappa_n(\theta)}
T^{\pm}_{L - n} \bigl(x^0, x^1; \theta + n \pi_0^{\sigma}\bigr),\qquad
n =1, \ldots, L,
\label{TLquasiper}
\end{align}
with $\check{\kappa}_n(\theta)$, $\kappa_n(\theta)$ from~\eqref{aux2}.
Specializing the second equation to $n =L$, $x^1=0$, shows that
\begin{gather}
T^{\pm}_{\text{\sc l}}\bigl(x^0, -2 \pi \text{\sc l};\theta\bigr) = \exp\{ Q \varkappa_{\text{\sc l}}(\theta)/2\},
\label{TLbc}
\end{gather}
consistent with the naive specialization of~\eqref{Tlimit5}. This entails
that \smash{$T_{\text{\sc l}}^{\pm}$} can be characterized as the solution of
\smash{$\partial_1 T^{\pm}_{\text{\sc l}} =T^{\pm}_{\text{\sc l}} L_1^{H,\pm}$} with alternative
boundary conditions
\[
T^{\pm}_{\text{\sc l}}\bigl(x^0,0;\theta\bigr) =
\exp\bigl\{ R_{\text{\sc l}}^{\pm}\bigl(x^0,\allowbreak\theta\bigr)\bigr\}
\]
 or~\eqref{TLbc}.

Next, we consider the $\text{\sc l} \rightarrow \infty$ limit of~\eqref{Tlimit5}.
For $\varkappa_{\text{\sc l}}(\theta)$, one has
\begin{gather}
\varkappa_{\text{\sc l}}(\theta) = \frac{1}{\pi_0^{\sigma}}
 [ \psi( 1 + \text{\sc l} + \theta/\pi_0^{\sigma} ) -
\psi( 1 + \theta/\pi_0^{\sigma} )] =
\frac{\ln \text{\sc l}}{\pi_0^{\sigma}}
- \frac{1}{\pi_0^{\sigma}} \psi\bigl( 1 + \theta/\pi_0^{\sigma}\bigr) +
O(1/\text{\sc l}).
\label{aux4}
\end{gather}
This $\ln \text{\sc l}$ divergence is the reason for subtracting
$- Q \varkappa_{\text{\sc l}}(\theta)/2$ from the \smash{$\int_{-2 \pi \text{\sc l}}^{x^1}$} integral.
Subject to a~restriction on $\theta$, the other sums have finite
limits
\begin{gather}
\lim_{\text{\sc l} \rightarrow \infty} a^{\pm}_\text{\sc l}(\theta) = a_{\infty}(\theta), \qquad
\lim_{\text{\sc l} \rightarrow \infty} b^{\pm}_\text{\sc l}(\theta) = b_{\infty}(\theta), \qquad
\pm \operatorname{Re}(\theta + \tilde{\rho}) >0,\qquad
 \rho < |\theta + \tilde{\rho}|.\!\!
\label{ablimit}
\end{gather}
This follows from the rewriting
\begin{gather}
\beta_{\pm}(\theta) = \pm \frac{s(\theta+ \tilde{\rho})}{
\sqrt{ 1 - \rho^2/(\theta + \tilde{\rho})^2}} -1,
 \qquad \rho < |\theta + \tilde{\rho}|,
\label{betapm}
\end{gather}
where \smash{$s(z) = {\rm e}^{{\rm i} \operatorname{Arg}(z)}
\bigl({\rm e}^{ 2{\rm i} \operatorname{Arg}(z)}\bigr)^{-1/2} = \operatorname{sign}(\operatorname{Re}(z))$}. Clearly,
cancellation of the constant term for large $\theta$ is necessary and
sufficient for the $b_{\text{\sc l}}^{\pm}(\theta)$ sum to have a~convergent
$\text{\sc l} \rightarrow \infty$ limit. Once this is ensured also the $a_{\text{\sc l}}(\theta)$
sum will have a~limit, by inspection of the first form of $\alpha_{\pm}(\theta)$.
Thus we choose $\pm \operatorname{Re}(\theta + \tilde{\rho}) >0$ for the
$\pm$ cases, respectively. With this choice the explicit sign dependence
drops out already in the finite sums,
and since the limits exist only then, we omit the $\pm$ superscripts
in the limits.

With~\eqref{ablimit} in place, Result~\ref{resultB1}\,(a) follows,
except for the fall-off. Note that $R_{\text{\sc l}}^{\pm}\bigl(x^0,\theta\bigr)$ has a~limit
under the same conditions, which we denote by $R_{\infty}\bigl(x^0,\theta\bigr)$.
Then~\eqref{Tlimit5} implies that~$U_{\pm}$ are solutions
of \smash{$\partial_1 U_{\pm} = U_{\pm} L_1^{H,\pm}$} with boundary
condition $U_{\pm}\bigl(x^0,0;\theta\bigr) = \exp R_{\infty}\bigl(x^0,\theta\bigr)$, see~\eqref{Ubc}.

The extension via the quasi-periodicity property (b) follows by
taking the $\text{\sc l} \rightarrow \infty$ limit of~\eqref{TLquasiper}. For
simplicity, we only record the $n=1$ case of the first line in
\eqref{TLquasiper}. The limit of the left-hand side is well-defined
for $x^1 \in [0,2\pi]$ by (a), so the right-hand side has a~$\text{\sc l} \rightarrow \infty$ limit as well, which is naturally taken to define
$U_{\pm}\bigl(x^0, x^1 + 2\pi;\theta\bigr)$, yielding Result~\ref{resultB1}\,(b).

Next, we analyze the limit in more detail. For $\pm \operatorname{Re}(\theta +
\tilde{\rho}) >0$, one has the convergent expansion
\begin{equation*}
\beta_{\pm}(\theta) = \sum_{j \geq 1} \binom{-1/2}{j} (-)^j
\left(\frac{\rho}{\theta + \tilde{\rho}} \right)^{2j}, \qquad
\rho < |\theta + \tilde{\rho}|,
\end{equation*}
and similarly for $\alpha_{\pm}(\theta)$. The sums $b_{\text{\sc l}}(\theta)$,
$a_{\text{\sc l}}(\theta)$ can therefore be performed termwise and in the
limit $\text{\sc l} \rightarrow \infty$ one encounters just two types of sums
\begin{gather}
 s_p(\theta) :=
\sum_{l=1}^{\infty} \frac{1}{\bigl(\theta + \tilde{\rho} + l \pi_0^{\sigma}\bigr)^p} =
\frac{(-)^p}{\bigl(\pi_0^{\sigma}\bigr)^p} \frac{1}{(p - 1)!}
\psi^{(p-1)}\left(1 + \frac{\theta + \tilde{\rho}}{\pi_0^{\sigma}} \right),
 \qquad 2 \leq p \in \mathbb{N},
\nonumber\\
 c_{\infty}(\theta) := \sum_{l =1}^{\infty}
\frac{\tilde{\rho}}{2 \bigl(\theta + l \pi_0^{\sigma}\bigr)
 \bigl(\theta + \tilde{\rho} + l \pi_0^{\sigma}\bigr)}
= \frac{1}{2 \pi_0^{\sigma}}
\psi\left(1 + \frac{\theta + \tilde{\rho}}{\pi_0^{\sigma}} \right)
- \frac{1}{2 \pi_0^{\sigma}} \psi\left(1 + \frac{\theta}{\pi_0^{\sigma}} \right),
\label{aux3}
\end{gather}
where $\psi^{(p-1)}(z) = \partial_z^p \ln \Gamma(z)$, $p \in \mathbb{N}$, denotes the
DiGamma function and its derivatives. In terms of them, one has
\begin{align*}
a_{\infty}(\theta) &= c_{\infty}(\theta) - \frac{1}{4} \rho^2 s_3(\theta) -
\frac{3}{16} \rho^4 s_5(\theta) + O\bigl(\rho^6\bigr),\\
b_{\infty}(\theta) &= \frac{1}{2} \rho^2 s_2(\theta) + \frac{3}{8} \rho^4
s_4(\theta) + O\bigl(\rho^6\bigr).
\end{align*}
Since $\psi(z) = \ln z - 1/(2 z) + O\bigl(z^{-2}\bigr)$, one has
$c_{\infty}(\theta) = O(1/\theta)$ and $s_p(\theta) = O\bigl(\theta^{-(p-1)}\bigr)$. Hence
\begin{equation*}
a_{\infty}(\theta) = O(1/\theta) + O\bigl(\rho^2/\theta^2\bigr), \qquad
\rho^{-1} b_{\infty}(\theta) = O(\rho/\theta),
\end{equation*}
which ensures
\[
R_{\infty}\bigl(x^0,\theta\bigr) := \lim_{\text{\sc l} \rightarrow \infty} R_{\text{\sc l}}^{\pm}\bigl(x^0,\theta\bigr)
= O(1/\theta)
\]
 pointwise in $x^0$, as $|\theta| \rightarrow \infty$, $\pm\operatorname{Re}(\theta) >0$.
For \smash{$L_1^{H,\pm}(\theta)$}, one has with the
above sign choice \smash{$L_1^{H,\pm}(\theta) = - J_0^H/(2\theta) + O\bigl(\theta^{-2}\bigr)$},
cf.\ \eqref{LHpm2}. This shows{\samepage
\[
\int_{0}^{x^1} {\rm d}x L_1^{H,\pm}\bigl(x^0,x;\theta\bigr) + R_{\infty}\bigl(x^0,\theta\bigr) =
O(1/\theta), \qquad |\theta| \rightarrow \infty, \qquad \pm \operatorname{Re}(\theta) >0,
\]
and hence the decay asserted in the Result~\ref{resultB1}\,(a).}

{\bf Gauge variation and $\boldsymbol{{\operatorname{tr}Q^2}}$ condition.}
It remains to establish that the $U_{\pm}$ in Result~\ref{resultB1}\,(a) has
the desired gauge
variation~\eqref{Ugauge}. This turns out to hold only subject to a~subsidiary condition on $Q$. Recall that $T_{\text{\sc l}}^{\pm}$ is characterized by
\begin{gather}
\partial_1 T_{\text{\sc l}}^{\pm} = T_{\text{\sc l}}^{\pm} L_1^{H,\pm}, \qquad
T_{\text{\sc l}}^{\pm}\bigl(x^0,-2\pi \text{\sc l};\theta\bigr) =
\exp\left\{\frac{Q}{2}\varkappa_{\text{\sc l}}(\theta)\right\}.
\label{Tlimit7}
\end{gather}
Adapting the derivation of~\eqref{THsol5}, one finds for the gauge transformations
\begin{align}
\delta_{\epsilon}^H T_{\text{\sc l}}^{\pm}\bigl(x^0;x^1;\theta\bigr) ={}&
T_{\text{\sc l}}^{\pm}\bigl(x^0;x^1;\theta\bigr)C_{\epsilon}^{\pm}\bigl(x^1,\theta\bigr)
\nonumber\\
 &- \exp\left\{\frac{Q}{2} \varkappa_{\text{\sc l}}(\theta)\right\}
 C^{\pm}_{\epsilon}(-2\pi \text{\sc l},\theta) \exp\left\{ -\frac{Q}{2}\varkappa_{\text{\sc l}}(\theta)\right\}
 T_{\text{\sc l}}^{\pm}\bigl(x^0;x^1;\theta\bigr).
\label{Tlimit8}
\end{align}
Assuming that $T_{\text{\sc l}}^{\pm}$ has the indicated limit, the crucial issue is
under what conditions the $\mathfrak{sl}(2,\mathbb{R})$ valued coefficient
in the second line vanishes for $\text{\sc l} \rightarrow \infty$. It is here that
the group theoretical results from Appendix~\ref{appb2} enter.

Since $Q$ is Lie algebra valued, we may expand it
according to $Q= q_0 \tau_0 + q_1 \tau_1 + q_2 \tau_2$,
where the $q_0$, $q_1$, $q_2$ are now given as integrals of suitable
combinations of the components of~\eqref{JHdef}. We thus apply
the terminology coined in~\eqref{sl2c} to $Q$ itself,
\begin{equation*}
\frac{1}{2} \operatorname{tr} Q^2 =
 \begin{cases} >0 ,& \mbox{boost-like},\\
 =0, & \mbox{null}, \\
 < 0 ,& \mbox{rotation-like}. \end{cases}
\end{equation*}
Further, the exponentials occurring in~\eqref{Tlimit8} can be evaluated
using~\eqref{sl2g} for $\alpha \propto \ln \text{\sc l}$. Since~$C_{\epsilon}^{\pm}(-2\pi \text{\sc l}, \theta)$ is Lie algebra, valued the adjoint action
of the group element will produce another element of the
Lie algebra, and we seek to determine its properties for large $\text{\sc l}$.
In terms of $\alpha_{\pm}$, $\beta_{\pm}$ from~\eqref{Tlimit2}, one has
\begin{equation*}
C_{\epsilon}^{\pm}\bigl(x^1,\theta\bigr) = - \frac{1}{2\theta}
\left[ \frac{\epsilon}{n} J_1^H + \epsilon^1 J_0^H \right] +
\alpha_{\pm}(\theta)
\left[ \frac{\epsilon}{n} J_1^H + \epsilon^1 J_0^H \right]
+ \beta_{\pm}(\theta) \frac{1}{2\rho}
\left[ \epsilon J_0^H + \epsilon^1 \frac{J_1^H}{n} \right].
\end{equation*}
For $x^1 = -2 \pi \text{\sc l}$, the terms in square brackets can be
identified with their values at $x^1 = 0$. The
\begin{gather*}
\alpha_{\pm}\bigl(x^0,-2\pi\text{\sc l},\theta\bigr) - 1/(2\theta) =
\alpha_{\pm}\bigl(x^0,0,\theta + \text{\sc l} \pi_0^{\sigma}\bigr) -1/\bigl(2\bigl(\theta + L \pi_0^{\sigma}\bigr)\bigr),\\
\beta_{\pm}\bigl(x^0,-2\pi\text{\sc l},\theta\bigr) = \beta_{\pm}\bigl(x^0,0,\theta + \text{\sc l} \pi_0^{\sigma}\bigr)
\end{gather*}
terms will have the previously utilized decay properties if $\pm \operatorname{Re}\bigl(\theta +
\tilde{\rho}\bigl(x^0,-2\pi \text{\sc l}\bigr)\bigr) = \pm \operatorname{Re}\bigl(\theta + \tilde{\rho}\bigl(x^0,0\bigr) +
\text{\sc l} \pi_0^{\sigma}\bigr) >0$, which for large $\text{\sc l}$ requires $\pm \pi_0^{\sigma} >0$.
Assuming this, the $\alpha_{\pm}\bigl(\theta+ L \pi_0^{\sigma}\bigr)
-1/\bigl(2\bigl(\theta + \text{\sc l}\pi_0^{\sigma}\bigr)\bigr)$, $\beta_{\pm}\bigl(\theta+ \text{\sc l} \pi_0^{\sigma}\bigr)$ terms
will decay like $O(1/\text{\sc l})$, $O\bigl(1/\text{\sc l}^2\bigr)$, respectively, for large $\text{\sc l}$.
In the rotation-like case, one therefore has
\begin{gather}
\lim_{\text{\sc l} \rightarrow \infty} \exp\left\{\frac{Q}{2} \varkappa_{\text{\sc l}}(\theta) \right\}
C_{\epsilon}^{\pm}(-2\pi \text{\sc l},\theta) \exp\left\{ -\frac{Q}{2} \varkappa_{\text{\sc l}}(\theta)\right\}
=0, \qquad \pm \pi_0^{\sigma} >0.
\label{Tlimit11}
\end{gather}
This extends to the null case where the exponentials reduce to $\mathbbm{1} \pm \varkappa_\text{\sc l}(\theta) Q/2$ and the decay of the quantity under the limit~\eqref{Tlimit11} is order $O\bigl((\ln\text{\sc l})^2/\text{\sc l}\bigr)$.

The boost-like case is more tricky. Using the first line
of~\eqref{sl2g} and~\eqref{aux4}, one finds
\begin{gather}
 \exp\left\{\frac{Q}{2} \varkappa_{\text{\sc l}}(\theta)\right\}
C_{\epsilon}^{\pm}(-2\pi \text{\sc l},\theta) \exp\left\{ -\frac{Q}{2} \varkappa_{\text{\sc l}}(\theta)\right\}
\label{Tlimit12}\\
 \qquad
= \exp\left\{\left(\ln \text{\sc l} - \psi\left(1 + \frac{\theta}{\pi_0^{\sigma}}\right)
\right)
\frac{\sqrt{q \cdot q}}{|\pi_0^{\sigma}|} \right\}
P_{\pm} C_{\epsilon}^{\pm}(- 2 \pi \text{\sc l}, \theta) P_{\mp}
+ O(1/\text{\sc l}), \qquad  \pm \pi_0^{\sigma} >0.\nonumber
\end{gather}
Since the $P_{\pm}$ are orthogonal projectors, the matrix
$P_{\pm} C_{\epsilon}^{\pm}(- 2 \pi \text{\sc l}, \theta) P_{\mp}$ is nilpotent. As such
it has rank $1$ and is of the form
\begin{gather}
P_{\pm} C_{\epsilon}^{\pm}(- 2 \pi \text{\sc l}, \theta) P_{\mp} =
\begin{pmatrix} \nu_{1,\epsilon} & \phantom{-}\nu_{2,\epsilon} \\
 \nu_{3,\epsilon} & - \nu_{1,\epsilon}\end{pmatrix},
 \qquad \nu_{2,\epsilon} \nu_{3,\epsilon} = - \nu_{1,\epsilon}^2.
\label{Tlimit13}
\end{gather}
Nevertheless, the matrix will in general not be the zero matrix. A~sufficient condition for the displayed term in~\eqref{Tlimit12}
to vanish as $\text{\sc l} \rightarrow \infty$ is
that $0 < \sqrt{q\cdot q}/|\pi_0^{\sigma}| < 1$, i.e., \begin{gather}
0 < \operatorname{tr} Q^2 < 2 \bigl(\pi_0^{\sigma}\bigr)^2.
\label{Tlimit14}
\end{gather}
This establishes the Result~\ref{resultB1}\,(c).

\begin{Remark}\quad
\begin{enumerate}\itemsep=0pt
\item[(i)] The AVD property for classical solutions is known to constrain
the energy density on the reduced phase space, see~\eqref{qprop2}.
However, even on-shell~\eqref{Tlimit14} is related to
some-such restriction only in a~special case. Below,
we therefore aim at a~result valid without the constraint~\eqref{Tlimit14}.
\item[(ii)] In the abelian case no restriction is needed. The off-shell
version of the transition matrix reads
\begin{equation*}
t^{H,\pm}\bigl(x^0;x^1,y^1;\theta\bigr) = \exp\left\{ \int_{y^1}^{x^1} {\rm d}x
\ell_1^{H,\pm}\bigl(x^0,x;\theta\bigr)\right\},
\end{equation*}
where $\ell_1^{H,\pm}$ is defined as $L_1^{H,\pm}$ in~\eqref{LHdef} with
the replacements, $J_0^H \leadsto \jmath_0^H := \pi^{\phi}$,
$J_1^H \leadsto \jmath_1^H := \rho n \partial_1 \phi$. The Noether charge
is \smash{$Q = \int_0^{2 \pi} {\rm d}x \pi^{\phi}$} and generates
infinitesimal constant shifts in $\phi \mapsto \phi + \alpha$.
This suggests to define $u_{\pm}$ as
\begin{gather}
u_{\pm}\bigl(x^0,x^1;\theta\bigr) = \lim_{\text{\sc l} \rightarrow \infty}
\exp\left\{ \frac{Q}{2} \varkappa_{\text{\sc l}}(\theta)\right\}
t^{H,\pm}\bigl(x^0;x^1;- 2\pi \text{\sc l};\theta\bigr).
\label{tlimit2}
\end{gather}
Denoting the quantity under the limit by $t_{\text{\sc l}}^{\pm}\bigl(x^0,x^1;\theta\bigr)$, the
gauge variation can be computed along the lines of
\eqref{THsol5}, \eqref{Tlimit8}. The result is
\begin{equation*}
\delta_{\epsilon}^H t_{\text{\sc l}}^{\pm}\bigl(x^0;x^1;\theta\bigr) =
\bigl[c_{\epsilon}\bigl(x^1,\theta\bigr) - c_{\epsilon}(-2\pi \text{\sc l},\theta)\bigr]
t_{\text{\sc l}}^{\pm}\bigl(x^0;x^1;\theta\bigr),
\end{equation*}
with $c_{\epsilon}^{\pm}\bigl(x^1,\theta\bigr) = (\epsilon/n) \ell_0^{H,\pm} +
\epsilon^1 \ell_1^{H,\pm}$ and
$\delta_{\epsilon}^H \ell_1^{H,\pm} = \partial_1 c_{\epsilon}^{\pm}$. Hence the
limit $\text{\sc l} \rightarrow \infty$
can be taken without problems and produces the desired gauge variation of
$u_{\pm}$ in~\eqref{tlimit2}, i.e.,~$\delta_{\epsilon}^H u_{\pm} = c_{\epsilon}^{\pm}
u_{\pm}$.
\item[(iii)] So far, the analysis was off-shell and without gauge fixing. As seen
in the $\mathbb{R} \times T^2$ system, the $\epsilon =n$, $\epsilon^1 =0$ specialized gauge
variations consistently generate the time evolution for~$\mathcal{O}_T$. In the $T^3$ system, this should be re-examined using the regularized current~\eqref{Jregdef}.
From~\eqref{Tlimit8}, one expects $T_{\text{\sc l}}^{\pm}$'s on-shell time
evolution equation to be
\begin{gather}
e_0\bigl(T^{\pm}_{\text{\sc l}}\bigl(x^0;x^1;\theta\bigr)\bigr) \nonumber\\
 \qquad= T_{\text{\sc l}}^{\pm}\bigl(x^0;x^1;\theta\bigr)
L_0^{H,\pm}\bigl(x^1,\theta\bigr)\nonumber\\
 \qquad\phantom{=}{}
- \exp\left\{\frac{Q}{2} \varkappa_{\text{\sc l}}(\theta)\right\}
 L_0^{H,\pm}(-2\pi \text{\sc l},\theta) \exp\left\{ -\frac{Q}{2} \varkappa_{\text{\sc l}}(\theta)\right\}
 T_{\text{\sc l}}^{\pm}\bigl(x^0;x^1;\theta\bigr).\label{TNevol}
\end{gather}
This can indeed be derived directly along the by now familiar lines:
omitting the $\pm$ superscripts momentarily, we define
$A_{\text{\sc l}} := e_0(T_{\text{\sc l}}) - T_{\text{\sc l}} L_0^H\bigl(x^1\bigr)$ and compute
$\partial_1 A_{\text{\sc l}} = A_{\text{\sc l}} L_1^H$, using~${\partial_1 e_0(T_{\text{\sc l}}) = e_0(\partial_1 T_{\text{\sc l}})}$
and $e_0\bigl(L_1^H\bigr) - \partial_1 L_0^H + \bigl[L_0^H, L_1^H\bigr] =0$. Next, this is
re-integrated with boundary condition~${A_{\text{\sc l}}\bigl(x^0,x^1 = - 2\pi \text{\sc l};\theta\bigr) =
- \exp\{ Q \varkappa_{\text{\sc l}}(\theta)/2\} L_0^H(-2\pi \text{\sc l},\theta)}$, where
\linebreak $e_0(Q) =0 =
e_0\bigl(\pi_0^{\sigma}\bigr)$ enters. This fixes $A_{\text{\sc l}}$ to equal
the second line of~\eqref{TNevol} and by $A_{\text{\sc l}}$'s definition yields~\eqref{TNevol}. Concerning the~${\text{\sc l} \rightarrow \infty}$ limit,
one encounters once more~\eqref{Tlimit11} in the rotation-like
and~\eqref{Tlimit12}, \eqref{Tlimit14} in the boost-like case.
This is because $L_0^H(-2\pi \text{\sc l}, \theta)$ will decay like $O(1/\text{\sc l})$
for large $\text{\sc l}$, with coefficient proportional to $J_1^H\bigl(x^0,0\bigr)$,
see~\eqref{LHpm2}.
\end{enumerate}
\end{Remark}

\subsection[SL(2,R) orbits and projectors]{$\boldsymbol{{\rm SL}(2,\mathbb{R})}$ orbits and projectors}\label{appb2}

Recall the standard basis of $\mathfrak{sl}(2,\mathbb{R})$
$
\tau_0 = \bigl(\begin{smallmatrix} 0 & 1\\ -1 & 0 \end{smallmatrix}\bigr)$, $
\tau_1 = \bigl(\begin{smallmatrix} 1 & 0\\ 0 & -1 \end{smallmatrix}\bigr)$, $
\tau_2 = \bigl(\begin{smallmatrix} 0 & 1\\ 1 & 0 \end{smallmatrix}\bigr)$.
Upon exponentiation~$\tau_0$,~$\tau_1$, $\tau_2$ generate, respectively,
rotations, rescalings, and boosts.
\begin{gather}
q_0 \tau_0 + q_1 \tau_1 + q_2 \tau_2 =
\begin{pmatrix} q_1 & q_0 + q_2 \\
 - q_0 + q_2 & - q_1
\end{pmatrix}.
\label{sl2b}
\end{gather}
Its exponential will involve the Lorentzian norm
\begin{gather}
q \cdot q :=
\frac{1}{2} \operatorname{tr} (q_0 \tau_0 + q_1 \tau_1 + q_2 \tau_2)^2
= - q_0^2 + q_1^2 + q_2^2 =
\begin{cases} >0, & \mbox{boost-like},\\
=0 ,& \mbox{null},\\
 < 0, & \mbox{rotation-like}. \end{cases}
\label{sl2c}
\end{gather}
One finds in the boost-like case
\begin{gather*}
 \exp \begin{pmatrix} q_1 & q_0 + q_2 \\
 - q_0 + q_2 & - q_1
\end{pmatrix}\\
\qquad =
\cosh \sqrt{q\cdot q} \begin{pmatrix} 1 & 0 \\ 0 & 1 \end{pmatrix}
+ \frac{\sinh \sqrt{q \cdot q}}{\sqrt{q \cdot q}}
\begin{pmatrix} q_1 & q_0 + q_2 \\
 - q_0 + q_2 & - q_1
\end{pmatrix} = {\rm e}^{\sqrt{q \cdot q}} P_+ + {\rm e}^{-\sqrt{q \cdot q}} P_-,
 \\
P_{\pm} := \frac{1}{2} \begin{pmatrix} 1 \pm \frac{q_1}{\sqrt{q\cdot q}} &
 \pm \frac{q_0 + q_2}{\sqrt{ q \cdot q}} \\
 \pm \frac{-q_0 + q_2}{\sqrt{ q \cdot q}} &
 1 \mp \frac{q_1}{\sqrt{q\cdot q}} \end{pmatrix}.
\end{gather*}
In the rotation-like case one has similarly
\begin{gather*}
 \exp \begin{pmatrix} q_1 & q_0 + q_2 \\
 - q_0 + q_2 & - q_1
\end{pmatrix}\\
 \qquad=
\cos \sqrt{-q\cdot q} \begin{pmatrix} 1 & 0 \\ 0 & 1 \end{pmatrix}
+ \frac{\sin \sqrt{-q \cdot q}}{\sqrt{-q \cdot q}}
\begin{pmatrix} q_1 & q_0 + q_2 \\
 - q_0 + q_2 & - q_1
\end{pmatrix}
 = {\rm e}^{{\rm i}\sqrt{-q \cdot q}} P_+ + {\rm e}^{-{\rm i}\sqrt{-q \cdot q}} P_-,
 \\
P_{\pm} := \frac{1}{2} \begin{pmatrix} 1 \pm \frac{q_1}{{\rm i}\sqrt{-q\cdot q}} &
 \pm \frac{q_0 + q_2}{{\rm i}\sqrt{-q \cdot q}} \\
 \pm \frac{-q_0 + q_2}{{\rm i}\sqrt{-q \cdot q}} &
 1 \mp \frac{q_1}{{\rm i}\sqrt{-q\cdot q}} \end{pmatrix}.
\end{gather*}
As suggested by the notation, the $P_{\pm}$ matrices are orthogonal projectors
in both cases
$
P_{\pm}^2 = P_{\pm}$, $P_+ P_- = 0$, $ P_+ + P_- =\mathbbm{1}$.
 Including the null case, one has for all $\alpha \in \mathbb{R}$
\begin{gather}
\exp \alpha \begin{pmatrix} q_1 & q_0 + q_2 \\
 - q_0 + q_2 & - q_1
\end{pmatrix} =
\begin{cases}
 {\rm e}^{\alpha \sqrt{q \cdot q}} P_+ + {\rm e}^{-\alpha \sqrt{q \cdot q}} P_-, & \mbox{boost-like}, \\
 \mathbbm{1} + \alpha \begin{pmatrix} q_1 & q_0 + q_2 \\
 - q_0 + q_2 & - q_1
\end{pmatrix}, & \mbox{null}, \\
{\rm e}^{{\rm i} \alpha \sqrt{-q \cdot q}} P_+ + {\rm e}^{-{\rm i} \alpha \sqrt{-q \cdot q}} P_-, & \mbox{rotation-like}.
\end{cases}
\label{sl2g}
\end{gather}
Initially this follows for $\alpha>0$ as the $P_\pm$ are invariant under constant rescalings $q_j \mapsto \alpha q_j$, $j =0,1,2$. It extends to $\alpha<0$ because the inverse matrices have the roles of $P_+$ and $P_-$ interchanged.
Note that for $\alpha \rightarrow \pm \infty$ the $P_{\pm}$ term dominates,
respectively. Later on, $\alpha \propto \ln \text{\sc l}$ will be taken large.

In the boost-like case, further analysis is needed, for which
we consider the maps
\begin{gather}
\mathrm{Nil}_{\pm}\colon\ \mathfrak{sl}(2,\mathbb{R})
\rightarrow \mathrm{Nil}(2,\mathbb{R}), \qquad
a \mapsto P_{\pm} a P_{\mp},
\label{nil1}
\end{gather}
where $\mathrm{Nil}(2,\mathbb{R})$ is the set of nilpotent real $2\times 2$
matrices. The general such matrix is parameterized by two real
numbers via
\begin{gather}
\nu \in \mathrm{Nil}(2,\mathbb{R}) \mathrm{iff}
\nu = \begin{pmatrix} \nu_1 & \phantom{-}\nu_2 \\
 \nu_3 & - \nu_1\end{pmatrix},
 \qquad \nu_2 \nu_3 = - \nu_1^2.
\label{nil2}
\end{gather}
Using this, one finds that $\mathrm{Nil}(2,\mathbb{R})$ is not a~vector space.
Only the subset, where $\nu_1/\nu_2$
(and hence~$\nu_1/\nu_3)$ is held fixed forms a~one-dimensional
vector space, which turns out to coincide with the
range of $\mathrm{Nil}$. In particular,
$
\dim \mathrm{Ker} \mathrm{Nil}_{\pm} =2$, $
\dim \mathrm{Ran} \mathrm{Nil}_{\pm} =1$,
which can be seen in several ways. Since this also provides
an explicit description of the range, we proceed by diagonalizing
the initial matrix~\eqref{sl2b}. Set
\begin{equation*}
C_q:= \begin{pmatrix} \gamma(q_0-q_2) & \gamma \bigl(q_1 - \sqrt{q\cdot q}\bigr) \\
 \check{\gamma} \bigl(q_1 - \sqrt{q\cdot q}\bigr)& \check{\gamma}(q_0+q_2) \end{pmatrix}
, \qquad
\gamma\check{\gamma} = \frac{1}{2 \sqrt{q \cdot q}\bigl( q_1 - \sqrt{q \cdot q}\bigr)},
\end{equation*}
where the condition on $\gamma \check{\gamma}$ is only needed to ensure
$\det C_q =1$. Then
\begin{equation*}
C_q \begin{pmatrix} q_1 & q_0 + q_2 \\ - q_0 + q_2 & -q_1 \end{pmatrix}
C_q^{-1} = \begin{pmatrix} \sqrt{q \cdot q} & 0 \\ 0 & - \sqrt{q \cdot q}
\end{pmatrix}.
\end{equation*} A~similar result was formulated in~\cite[Lemma 8.2]{RingstroemWave}
for different purposes. The projectors simplify according to
\begin{equation*}
C_q P_+ C_q^{-1} = \begin{pmatrix} 1 & 0 \\ 0 & 0 \end{pmatrix}, \qquad
C_q P_- C_q^{-1} = \begin{pmatrix} 0 & 0 \\ 0 & 1 \end{pmatrix}.
\end{equation*}
Thus
\begin{align}
& C_q \mathrm{Nil}_+(a) C_q^{-1} =
\begin{pmatrix} 1 & 0 \\ 0 & 0 \end{pmatrix} C_q
\begin{pmatrix} a_1 & a_2 \\ a_3 & - a_1 \end{pmatrix}
C_q^{-1}
\begin{pmatrix} 0 & 0 \\ 0 & 1 \end{pmatrix} =
\begin{pmatrix} 0 & \frac{\gamma}{\check{\gamma}}
 b_{12} \\ 0 & 0 \end{pmatrix},
\nonumber\\
& C_q \mathrm{Nil}_-(a) C_q^{-1} =
\begin{pmatrix} 0 & 0 \\ 0 & 1 \end{pmatrix} C_q
\begin{pmatrix} a_1 & a_2 \\ a_3 & - a_1 \end{pmatrix}
C_q^{-1}
\begin{pmatrix} 1 & 0 \\ 0 & 0 \end{pmatrix} =
\begin{pmatrix} 0 & 0 \\
 \frac{\check{\gamma}}{\gamma}
 b_{21} & 0 \end{pmatrix},
\label{sim5}
\end{align}
where
\begin{align*}
& b_{12} := \frac{\check{\gamma}}{\gamma}
\bigl[C_q a C_q^{-1}\bigr]_{12} =
-\frac{q_0 - q_2}{2 \sqrt{q\cdot q}}
 \left( 2 a_1 - a_2 \frac{q_0 - q_2}{q_1 - \sqrt{q\cdot q}}
+ a_3 \frac{q_1 - \sqrt{q\cdot q}}{q_0 - q_2} \right),\\
& b_{21} := \frac{\gamma}{\check{\gamma}}
\bigl[C_q a C_q^{-1}\bigr]_{21} =
\frac{q_0 + q_2}{2 \sqrt{q\cdot q}}
\left( 2 a_1 - a_2
\frac{q_1 - \sqrt{q\cdot q}}{q_0 + q_2}
+ a_3 \frac{q_0 + q_2}{q_1 - \sqrt{q\cdot q}} \right).
\end{align*}
The range of $\mathrm{Nil}_{\pm}$ is therefore proportional to a~specific linear combination of the matrix elements~${(a_1,a_2,a_3)}$ of $a \in \mathfrak{sl}(2,\mathbb{R})$. Explicitly,
\begin{gather}
\mathrm{Nil}_+(a) = C_q^{-1} \begin{pmatrix} 0 & \frac{\gamma}{\check{\gamma}}
 b_{12} \\ 0 & 0 \end{pmatrix} C_q
= b_{12}\frac{q_0+q_2} {2 \sqrt{q \cdot q}}
\begin{pmatrix} 1 & \dfrac{q_0 + q_2}{q_1 - \sqrt{q\cdot q}}
\\
- \dfrac{q_1 - \sqrt{q\cdot q}}{q_0 + q_2} & -1
\end{pmatrix},
\\
\mathrm{Nil}_-(a) = C_q^{-1} \begin{pmatrix} 0 & 0
\\ \frac{\check{\gamma}}{\gamma} b_{21} & 0 & \end{pmatrix} C_q
= -b_{21}\frac{q_0-q_2} {2 \sqrt{q \cdot q}}
\begin{pmatrix} 1 &
- \dfrac{q_1 - \sqrt{q\cdot q}}{q_0 - q_2}
\\
 - \dfrac{q_0 - q_2}{q_1 - \sqrt{q\cdot q}} &-1
\end{pmatrix}.\label{sim6}
\end{gather}
The kernel is described by the two-dimensional subspace of $\mathfrak{sl}(2,\mathbb{R})$
parameterized by $(a_1,a_2,a_3)$ satisfying $b_{12} =0$ and $b_{21} =0$,
respectively. Comparing~\eqref{sim6} with~\eqref{nil2}, one sees that the
range is precisely of that form with $\nu_1 = b_{12}
(q_0 + q_2)/\bigl(2 \sqrt{q\cdot q}\bigr)$ and $\nu_1 = -b_{21}
(q_0 - q_2)/\bigl(2 \sqrt{q\cdot q}\bigr)$, respectively, while the
ratios $\nu_2/\nu_1$, $\nu_3/\nu_1$ are parameter-free combinations of the
$q_0$, $q_1$, $q_2$. This provides an alternative
description of the range of $\mathrm{Nil}_{\pm}$ as a~subspace of
$\mathrm{Nil}(2,\mathbb{R})$.
Later on, we shall encounter the related problem of finding
the $\nu \in \mathrm{Nil}(2,\mathbb{R})$ matrices that solve
\begin{gather}
[P_{\pm} c P_{\mp}, \nu] \stackrel{\displaystyle{!}}{=} 0
 \qquad \mbox{for all} \quad c \in \mathfrak{sl}(2,\mathbb{R}).
\label{nil4}
\end{gather}
By direct inspection, one sees that there can be no solutions
that are not nilpotent. Clearly, elements $\nu = P_{\pm} a P_{\mp}$ in
the range of $\mathrm{Nil}_{\pm}$ are solutions and since the solutions
form a~vector space these must be all solutions.

\subsection{Results for the current and its gauge variation}

We proceed with deriving Result~\ref{resultJtheta} for the current.
Beginning with part (a), recall the definition~\eqref{Jregdef} of the
regularized current
components in terms of $T_{\text{\sc l}}^{\pm}$, i.e., $\smash{\mathcal{J}_{\mu}^{H,\pm}(\theta)_{\text{\sc l}}}
:=\smash{ T_{\text{\sc l}}^{\pm}(\theta) K_{\mu}^{H,\pm}(\theta) T_{\text{\sc l}}^{\pm}(\theta)^{-1}}$, $\mu =0,1$.
As seen before, $T_{\text{\sc l}}^{\pm}$ has a~well-defined limit as $\text{\sc l} \rightarrow \infty$,
so the limits
\begin{equation*}
\mathcal{J}_{\mu}^{H,\pm}(\theta) = \lim_{\text{\sc l} \rightarrow \infty}\mathcal{J}_{\mu}^{H,\pm}(\theta)_{\text{\sc l}} =
U_{\pm}(\theta) K_{\mu}^{H,\pm}(\theta) U_{\pm}(\theta)^{-1},
 \qquad \pm \operatorname{Re}(\theta + \tilde{\rho}) >0,
\end{equation*}
exist pointwise, initially for $x^1 \in [0,2\pi]$. Using the
quasi-periodicity~\eqref{quasiperB1} of
$U_\pm$ the analogous property~\eqref{Jquasiper} of \smash{$\mathcal{J}_{\mu}^{H,\pm}(\theta)$}
follows, allowing one to extend \smash{$\mathcal{J}_{\mu}^{H,\pm}(\theta)$} to other
$x^1$ intervals. In order to determine the gauge variation of
\smash{$\mathcal{J}_0^{H,\pm}(\theta)$}, we return to the regularized version,
\smash{$\mathcal{J}_{\mu}^{H,\pm}(\theta)_{\text{\sc l}}$}.
Repeating the computation leading to~\eqref{JHthetaresult}
and taking into account the boundary condition~${T_{\text{\sc l}}^{\pm}\bigl(x^0,-2 \pi \text{\sc l},\theta\bigr)
= \exp\{ Q \varkappa_{\text{\sc l}}(\theta)/2\}}$, one finds
\begin{align}
\delta_{\epsilon}^H \mathcal{J}_0^{H,\pm}(\theta)_{\text{\sc l}} ={}&
\partial_1\left( \frac{\epsilon}{n} \mathcal{J}_1^{H,\pm}(\theta)_{\text{\sc l}} +
\epsilon^1 \mathcal{J}_0^{H,\pm}(\theta)_{\text{\sc l}} \right)
\nonumber\\
& + \left[ \mathcal{J}_0^{H,\pm}(\theta)_{\text{\sc l}},
\exp\left\{\frac{Q}{2} \varkappa_{\text{\sc l}}(\theta) \right\}
C_{\epsilon}^{\pm}(-2\pi \text{\sc l},\theta) \exp\left\{ -\frac{Q}{2} \kappa_{\text{\sc l}}(\theta)\right\}
\right].
\label{JHNthetavar}
\end{align}
Assuming $\pm \operatorname{Re}(\theta + \tilde{\rho})>0$, $\pm \pi_0^{\sigma} >0$,
the analysis of the $\text{\sc l} \rightarrow \infty$ limit is similar as before.
For rotation-like $Q$, the second line in~\eqref{JHNthetavar}
always vanishes on account of~\eqref{Tlimit11}. In the boost-like
case one might hope that the commutator produces a~slightly weaker
vanishing condition than~\eqref{Tlimit14}. However, this is not the case:
using~\eqref{Tlimit12} the coefficient of $\exp\{ \bigl(\ln \text{\sc l}
- \psi\bigl(1 + \theta/\pi_0^{\sigma}\bigr)\bigr)\sqrt{q\cdot q} /|\pi_0^{\sigma}|\}$
is $\bigl[ \mathcal{J}_0^H(\theta)_{\text{\sc l}}, P_{\pm} C_{\epsilon}^{\pm}(- 2 \pi \text{\sc l}, \theta)
 P_{\mp}\bigr]$. Even
with one of the terms of the form~\eqref{Tlimit13} this commutator
does in general not vanish on algebraic grounds, and one is led back
to~\eqref{Tlimit14}. This yields Result~\ref{resultJtheta}\,(b).

For part (c), we try to remove the undesired condition on $\operatorname{tr} Q^2$
by considering the trace of~\eqref{JHNthetavar} with a~suitable matrix
$\nu$. For this to work, $\nu$ must be spatially constant
and separately gauge invariant, $\delta_{\epsilon}^H \nu =0$.
The basic variational equation~\eqref{JHNthetavar}
then carries over to the trace just with the second line
replaced by
\begin{gather*}
 \operatorname{tr} \left\{ \mathcal{J}_0^{H,\pm}(\theta)_{\text{\sc l}} \left[
\exp\left\{\frac{Q}{2} \varkappa_{\text{\sc l}}(\theta)\right\}
C_{\epsilon}^{\pm}(-2\pi \text{\sc l},\theta) \exp\left\{ -\frac{Q}{2} \varkappa_{\text{\sc l}}(\theta)\right\}
, \nu \right]\right\}\\
 \qquad
= \exp\left\{\left(\ln \text{\sc l} - \psi\left(1 + \frac{\theta}{\pi_0^{\sigma}}\right)
\right) \frac{\sqrt{q \cdot q}}{|\pi_0^{\sigma}|} \right\}
 \operatorname{tr} \bigl\{ \mathcal{J}_0^{H,\pm}(\theta)_{\text{\sc l}}
\big[ P_{\pm} C_{\epsilon}^{\pm}(- 2\pi \text{\sc l}) P_{\mp}, \nu \big] \bigr\} + O(1/\text{\sc l}).
\end{gather*}
Since we cannot assume \smash{$\mathcal{J}_0^{H,\pm}(\theta)_{\text{\sc l}}$} to have a~specific
$\mathfrak{sl}(2,\mathbb{R})$ matrix structure and $C_{\epsilon}^{\pm}(- 2\pi \text{\sc l},\theta)$ must
likewise be allowed
to be generic, the only possibility for the trace to vanish identically
is when $\nu$ is such that $[P_{\pm} c P_{\mp}, \nu] =0$, for
all $c \in \mathfrak{sl}(2,\mathbb{R})$. In this situation, the condition~${0 < \sqrt{q \cdot q}/|\pi_0^{\sigma}| < 1}$ is not needed in order to obtain a~well-defined $\text{\sc l} \rightarrow \infty$ limit of the~${\operatorname{tr}\{\nu (\cdots)\}}$
version of~\eqref{JHNthetavar}. The solution of the relevant condition
\eqref{nil4} has been prepared earlier and was found to consist
precisely of matrices of the form $\nu = P_{\pm} a P_{\mp}$, with
generic~${a \in \mathfrak{sl}(2,\mathbb{R})}$. At first sight this seems to produce a~three-parameter family of solutions. Closer examination shows, however,
that two parameters' worth just produce $\nu =0$ solutions
and only a~single parameter combination yields a~nontrivial
$0 \neq \nu_{\pm} = P_{\pm} a P_{\mp}$, see~\eqref{sim5}, \eqref{sim6}.
Taking $\nu$ to be of this form, one has
\begin{gather}
\delta_{\epsilon}^H \operatorname{tr}\bigl\{ \nu_{\pm} \mathcal{J}_0^{H,\pm}(\theta)_{\text{\sc l}} \bigr\} =
\partial_1\left( \frac{\epsilon}{n}
 \operatorname{tr}\bigl\{\nu_{\pm} \mathcal{J}_1^{H,\pm}(\theta)_{\text{\sc l}} \bigr\}+ \epsilon^1
 \operatorname{tr} \bigl\{ \nu_{\pm} \mathcal{J}_0^{H,\pm}(\theta)_{\text{\sc l}}\bigr\} \right) + O(1/\text{\sc l}).
\label{traceJ2}
\end{gather}
Explicitly, $\nu_{\pm} = \mathrm{Nil}_{\pm}(a)$ is of the form~\eqref{sim6}
with $b_{12}$, $b_{21}$ from~\eqref{sim5}. Since the projectors~$P_{\pm}$
are constructed from the original $Q$'s components, they are gauge
invariant as well, and so is~${\nu_\pm = \mathrm{Nil_\pm}(a)}$, as required.
Taking the $\text{\sc l} \rightarrow \infty$ limit in~\eqref{traceJ2} gives
the Result~\ref{resultJtheta}\,(c).

\textbf{$\boldsymbol{\mathrm{\operatorname{tr}\bigl\{\nu J_0^H(\theta)_{\text{\bf \sc l}}\bigr\}}}$ gauge variation and global $\boldsymbol{\mathrm{SL}(2,\mathbb{R})}$ transformations.}
The result~\eqref{traceJ2} can alternatively be understood
in terms of global $\mathrm{SL}(2,\mathbb{R})$ transformations originating
in the isometries of the generic Gowdy metric~\eqref{2Kmetric}.
For simplicity, we consider only the $+$ case in the following;
the $-$ case is analogous.
Recall that the matrix $M$ in~\eqref{Mmat1} transforms according to
$M \mapsto C M C^{\mathsf T}$, $C \in \mathrm{SL}(2,\mathbb{R})$. The basic currents
$J_0^H$, $J_1^H$ are thus subjected to a~similarity transformation
\begin{gather}
J_0^H \mapsto C J_0^H C^{-1}, \qquad
J_1^H \mapsto C J_1^H C^{-1},
\label{sim1}
\end{gather}
with an $x^1$-independent gauge invariant $\mathrm{SL}(2,\mathbb{R})$ valued matrix,
$\delta_{\epsilon}^H C =0$. Then, $Q \mapsto C Q C^{-1}$ and
$L_{\mu}^{H,+} \mapsto C L_{\mu}^{H,+} C^{-1}$.
The defining relation~\eqref{Tlimit5} gives
$T_{\text{\sc l}}^+ \mapsto C T_{\text{\sc l}}^+ C^{-1}$, and the key relations
\eqref{Tlimit7}, \eqref{Tlimit8} are preserved.
Finally, from~\eqref{Jregdef} one sees
\smash{$\mathcal{J}_{\mu}^{H,+}(\theta)_{\text{\sc l}} \mapsto C \mathcal{J}_{\mu}^{H,+}(\theta)_{\text{\sc l}} C^{-1}$},
and~\eqref{JHNthetavar} transforms covariantly with
$C^+_{\epsilon}(-2\pi \text{\sc l},\theta) \mapsto C C^+_{\epsilon}(-2\pi \text{\sc l},\theta) C^{-1}$.
Similarly for~\eqref{TNevol}.

Since $C_q$ has been constructed from the components of the
original $Q$ it is gauge invariant as well,
$\delta^H_{\epsilon} C_q =0$. It is therefore a~legitimate choice
for use in~\eqref{sim1} and the subsequent chain
of transformations. In the context of our search for a~benign
$\text{\sc l} \rightarrow \infty$ limit in~\eqref{Tlimit12} and the ensued
\eqref{Tlimit8}, \eqref{JHNthetavar}, \eqref{TNevol}, the net
effect is that $Q$ can be replaced by the diagonal matrix~${Q_q = \sqrt{q\cdot q} \tau_1}$. This, of course,
does not change the sufficient decay condition~\eqref{Tlimit14}. It
does affect, however, the situation where it is necessary. Indeed,
for $Q_q = \sqrt{q\cdot q} \tau_1$ the projectors
simplify to $P_+ = (\mathbbm{1} +\tau_1)/2$, $P_- = (\mathbbm{1} - \tau_1)/2$.
If we expand
$
C_qC_{\epsilon}^+(-2 \pi \text{\sc l},\theta) C_q^{-1}
= g_{\epsilon, 0}(-2 \pi \text{\sc l}, \theta) \tau_0 +
g_{\epsilon, 1}(-2 \pi \text{\sc l},\theta) \tau_1 + g_{\epsilon, 2}(-2 \pi \text{\sc l}, \theta) \tau_2$,
the matrix entering~\eqref{Tlimit12} simplifies to
\begin{align}
P_+ C_qC^+_{\epsilon}(-2 \pi \text{\sc l},\theta) C_q^{-1} P_- &=
\begin{pmatrix} 0 & g_{\epsilon, 0} + g_{\epsilon, 2} \\ 0 & 0
\end{pmatrix}.
\label{sim8}
\end{align}
This is still of the form~\eqref{Tlimit13} but now with
$\nu_{1,\epsilon} =0= \nu_{3,\epsilon}$. Multiplying~\eqref{sim8} from the
right with $T_{\text{\sc l}}$, the second row of the resulting matrix
vanishes identically. As a~consequence, the decay
condition~\eqref{Tlimit14} is now only needed for the
upper row of the matrix components of the~$+$ version of~\eqref{Tlimit8},
i.e.,~the $(1,1)$ and $(1,2)$ components. The same applies to~\eqref{TNevol} but not to~\eqref{JHNthetavar}. This is because
multiplying~\eqref{sim8} from the left with a~generic
$\mathrm{SL}(2,\mathbb{R})$ valued matrix produces one whose first
column vanishes. After the transformation therefore only
the lower left component of the second line in the~$+$ version of~\eqref{JHNthetavar} vanishes algebraically. Nevertheless, for
the transformed currents
\begin{gather}
\delta_{\epsilon}^H \bigl[C_q \mathcal{J}_0^{H,+}(\theta)_{\text{\sc l}} C_q^{-1}\bigr]_{21}\nonumber \\
\qquad=
\partial_1\left( \frac{\epsilon}{n} \bigl[C_q \mathcal{J}_1^{H,+}(\theta)_{\text{\sc l}} C_q^{-1}\bigr]_{21} +
\epsilon^1 \bigl[C_q \mathcal{J}_0^{H,+}(\theta)_{\text{\sc l}} C_q^{-1}\bigr]_{21} \right) + O(1/\text{\sc l}),
\label{JHNthetavar2}
\end{gather}
holds without constraints on $q\cdot q$. For the two
other independent matrix components
\begin{equation*}
\bigl[C_q \mathcal{J}_0^H(\theta)_{\text{\sc l}} C_q^{-1}\bigr]_{11}, \qquad
\bigl[C_q \mathcal{J}_0^H(\theta)_{\text{\sc l}} C_q^{-1}\bigr]_{12},
\end{equation*}
the condition~\eqref{Tlimit14}
is still needed.

The relation~\eqref{JHNthetavar2} has the same structure as
\eqref{traceJ2}, and indeed both are equivalent. To see this,
we interpret the trace as
\begin{align*}
\operatorname{tr}\bigl\{ \nu_+ \mathcal{J}_0^{H,+}(\theta)_{\text{\sc l}} \bigr\} ={}&
\operatorname{tr}\bigl\{ C_q \nu_+ C_q^{-1} C_q \mathcal{J}_0^{H,+}(\theta)_{\text{\sc l}} C_q^{-1} \bigr\} =\\
={}& \operatorname{tr} \bigg\{ \begin{pmatrix} 0 &
\frac{\gamma}{\check{\gamma}} b_{12} \\ 0 & 0 \end{pmatrix}
C_q \mathcal{J}_0^{H,+}(\theta)_{\text{\sc l}} C_q^{-1} \bigg\} = \bigl[C_q a C_q^{-1}\bigr]_{12}
\bigl[ C_q \mathcal{J}_0^{H,+}(\theta)_{\text{\sc l}} C_q^{-1} \bigr]_{21},
\end{align*}
using~\eqref{sim5}. Since $\bigl[C_q a C_q^{-1}\bigr]_{12}$ is spatially
constant and separately gauge invariant, \eqref{JHNthetavar2} and~\eqref{traceJ2} are equivalent. The $-$ case runs entirely parallel.

\textbf{Periodic extension.} Finally, we derive Result~\ref{resultT3obs}.
In the periodic extension, the spatial translates of
$\operatorname{tr}\bigl\{\nu \mathcal{J}_0^H(\theta)\bigr\}$ are needed. Either directly
from~\eqref{Jquasiper} or by taking limits in~\eqref{TLquasiper},
one finds for the left and right $n \in \mathbb{N}$ translates
\begin{align*}
\mathcal{J}_{\mu}^H\bigl(x^1 + 2 \pi n;\theta\bigr) &=
{\rm e}^{ - \frac{Q}{2} \check{\varkappa}_n(\theta) }
\mathcal{J}_{\mu}^H\bigl(x^1;\theta - n \pi_0^{\sigma}\bigr)
{\rm e}^{ \frac{Q}{2} \check{\varkappa}_n(\theta) }, \qquad\\
\mathcal{J}_{\mu}^H\bigl(x^1 - 2 \pi n;\theta\bigr) &=
{\rm e}^{ \frac{Q}{2} \varkappa_n(\theta) }
\mathcal{J}_{\mu}^H\bigl(x^1;\theta + n \pi_0^{\sigma}\bigr)
{\rm e}^{ -\frac{Q}{2} \varkappa_n(\theta) }.
\end{align*}
Since $\check{\varkappa}_n(\theta)$, $\varkappa_n(\theta)$ diverge like $\ln n$ for large
$n$, the convergence of the partial sum
\begin{equation*}
\sum_{n=-N}^N \mathcal{J}_{\mu}^H\bigl(x^1 + 2 \pi n;\theta\bigr)
\end{equation*}
for $N \rightarrow \infty$
would again require a~condition on $\operatorname{tr}Q^2$. This changes if the
trace with $\nu$ is considered. As
\begin{gather}
 \sum_{n=-N}^N \operatorname{tr}\bigl\{ \nu \mathcal{J}_{\mu}^H\bigl(x^1 + 2 \pi n;\theta\bigr) \bigr\}
= \operatorname{tr}\bigl\{ \nu \mathcal{J}_{\mu}^H\bigl(x^1;\theta\bigr) \bigr\} + \sum_{n=1}^N \operatorname{tr}\bigl\{ \check{\nu}(n,\theta)
\mathcal{J}_{\mu}^H\bigl(x^1;\theta - n \pi_0^{\sigma}\bigr) \bigr\} \nonumber
\\
\phantom{ \sum_{n=-N}^N \operatorname{tr}\bigl\{ \nu \mathcal{J}_{\mu}^H\bigl(x^1 + 2 \pi n;\theta\bigr) \bigr\}
= }{}+ \sum_{n=1}^N \operatorname{tr}\bigl\{ \nu(n,\theta)
\mathcal{J}_{\mu}^H\bigl(x^1;\theta + n \pi_0^{\sigma}\bigr) \bigr\},
\nonumber\\
\check{\nu}(n,\theta) := {\rm e}^{ \frac{Q}{2} \check{\varkappa}_n(\theta) }
\nu {\rm e}^{ -\frac{Q}{2} \check{\varkappa}_n(\theta) },\qquad
\nu(n,\theta) := {\rm e}^{ -\frac{Q}{2} \varkappa_n(\theta) }
\nu {\rm e}^{\frac{Q}{2} \varkappa_n(\theta) },
\label{Jthetaper2}
\end{gather}
the decay of $\check{\nu}(n,\theta)$, $\nu(n,\theta)$ in the boost-like case
$q \cdot q >0$ is relevant, see~\eqref{sl2c}. Restoring momentarily
the $\pm$ case distinction, all quantities in~\eqref{Jthetaper2} carry
$\pm$ sub- or superscripts. In particular, we write
$\nu_{\pm} = P_{\pm} a P_{\mp}$ for the two images in~\eqref{sim6}.
Using~\eqref{sl2g}, one finds
\begin{gather}
\check{\nu}_{\pm}(n,\theta) = {\rm e}^{\pm \check{\varkappa}_n(\theta) \sqrt{q\cdot q}}
\nu_{\pm},
\nu_{\pm}(n,\theta) = {\rm e}^{\mp \varkappa_n(\theta) \sqrt{q\cdot q}} \nu_{\pm}.
\label{nun}
\end{gather}
Since $\check{\varkappa}_n(\theta) = - (\ln n)/\pi_0^{\sigma} + O\bigl(n^0\bigr)$,
$\varkappa_n(\theta) = (\ln n)/\pi_0^{\sigma} + O\bigl(n^0\bigr)$, we take
$\pm \pi_0^{\sigma} >0$ (in line with Result~\ref{resultJtheta}\,(c))
so that both quantities in~\eqref{nun}
decay like $\exp\{ - (\ln n) \sqrt{q\cdot q}/|\pi_0^{\sigma}|\}$
for large~${n \in \mathbb{N}}$. Since the
\smash{$\mathcal{J}_{\mu}^{H,\pm}\bigl(x^1, \theta - n \pi_0^{\sigma}\bigr)$} and
\smash{$\mathcal{J}_{\mu}^{H,\pm}\bigl(x^1, \theta + n \pi_0^{\sigma}\bigr)$} also decay like $1/n^2$
(irrespective of sign choices for $\operatorname{Re}(\theta + \tilde{\rho})$ or
$\pi_0^{\sigma}$), one concludes that~\eqref{Jthetaper2} has
pointwise in $\bigl(x^0,x^1;\theta\bigr)$ a~finite $N \rightarrow \infty$ limit.
This is the first part of the assertion in Result~\ref{resultT3obs}.

It remains to examine the gauge variation of the limit. A~straightforward
computation gives
\begin{gather}
 \delta_{\epsilon}^H \sum_{n=-N}^N \int_0^{2\pi} {\rm d}x
 \operatorname{tr} \bigl\{ \nu_{\pm} \mathcal{J}_0^{H,\pm}(x + 2\pi n;\theta) \bigr\}
\nonumber\\
 \qquad = \left(\frac{\epsilon}{n}\right)(0) \operatorname{tr} \bigl\{
\check{\nu}_{\pm}(N + 1,\theta)\mathcal{J}_1^{H,\pm}\bigl(0, \theta - (N + 1) \pi_0^{\sigma}\bigr)
-\nu_{\pm}(N,\theta)\mathcal{J}_1^{H,\pm}\bigl(0, \theta + N \pi_0^{\sigma}\bigr) \bigr\}
\nonumber\\
\phantom{=}{} \qquad +\epsilon^1(0) \operatorname{tr} \bigl\{
\check{\nu}_{\pm}(N + 1,\theta)\mathcal{J}_0^{H,\pm}\bigl(0, \theta - (N + 1) \pi_0^{\sigma}\bigr)\nonumber\\
\phantom{=}{}\qquad
-\nu_{\pm}(N,\theta)\mathcal{J}_0^{H,\pm}\bigl(0, \theta + N \pi_0^{\sigma}\bigr) \bigr\}.
\label{Jthetaper3}
\end{gather}
The $\mathcal{J}_{\mu}^{H,\pm}\bigl(0,\theta - (N + 1) \pi_0^{\sigma}\bigr)$ and
$\mathcal{J}_{\mu}^{H,\pm}\bigl(0,\theta + N \pi_0^{\sigma}\bigr)$ are $O\bigl(N^{-2}\bigr)$
for large $N$, irrespective of sign choices for
$\operatorname{Re}(\theta +\tilde{\rho})$ or $\pi_0^{\sigma}$. Further,
as seen after~\eqref{nun}, for $\pm \pi_0^{\sigma} >0$ the
$\check{\nu}_{\pm}(N + 1,\theta)$ and $\nu_{\pm}(N,\theta)$ decay
like $\exp\{ - (\ln N) \sqrt{q\cdot q}/|\pi_0^{\sigma}|\}$.
Hence the right-hand side~\eqref{Jthetaper3} goes to zero as $N \rightarrow \infty$.
As seen before, the left-hand side has a~finite limit and we arrive
at the gauge invariance of the periodic extension,
i.e.,~the second part of the assertion in Result~\ref{resultT3obs}.

\section{Off-shell monodromy matrix}\label{appC}

The monodromy matrix is part of the standard toolbox of flat space
integrable field theories, see, e.g.,~\cite{FT}. In the context of the
2-Killing vector reductions, its status is somewhat different and depends
on the subsector considered. For stationary axisymmetric systems, it
goes back to~\cite{BM}, see also the references in~\cite{Alekseev}.
For the $\mathbb{R} \times T^2$ Gowdy systems a~variant enters prominently
in~\cite{KSGowdyNC,KSYangian}. In particular, their monodromy matrix
evaluates to $M(t=0,\zeta=-\theta)$ in the adapted coordinates~\eqref{coords}. This is clearly not applicable in our framework as we
seek to maintain generic data, where~$M$ is singular as $t \rightarrow 0^+$.
There are also technical differences: the construction in
\cite{KSGowdyNC,KSYangian} relies on analyticity assumptions in the
``moving spectral parameter'' and the monodromy matrix is eventually
defined on the branch cut in Figure~\ref{fig1} only. Finally, the~$G/H$ coset
formalism is used in~\cite{KSGowdyNC,KSYangian} while ours is based on
the nonredundant version. In the following, we first introduce a~natural
notion of a~monodromy matrix for the Gowdy systems in the present framework
and then argue that it does, unfortunately, not lead to interesting Dirac
observables. While this is primarily a~negative result, it may be worth
recording here to guide future investigations. As usual, the discussion
is fairly different for the $\mathbb{R} \times T^2$ and $T^3$ Gowdy systems.

\textbf{Monodromy matrix in $\boldsymbol{\mathbb{R} \times T^2}$ Gowdy.}
We define the off-shell monodromy matrix by
\begin{gather}
\mathcal{M}\bigl(x^0,\theta\bigr) := U_{+}\bigl(x^0,x^1;\theta\bigr)
M\bigl(x^0,x^1\bigr) U_{-}\bigl(x^0,x^1;\theta\bigr)^{\mathsf T}, \qquad \operatorname{Im}(\theta) \neq 0.
\label{Moffdef}
\end{gather}
Here $U_{\pm}$ is defined, respectively, as the limit~\eqref{UdefNC}
of $T^{H,\pm}$ built from the $L_1^{H,\pm}$ in~\eqref{LHdef}.
As anticipated in the notation, the right-hand side is in fact
independent of $x^1$. Indeed,
\begin{gather}
\partial_1 \mathcal{M}\bigl(x^0,\theta\bigr) =0,
\label{Moffxindep}
\end{gather}
holds on purely algebraic grounds. The derivation of~\eqref{Moffxindep}
starts from the relations
\begin{gather}
\bigl[J_1^H\bigr]^{\mathsf T} = M^{-1} J_1^H M, \qquad
 \bigl[J_0^H\bigr]^{\mathsf T} = M^{-1} J_0^H M.
\label{Mderiv1}
\end{gather} A~direct way of verifying those is by noting that for an
arbitrary $\mathfrak{sl}(2,\mathbb{R})$ matrix the relevant condition has a~two-parameter solution
\begin{equation*}
\begin{pmatrix} a & \phantom{-}b \\
 c & - a \end{pmatrix}^{\mathsf T} -
M^{-1} \begin{pmatrix} a & \phantom{-}b \\
 c & - a \end{pmatrix} M =0, \qquad \mbox{for} \quad b = c\bigl(\Delta^2 + \psi^2\bigr) - 2 a \psi.
\end{equation*}
Using this for the $a$, $b$, $c$ read off from~\eqref{JHdef}, one finds
\eqref{Mderiv1}. From the form~\eqref{LHdef} of the linear system,
it is clear that these relations carry over to $L_{1}^{H,\pm}(\theta)$,
separately for the upper/lower sign. In addition, there is a~variant
where the signs are flipped
\begin{align}
M L_0^{H,\pm}(\theta)^{\mathsf T} M^{-1} &= -
L_0^{H, \mp}(\theta) - \frac{\lambda_{\text{\sc n}} n}{\rho} J_0^H,
\nonumber\\
M L_1^{H,\pm}(\theta)^{\mathsf T} M^{-1} &= -
L_1^{H, \mp}(\theta) - \frac{\lambda_{\text{\sc n}}}{n\rho} J_1^H.
\label{Mderiv3}
\end{align}
On the other hand, differentiating~\eqref{Moffdef} one finds
\smash{$L_1^{H,+} M + \partial_1 M + M \bigl(L_1^{H,-}\bigr)^{\mathsf T}$} sandwiched between $U_+$
and $U_{-}^{\mathsf T}$. Since $\lambda_{\text{\sc n}} J_1^H/(\rho n) = \partial_1 M M^{-1}$
by definition, the assertion~\eqref{Moffxindep} follows.

For the $\mathbb{R} \times T^2$ systems, one can therefore evaluate $\mathcal{M}\bigl(x^0,\theta\bigr)$
by taking the limit $x^1 \rightarrow - \infty$. On account of the boundary
conditions~\eqref{Ubc}, the value of $\mathcal{M}\bigl(x^0,\theta\bigr)$ must
coincide with the limit of $M\bigl(x^0,x^1\bigr)$ for $x^1 \rightarrow - \infty$.
For the $\mathbb{R} \times T^2$ systems, the standard fall-off conditions on the
fields~\cite{TorreRom} entail $\ln \Delta \rightarrow 0$, $\psi \rightarrow 0$,
as $\big|x^1\big| \rightarrow \infty$. Together,
\begin{gather}
\mathcal{M}\bigl(x^0,\theta\bigr) =\mathbbm{1},
\label{Mofftriv}
\end{gather}
so the otherwise natural definition~\eqref{Moffdef} does not
produce anything interesting. On the other hand, the right-hand side
of~\eqref{Moffdef} can also be evaluated for
$x^1 \rightarrow \infty$, where by Result~\ref{result0GowdyNC} the
respective $\mathcal{O}_T^{\pm}\bigl(x^0,\theta\bigr)$ occur. We conclude
\smash{$
\mathcal{O}_T^-\bigl(x^0,\theta\bigr) = \bigl[\mathcal{O}_T^+\bigl(x^0,\theta\bigr)^{\mathsf T}\bigr]^{-1}$},
where (not incidentally) the $\mathrm{SL}(2,\mathbb{R})$ involution \smash{$C \mapsto
\bigl(C^{\mathsf T}\bigr)^{-1}$} occurs. Componentwise the $\mathcal{O}^{\pm}_T\bigl(x^0,\theta\bigr)$ thus give
rise to three real-valued one-parameter families of Dirac observables.

\textbf{Monodromy matrix in $\boldsymbol{T^3}$ Gowdy.} A~candidate monodromy
matrix for the $T^3$ system is~\eqref{Moffdef} with $U_{\pm}$
read as the `renormalized' solutions from Result~\ref{resultB1}.
The computation leading to $\partial_1 \mathcal{M}\bigl(x^0,\theta\bigr) =0$ carries over,
but does not now entail~\eqref{Mofftriv}. This is because the
`renormalized'~${U_{\pm}\bigl(x^0,x^1;\theta\bigr)}$ are naturally defined
for $x^1 \in [0,2\pi]$, and can be extended to other $x^1$ intervals
via the quasi-periodicity relation in~\eqref{quasiperB1}.
However, the attempt to push $x^1$ to $-\infty$ produces a~divergent
result. Treating~\eqref{Moffdef} with $U_{\pm}$ from Result
\ref{resultB1} as nontrivial, one formally finds it to be gauge
invariant as well,
\begin{gather}
\delta_{\epsilon}^H \mathcal{M}\bigl(x^0,\theta\bigr) =0 \qquad (\mathrm{formally}).
\label{Moffvar}
\end{gather}
Indeed, assuming that both gauge variations
\[
\delta^H_\epsilon U_{\pm}\bigl(x^1,\theta\bigr)
= U_{\pm}\bigl(x^1,\theta\bigr) C^{\pm}_\epsilon\bigl(x^1,\theta\bigr)
\]
apply one obtains
\begin{equation*}
\delta_{\epsilon}^H \mathcal{M} =
U_+ \bigl\{ C_{\epsilon}^+ + \delta_{\epsilon}^H M M^{-1} + M \bigl(C_{\epsilon}^-\bigr)^{\mathsf T} M^{-1}
\bigr\} M U_-^{\mathsf T},
\end{equation*}
with the arguments omitted. The expression in curly bracket vanishes on account of~\eqref{Upm2} and~\eqref{Upm3}, giving~\eqref{Moffvar}.

Unfortunately, the `fine print' in Result~\ref{resultB1} successively
restricts the possible $\theta$ dependence as follows. First, in order
for the `renormalized' limits $U_{\pm}$ to exist one needs
$\pm \operatorname{Re}(\theta + \tilde{\rho}) >0$. For both conditions to be
met for the same $\theta$, in a~limiting sense
$\operatorname{Re}(\theta + \tilde{\rho}) =0$ must hold. Evaluating
\begin{equation*}
\beta_{\pm}(\theta) \big|_{\operatorname{Re}(\theta + \tilde{\rho}) =0} =
 \pm \operatorname{sign}( \operatorname{Im} \theta) \frac{1}{\big[ 1 + (\rho/\operatorname{Im}(\theta))^2
 \big]^{1/2} } -1,
\end{equation*}
one also needs $\pm \operatorname{Im}(\theta) >0$ for the unwanted constant terms
to cancel for large $|{\operatorname{Im}(\theta)}|$. Again, in order to meet both
conditions at the same time a~distributional limit
$\operatorname{Im}(\theta) \rightarrow 0^{\pm}$ is needed. With~${\operatorname{Re}(\theta) = -\tilde{\rho}
+ 0^{\pm}}$ also fixed not much of a~one-parameter family is left. A~second
caveat concerns the $\pm \pi_0^{\sigma} >0$ condition from
Result~\ref{resultB1}\,(c). For both to apply, $\pi_0^{\sigma} = 0^{\pm}$ is
needed, but $\pi_0^{\sigma} =0$ (in combination with the velocity
momentum relation in~\eqref{Momenta}) means that $\rho$ can no longer
be used as a~temporal function. In summary, a~well-defined limit
of~\eqref{Moffdef} based on Result~\ref{resultB1}\,(a) does not seem to
give rise to Dirac observables associated with the
$T^3$ Gowdy system proper. These problems do not arise in
Section \ref{sec3} because all constructions utilize either~$U_+$ or $U_-$,
while the monodromy matrix tries to capitalize on the simultaneous
use of both.

\subsection*{Acknowledgements}
We would like to thank Javier Peraza for discussions
and Cory Ives for participating in other aspects of this project.
We also appreciate the referees' input, in suggesting improvements
in the exposition and inquiring about several of the open problems raised
in the conclusions. This work was supported in part by the PittPACC initiative.

\pdfbookmark[1]{References}{ref}
\LastPageEnding

\end{document}